\documentclass[fleqn,3p]{elsarticle}

\usepackage{graphicx}
\usepackage{subfig}
\usepackage{color}
\usepackage{multicol}
\usepackage[author={Ali Nasseri}, final]{pdfcomment}

\usepackage{float}
\usepackage{epstopdf}
\usepackage{amsmath}
\usepackage{array}
\usepackage{placeins}
\usepackage{setspace}
\usepackage{rotating}
\usepackage{hyperref}
\usepackage{multirow}

\newcommand*{\doi}[1]{\href{http://dx.doi.org/#1}{doi: #1}}

\everymath{\rm}

\journal{Journal of Magnetism and Magnetic Materials}

\begin{document}
	
	\begin{frontmatter}

\title{Collective Coordinate Descriptions of Magnetic Domain Wall Motion in Perpendicularly Magnetized Nanostructures under the Application of In-plane Fields}



\author[ISI,Polito]{S. Ali Nasseri\corref{cor1}}
\ead{ali.nasseri@isi.it}
\author[USAL]{Eduardo Martinez}
\author[ISI, INRiM]{Gianfranco Durin}
\cortext[cor1]{Corresponding author}

\address[ISI]{ISI Foundation, Via Chisola 5 10126 Torino, Italy}
\address[Polito]{Politecnico di Torino - Corso Duca degli Abruzzi 24, 10129 Torino, Italy}
\address[USAL]{University of Salamanca, Dpto Fisica Aplicada, Plaza de los Caidos s/n. E37008, Salamanca, Spain}
\address[INRiM]{Istituto Nazionale di Ricerca Metrologica (INRiM), Strada delle Cacce, 91 10135 Torino, ITALY}

\begin{abstract}
Manipulation of magnetic domain walls can be used to improve the capabilities of the next generation of memory and sensing devices. Materials of recent interest for such devices include heterostructures of ultrathin ferromagnets sandwiched between a heavy metal and an oxide, where spin-orbit coupling and broken inversion symmetry give rise to the Dzyaloshinskii-Moriya interaction (DMI), stabilizing chiral domain walls. The efficiency of the motion of these chiral domain walls may be controlled using in-plane magnetic fields. This property has been used for measurement of DMI strength. While micromagnetic simulations are able to accurately predict domain wall motion under in-plane fields in these materials, collective coordinate models such as the $q-\phi$ and $q-\phi-\chi$ models fail to reproduce the micromagnetic results. In this theoretical work, we present a set of extended collective coordinate models including canting in the domains, which better reproduce micromagnetic results, and helps us better understand the effect of in-plane fields on magnetic domain walls. These models are used in conjunction with micromagnetic simulations to identify critical points observed in the motion of the domain walls driven by out-of-plane magnetic fields, and electric current under magnetic in-plane fields. Our new models and results help in the development of future domain wall based devices based on perpendicularly magnetized materials. 
\end{abstract}

\begin{keyword}
	Domain Wall Motion \sep Spintronics \sep Collective Coordinate Modeling
\end{keyword}

\end{frontmatter}

\section{Introduction}

Manipulating magnetic domain walls (DWs) within nanostructures has been linked with the development of spintronic logic \cite{ALL-05, THO-12}, memory \cite{CHA-07,  PAR-08, PAR-15, MOO-15} and sensing \cite{KAT-08} devices. The next generation of magnetic memory and storage devices could rely on DWs moving along magnetic tracks or wires, with different principles for such devices being explored to achieve mass storage without the need for mechanical moving parts \cite{PAR-08, PAR-15}.  

Simulation capabilities are key to better understand the underlying processes in these systems, and to assess different design concepts. The main computational framework to analyze these ferromagnetic systems is based on the Landau-Lifshitz-Gilbert (LLG) equation which is applicable to a wide range of problems in magnetism, including DW motion. However, the use of micromagnetic simulations specially for large samples is computationally costly and time consuming, as the numerical solution for the magnetic configuration needs to be determined both spatially and temporally. 

Alternatively, simpler models may be extracted from the LLG equation to analyze the motion of specific topological defects of interest, such as vortices and DWs \cite{SLO-72, THI-73, THI-73b,THI-04,THI-05,THI-06,BOU-13, VAN-16, VAN-17, NAS-17}. The simplified nature of these collective coordinate models (CCMs) is due to the introduction of an ansatz which characterizes the structure of the spin texture of interest. 

In 1972, Slonczewski used a Lagrangian approach to propose a CCM for analyzing DW motion in perpendicularly magnetized materials (the $q-\phi$ model) \cite{SLO-72}. This model relates the DW position ($q$), and the DW's supposedly uniform magnetization ($\phi$) to the different interactions affecting the system. 

Thiaville and Nakatani later extended the $q-\phi$ model to in-plane systems and introduced the DW width parameter ($\Delta$) as an additional time varying coordinate, leading to the $q-\phi-\Delta$ model \cite{THI-06}. However, their findings showed that the evolution of $\Delta$ has minimal effect on the dynamics and could be neglected. Due to interest in current-driven DW motion at the time, the spin-transfer torque (STT) mechanism was also implemented as part of these newer models \cite{THI-04,THI-05}.

Recent studies on DW motion have focused on perpendicular magnetic anisotropy (PMA) heterostructures in which ultrathin ferromagnets are sandwiched between a heavy metal layer and an oxide ($HM/FM/Ox$). In these structures, spin-orbit coupling (SOC) and broken inversion symmetry (BIC) affect the static structure of the DW and contribute to DW motion \cite{NGU-07, MAN-08a, MAT-09,CHE-09, HAA-13}. Specifically, the Dzyaloshinski-Moriya interaction (DMI) present in these systems stabilizes N\'eel DW structures of specific chirality. SOC has also been linked to enhanced current induced DW motion, with the spin Hall effect (SHE) suggested as the dominant mechanism for this observation \cite{HAA-13}. Moving DWs tend to tilt in the plane of the sample in these systems, with the $q-\phi-\chi$ CCM (where $\chi$ denotes the DW tilting) developed to describe DW motion in these systems \cite{BOU-13}.

The efficiency of DW motion depends on the internal magnetic structure of the DW. As such, applied fields in-plane of the sample can be used to control DW chirality, enhancing the efficiency of current-driven DW motion \cite{MAR-13, MAR-14, BOU-12, BOU-13, BOU-14, EMO-13, EMO-14, NAS-15, NAS-17}. While micromagnetic simulations of this problem are in agreement with experiments, conventional CCMs (such as the $q-\phi$\cite{SLO-72} and $q-\phi-\chi$ \cite{BOU-13}) fail to reproduce the micromagnetic results \cite{MAR-14, NAS-17}. 

Despite this shortcoming, equations derived from the $q-\phi-\chi$ model are used in two of the most prominent methods of assessing the strength of the DMI in which the DW is manipulated under in-plane magnetic fields. In the most common method of assessing DMI strength, magnetic bubbles are expanded in the thin film of interest under the application of in-plane and out-of-plane fields in the creep regime \cite{JE-13, HRA-14}. This method of assessing the DMI strength assumes that the points with significant N\'eel character are located on the axis of the in-plane field applied, and the DMI field is assumed to be equal to the in-plane field which reverses the chirality of the DW. A less common method of assessing DMI strength uses a critical longitudinal field which can be identified in current-driven DW motion in nanowires with DMI; at the critical point the DW is locked in place irrespective of applied current, and the value of the longitudinal at that point is related to the DMI strength \cite{NAS-17}. While most experimentalists rely on the $q-\phi-\chi$ model in DMI strength measurements using the methods above, as mentioned previously, these models seem to not be accurate as they cannot reproduce the micromagnetic results. This calls for improvements in collective coordinate modeling of DW motion, both to reproduce micromagnetic results and to help in the assessment of DMI strength in material stacks.

In our previous work, we developed an extended collective coordinate model which better reproduced micromagnetic results in the case of current-driven DW motion in PMA systems with strong DMI under the application of in-plane fields. This model was developed based on the Bloch profile and had four collective coordinates \cite{NAS-17}. The increased accuracy of the model was attributed to inclusion of an approximation of canting in the domains as an additional parameter. Canting in the domains arises from the application of in-plane fields to the system, and affects the limits of integration when deriving the CCM.

In this paper, we present a new extended CCM based on an inherently canted ansatz to describe DW motion in PMA systems with DMI. We compared this model in mathematical form to past models presented in the literature \cite{SLO-72, THI-06, BOU-13, NAS-17}. The CCMs presented in this work are used to study two material stacks, which differ in the strength of DMI and uniaxial anisotropy. Specifically, we find that while our past studies showed that only a four coordinate model can predict the characteristic shape of the DW velocity curve \cite{NAS-17}, our new model maintains higher accuracy when only the DW position and magnetization angle are used as coordinates. This highlights the rigidity of the DW during motion, and the fact that canting in the domains plays an important role in magnetic DW motion under in-plane fields. We also found that the anisotropy of the system plays an important role in the applicability of the models, with minimal difference observed between the different models in systems with high anisotropy (low canting and narrower DWs).

We also showcase in detail the impact of in-plane fields on field- and current-driven DW motion, identifying critical in-plane fields in DW dynamics which lead to effects such as no tilting, no movement or a Bloch DW structure.  Analytical solutions are proposed based on the CCMs for these critical points that shed some light on the physics at these points, and how these points could help in measuring the strength of various interactions. 

\section{Methods}

\subsection{Systems Under Study}
In this work, we study two $2.8 \mu m$ long, $160 nm$ wide nanowires  with the magnetic properties listed in Table \ref{materials} and a $0.6 nm$ thickness for the ferromagnetic layer. 
\pdfcomment{I used the same thickness for both, but maybe the PtNi sample should be thicker? in the actual system it is 4.5 nm thick; we could say we fixed dimensions to improve comparison. Eduardo suggested to leave as is.}

These samples were selected as they both have DMI and PMA; however, the DMI strength of the $Pt/CoFe/MgO$ sample is twice that of the $Pt/Co/Ni/Co/MgO/Pt$ sample, while its PMA constant is $1/3$ that of the later sample. This helps us better understand the impact of these two parameters on the structure and dynamics of DWs.

\begin{table}
	\caption{Material parameters of the two systems studied in this work. The DMI strength of the $Pt/CoFe/MgO$ sample is twice that of the $Pt/Co/Ni/Co/MgO/Pt$ sample, while its PMA constant is $1/3$ that of the later sample. This difference in material properties helps better understand their effects on the motion of the DW.}
	\label{materials}
	\centering
	\begin{tabular}{|l|c|c|}
		\hline
		& $Pt/CoFe/MgO$ \cite{EMO-13} & $Pt/Co/Ni/Co/MgO/Pt$ \cite{YOS-16} \\
		\hline
		Saturation magnetization $M_s$& 700 & 837 \\
		($kA/m$) & & \\
		\hline
		Exchange constant $A$& 10 & 10 \\
		($pJ/m$) & & \\
		\hline
		Uniaxial perpendicular anisotropy  &  \textbf{0.48 } & \textbf{1.310} \\
		constant $K_u$ ($MJ/m^3$) & & \\
		\hline
		DW width parameter (nm) & 7.2 & 3.39 \\
		$\Delta \sim \sqrt{\frac{A}{K_u - \mu_0 M_s^2/2}}$ & &\\
		\hline
		Gilbert damping $\alpha$ & 0.3 & 0.15 \\
		\hline
		SHE angle $\theta_{SH}$ & 0.07 &  0.07 (assumed)\\
		\hline
		DMI strength $D$ & \textbf{-1.2}  & \textbf{0.6}  \\
		($mJ/m^2$) & & \\
		\hline
	\end{tabular}
\end{table}

\subsection{Micromagnetic Simulations}
To understand the magnetization dynamics in these samples, we conducted micromagnetic simulations using the Mumax$^3$ software \cite{VAN-14} which numerically solves the Landau-Lifshitz-Gilbert (LLG) equation. A micromagnetic cell size of 1 nm $\times$ 1 nm $\times$ 0.6 nm was used for all micromagnetic simulations. 

As we are interested in magnetic DW dynamics under applied fields and currents in a perpendicularly magnetized heterostructure, the DMI \cite{BOG-89,BOG-01, THI-12}, spin-orbit torques (SOTs) \cite{HAN-13, GAR-13, KHV-13}, and the spin transfer torque (STT) mechanism \cite{ZHA-04, THI-05} were included in addition to the traditional interactions included in the effective field (exchange, anisotropy, magnetostatics, and the Zeeman energy). With these terms, the LLG will take the following form:

\begin{equation}
\begin{split}
\frac{d\vec{m}}{dt} & = - \gamma \, \vec{m} \times \vec{H}_{eff} \overbrace{+ \, \alpha \, \vec{m} \times \frac{d\vec{m}}{dt}}^{Damping \, term} \\
& \overbrace{-\left(\vec{u} \cdot \nabla \right)\vec{m}}^{adiabatic \,  STT} + \overbrace{\beta \, \vec{m} \times \left(\left(\vec{u} \cdot \nabla \right)\vec{m} \right)}^{non-adiabatic \,  STT} \\
& + \overbrace{\gamma \, H_{FL} \left(\vec{m} \times \hat{u}_{SOT}\right)}^{field-like \,  SOT} - \overbrace{\gamma \, H_{SL} \vec{m} \times \left(\vec{m} \times \hat{u}_{SOT}\right)}^{Slonczewski-like \,  SOT} 
\end{split}
\end{equation}
with the effective field linked with the internal energy of the system through $$\vec{H}_{eff} = \frac{1}{\mu_0 M_s} \frac{\delta E_{dens}}{\delta \vec{m}}$$. 

The internal energy density of the magnetic system can be written as:
\begin{equation}
E_{dens} = \overbrace{A \sum_{i=1}^{3}|\nabla m_i|^2}^{Exchange} + \overbrace{K_U \sin^2 \theta}^{Anisotropy} - \overbrace{\frac{\mu_0 M_s}{2} \vec{H_d} \cdot \vec{m}}^{Magnetostatics} - \overbrace{\mu_0 M_s \vec{H_a} \cdot \vec{m}}^{Zeeman} + \overbrace{D (m_z \nabla.\vec{m} - (\vec{m}.\nabla) m_z)} ^ {DMI}
\end{equation}
While the LLG equation can grasp the intricate details of the dynamics, the fact that magnetization has to be determined spatially and temporally at every point within the system translates to long computation times. 

\subsection{Collective Coordinate Modeling}
To better understand the underlying physics observed in the motion of DWs under in-plane fields in these materials, we developed extended collective coordinate models (CCMs) using the Euler Lagrange equation based on the Lagrangian and dissipation functions presented in the literature \cite{BOU-13,NAS-17}. 

\begin{figure}
	\centering
	\includegraphics[trim = 2cm 18cm 2cm 3cm , clip=true, scale = 0.5]{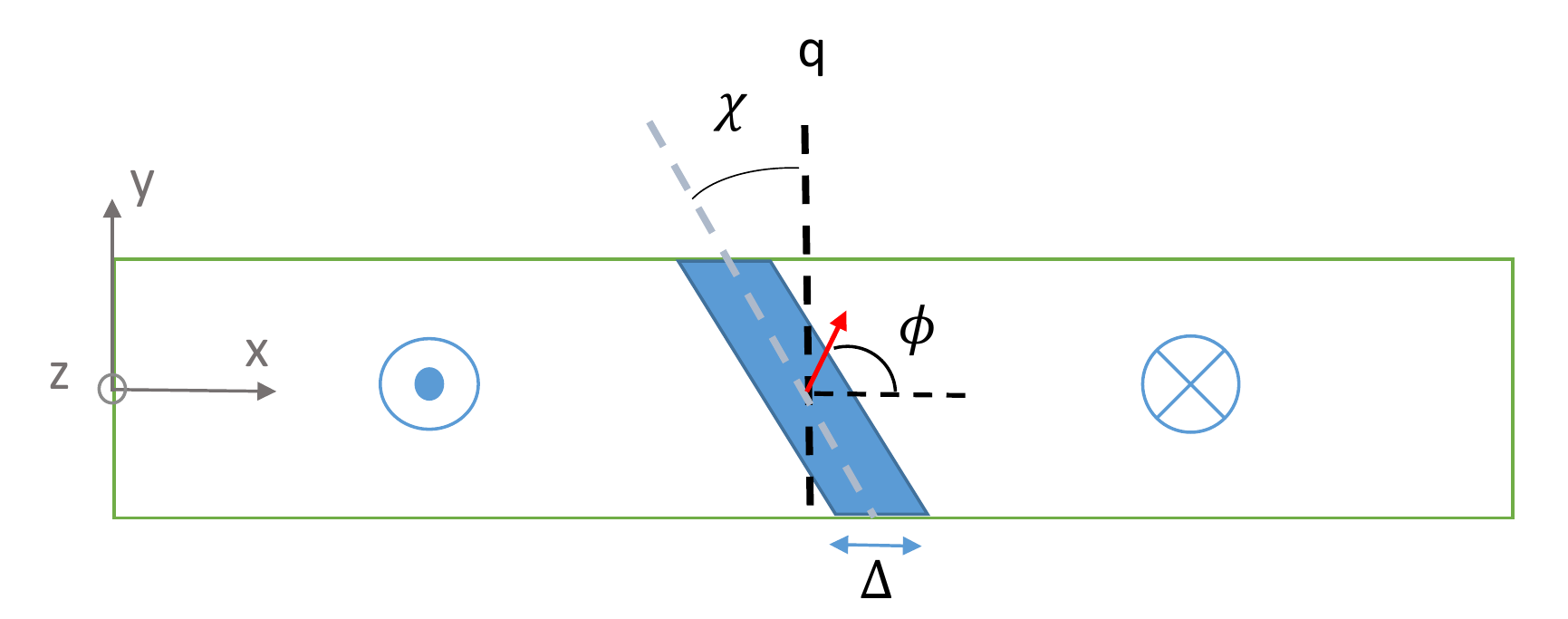}
	\caption[Collective coordinates used in this work.]{The collective coordinates used to describe the DW in this work.}
	\label{coll_coords}
\end{figure}

Based on micromagnetic simulations, experimental observations and previous work \cite{SLO-72,THI-06,BOU-13}, we selected four time dependent collective coordinates to describe the collective behavior of the DW:
\begin{enumerate}
	\item The position of the center of the DW ($q$);
	\item Magnetization angle at the center of the DW (assumed to be homogeneous) ($\phi$);
	\item The DW width parameter ($\Delta$);
	\item The geometric tilt angle of the DW ($\chi$).
\end{enumerate} 
Using these collective coordinates, the DW is modeled as a thin line with four degrees of freedom, as defined in Figure \ref{coll_coords}.a. As such, this description is not valid when the DW changes its shape.

The collective coordinates need to be linked with spherical coordinates of the magnetization in order to write the Lagrangian and dissipation functions in terms of these coordinates. We used two ansatzes in this work:
\begin{enumerate}
	\item Ansatz 1 (tilted Bloch profile \cite{BOU-13}): $\tan\frac{\theta}{2}=  \exp\left( Z \right)$
	\item Ansatz 2 (inherently canted profile, a continuous version of an ansatz previously used in the literature \cite{SOB-93, SOB-94, SOB-95, SOB-95b}):  $\tan\left(\frac{\theta + \theta_c}{2}\right) = \frac{\exp\left( Z \right) + \sin\theta_c}{\cos\theta_c}$.
\end{enumerate}
with $Z = \frac{(x-q) \cos\chi + y \sin\chi}{p_w \Delta}$.The value of the parameter $p_w = \pm 1$ is used to adjust for up-down vs down-up DWs (although other values could be used as prefactors for the DW width parameter). In both cases, we also assume that magnetization is constant along the DW, hence $\phi(r,t) = \phi(t)$. 

The difference between the two ansatzes is that ansatz 2 takes into account the canting in the domains due to the application of in-plane fields in the profile itself, a feature which was observed to play an important role on DW motion under in-plane fields \cite{NAS-17}. In the presence of an applied field, the magnetization in the domains may be described in spherical coordinates as  $\phi = \mathrm{atan}\left(\frac{H_y}{H_x}\right)$ and $\sin\theta_c = \frac{\mu_0 M_s \left( H_x \cos\phi + H_y \sin\phi \right)}{2 K_u + \mu_0 M_s^2 \left(N_x \cos^2\phi + N_y \sin^2\phi - N_z \right)}$ where $\theta_c$ is the value of the canting angle, and $N_x$, $N_y$, $N_z$ denote the demagnetizing factors felt by the spins in the domain  \cite{NAS-17}. This description may be derived through energy minimization in the domains. Note that, as we use  planar view of the system in our model, the sign of the canting angle $\theta_c$ also needs to be determined; we defined the angle to be negative for negative in-plane fields and positive for positive in-plane fields; comparison to cases where this sign was not taken into account later revealed the importance of this convention on the sign of the angle. 

\begin{figure}
	\centering
	\subfloat[DW profile of $Pt/CoFe/MgO$.]{\includegraphics[trim = 2cm 1cm 2cm 2cm , clip=true, scale = 0.32]{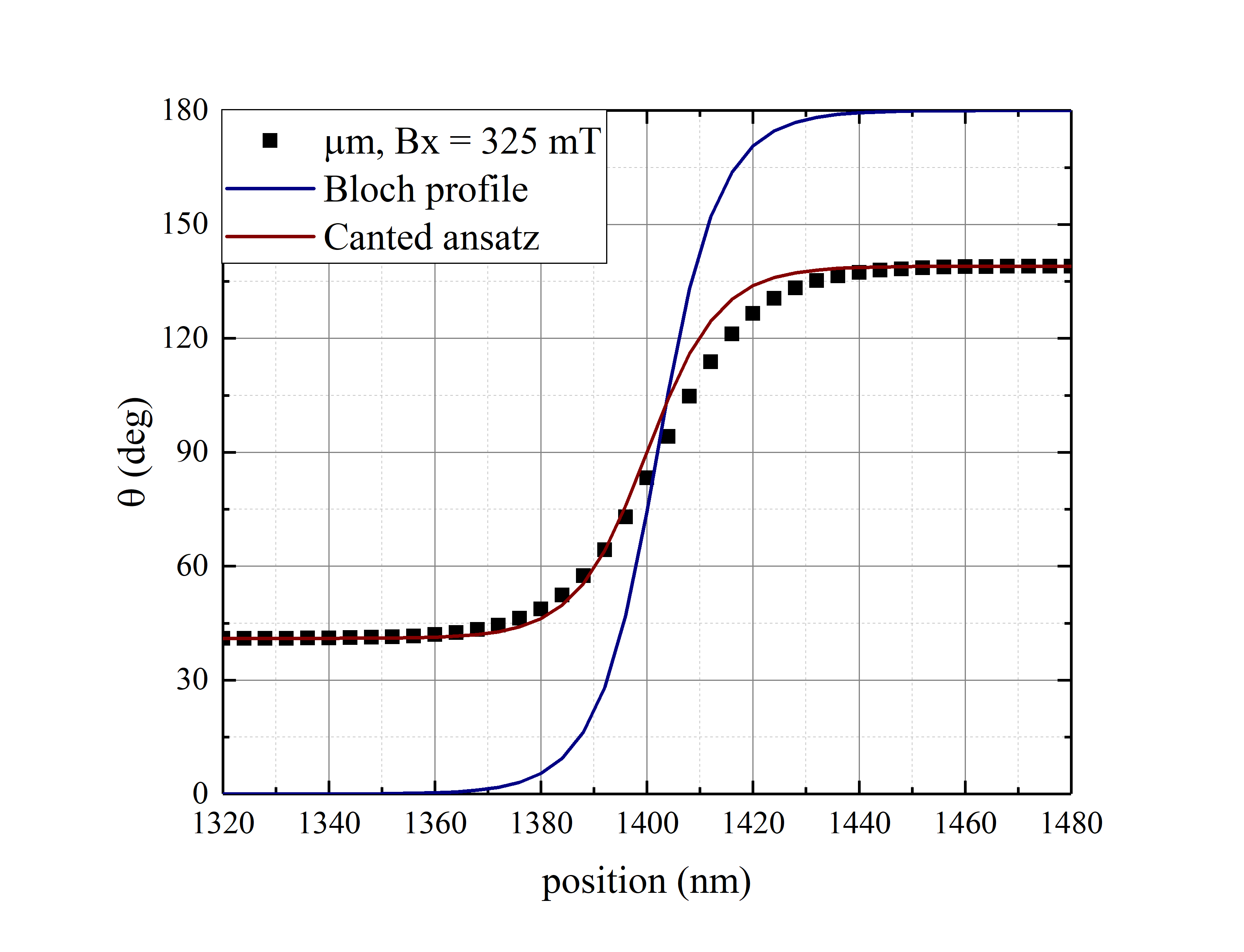}}
	\subfloat[DW profile of $Pt/Ni/Co/Ni/MgO/Pt$.]{\includegraphics[trim = 2cm 1cm 2cm 2cm , clip=true, scale = 0.32]{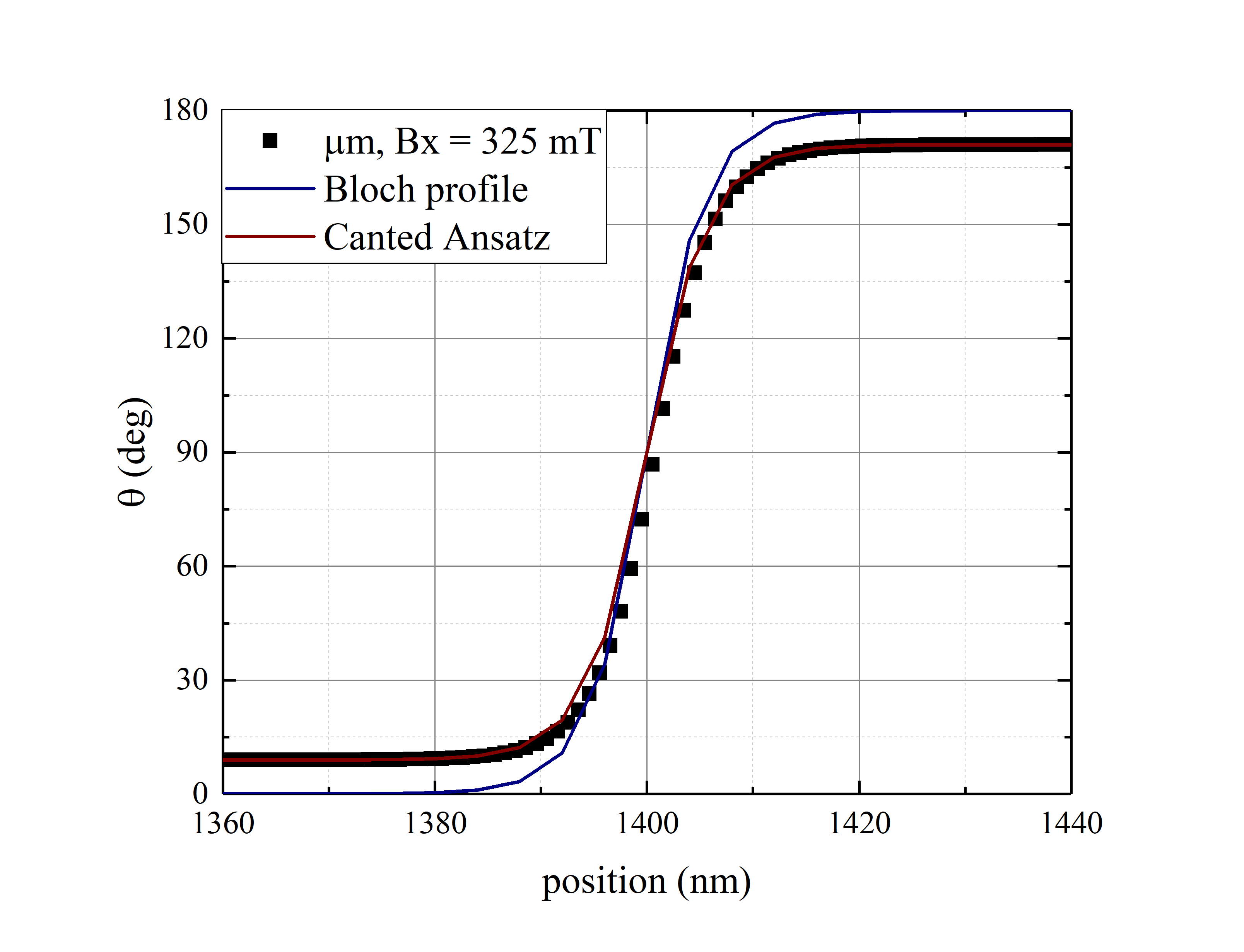}}
	\caption{Comparison of the Ansatzes used to model DW structure to micromagnetic simulation results under $B_x = 325 mT$ and static conditions for the (a) $Pt/CoFe/MgO$, and (b) $Pt/Co/Ni/Co/MgO/Pt$ samples. The magnetization angle $\theta = acos(m_z)$ was calculated at the middle of the wire.}
	\label{ansatz}
\end{figure}

Figure \ref{ansatz} shows the accuracy of these ansatzes in predicting the static structure of the DWs compared to micromagnetic simulations ($\mu m$). We know from past studies that the $\theta$ component of magnetization does not change significantly under dynamic conditions. 

Figure \ref{ansatz} shows that under the same in-plane field the domains in the $Pt/Co/Ni/Co/MgO/Pt$ show lower canting, which can be attributed to the higher uniaxial anisotropy of this system. We can deduce from this observation that canting will have a much lower impact on DW motion in this material, and the models including canting will not differ dramatically from those without canting. In contrast, in the $Pt/CoFe/MgO$ sample the lower anisotropy leads to high canting up to around 40 degrees under the same conditions. In addition, Figure \ref{ansatz} and similar simulations, showed us that in systems with high uniaxial anisotropy (narrower DWs), the width of the DW does not change dramatically under in-plane fields. 

Overall, these simulations show that systems with very high anisotropies are less affected by in-plane fields in general, and effects such as canting and DW width change which arise from the application of in-plane fields will not impact their dynamics as much as systems with lower anisotropy.

In order to evaluate DW dynamics using the collective coordinates, we need to rewrite the energy terms using these coordinates and the properties of the ansatz. The energy and dissipation function then need to be integrated along the length and width of the wire, and plugged into the Euler-Lagrange equation. This process will result in four equations: 

\begin{equation}
\alpha I_1 \frac{\dot{q}}{p_w \Delta} cos\chi + I_2 \dot{\phi} = \mu_0 \gamma \left(I_2 H_z - I_3 H_{SL} \left[\sin\phi u_{SOT,x} - \cos\phi u_{SOT,y}\right]\right) + \beta I_1 \frac{u}{p_w \Delta} \cos\chi
\label{eq1}
\end{equation}

\begin{equation}
\begin{split}
I_2 \frac{\dot{q}}{p_w\Delta} \cos\chi -  \alpha I_4 \dot{\phi}  & = I_4  \frac{\mu_0 \gamma M_s}{2}  \left(N_y - N_x \right) \sin 2(\phi-\chi) + I_5 \frac{u}{p_w \Delta} \cos\chi\\
& + I_6 \mu_0 \gamma \left[H_x \sin\phi - H_y \cos\phi \right] - I_3 \frac{\gamma D}{M_s p_w \Delta} \sin(\phi - \chi) 
\end{split}
\label{eq2}
\end{equation}

\begin{equation}
\alpha I_7  \left[\frac{\dot{\Delta}}{p_w \Delta} + \dot{\chi} \tan\chi \right] = \frac{\gamma}{M_s} \left[ I_1 \frac{A}{(p_w \Delta)^2} - I_4 p_w K \right] + I_6 \mu_0 \gamma p_w \left(H_x \cos\phi + H_y \sin\phi \right)
\label{eq3}
\end{equation}

\begin{equation}
\begin{split}
-  \alpha I_7 \left[ \frac{\dot{\Delta}}{p_w \Delta} \sin\chi + \frac{\dot{\chi}}{\cos\chi} \left[ \frac{I_1}{6 I_7}\left(\frac{w}{ p_w \Delta} \right)^2 + \sin^2\chi \right] \right] &=  \frac{\gamma}{M_s} \sin\chi \Bigg[ I_1 \frac{A}{(p_w \Delta)^2} \\
&+ I_4 K - I_6 \mu_0 M_s \left(H_x \cos\phi + H_y \sin\phi\right) \Bigg] \\
&+ I_4 \frac{\mu_0 \gamma  M_s}{2} \cos\chi (N_x - N_y) \sin 2\left(\phi - \chi\right)\\
& + I_3 \frac{\gamma D}{M_s p_w \Delta} \sin\phi 
\end{split}
\label{eq4}
\end{equation}

where $K = K_u + \frac{\mu_0 M_s^2}{2} \left[N_x \cos^2\left(\phi - \chi\right) + N_y \sin^2\left(\phi - \chi\right)-N_z\right]$ and $N_x$, $N_y$ and $N_z$ are the demagnetizing factors of the DW assumed to be of ellipsoidal form \cite{CRO-91}. For the cases of interest in this paper, we assumed that the spin Hall effect gives rise to a Slonczewski-like field (meaning $H_{FL} \sim 0$). The strength of the SHE fields can be calculated as $H_{SL} = \frac{\hbar \theta_{SHE} J}{2 \mu_0 e M_s t_f}$ where $t_f$ is the thickness of the ferromagnetic layer \cite{KHV-13}. 

Note that equations \ref{eq1}-\ref{eq4} above may be further simplified into two coordinate models (assuming $\chi = 0$ and a fixed $\Delta$ leads to a model similar to the $q-\phi$ model \cite{SLO-72}), or three coordinate models ($\chi=0$ leads to a model similar to the $q-\phi-\Delta$ \cite{THI-06} model, while assuming a fixed $\Delta$ leads to a model similar to the   $q-\phi-\chi$ \cite{BOU-13} model). 

In the equations above, $I_i$s are integration constants which depend on the ansatz used and relate to the amount of canting in the domains. Three classes of models were derived based on the ansatz used and how they were integrated:  
\begin{enumerate}
	\item Integrating ansatz 1 from  0 to $\pi$: This model does not take into account the canting in the domains, and was presented in one of our previous works \cite{VAN-17}.
	\item Integrating ansatz 1 from $\theta_c$ to $\pi - \theta_c$: This model approximates the effect of the canting in the domains and was presented in our past work \cite{NAS-17}.
	\item Integrating ansatz 2 from $\theta_c$ to $\pi - \theta_c$: We expect this model to be the most accurate, as it takes into account the effect of canting not just in the domains but on the DW structure.
\end{enumerate}

Equations \ref{eq1}-\ref{eq4} are rather interesting as they show that the ansatz used does not affect the functionality in terms of the collective coordinates. Instead, the effect of the ansatz (including canting) is taken into account in the $I_i$ parameters. 

Table \ref{integ_parameters} summarizes the value of the $I_i$s for the three different groups of models. For model 2, the closed form of the $I_7$ constant is:
\begin{equation}
\begin{split}
I_7 &= \bigg[\mathrm{Li}_2\left(-\cos\frac{\theta_c}{2}\right) - \mathrm{Li}_2\left(-\sin\frac{\theta_c}{2}\right) - \mathrm{Li}_2\left(1-\cos\frac{\theta_c}{2}\right) + \mathrm{Li}_2\left(1 - \sin\frac{\theta_c}{2}\right)\bigg]	\\
&- \cos^2\frac{\theta_c}{2} \left[1 - log(\cos\frac{\theta_c}{2})\right] + \sin^2\frac{\theta_c}{2} \left[1 - \log(\sin\frac{\theta_c}{2})\right]\\
&- \cos^2\frac{\theta_c}{2} \log(\sin^2\frac{\theta_c}{2}) \left[ \log(\cos\frac{\theta_c}{2}) - \frac{1}{2}\right] + \sin^2\frac{\theta_c}{2} \log(\cos^2\frac{\theta_c}{2}) \left[\log(\sin\frac{\theta_c}{2}) - \frac{1}{2}\right]\\
&- \frac{1}{2} \log(\sin^2\frac{\theta_c}{2}) + \frac{1}{2} \log(\cos^2\frac{\theta_c}{2}) + \log(\cos\frac{\theta_c}{2}) \log(1 + \cos\frac{\theta_c}{2}) \\
&- \log(\sin\frac{\theta_c}{2}) \log(1 + \sin\frac{\theta_c}{2}) + \cos\theta_c + (\cos\theta_c + 1) \ln^2(\cos \frac{\theta_c}{2})+ (\cos\theta_c - 1) \ln^2(\sin \frac{\theta_c}{2})
\end{split}
\label{I_7_model 2}
\end{equation}
where $\mathrm{Li}_2$ is the polylogarithm function of order 2.

A closed form for $I_7$ could not be derived for model 3. Instead, the integral was numerically solved and fitted to the following polynomial function ($R^2 = 1$, $RMSE = 3.82 \times 10^{-4}$)

\begin{equation}
I_7 = 0.568 \theta_c ^3 - 0.4232 \theta_c^2 -1.47 \theta_c + 1.649
\label{I_7_model 3}
\end{equation}

\begin{figure}
	\begin{center}
		\subfloat[$I_1$.]{\includegraphics[trim=1cm 1cm 1cm 2cm, clip=true,scale=0.32]{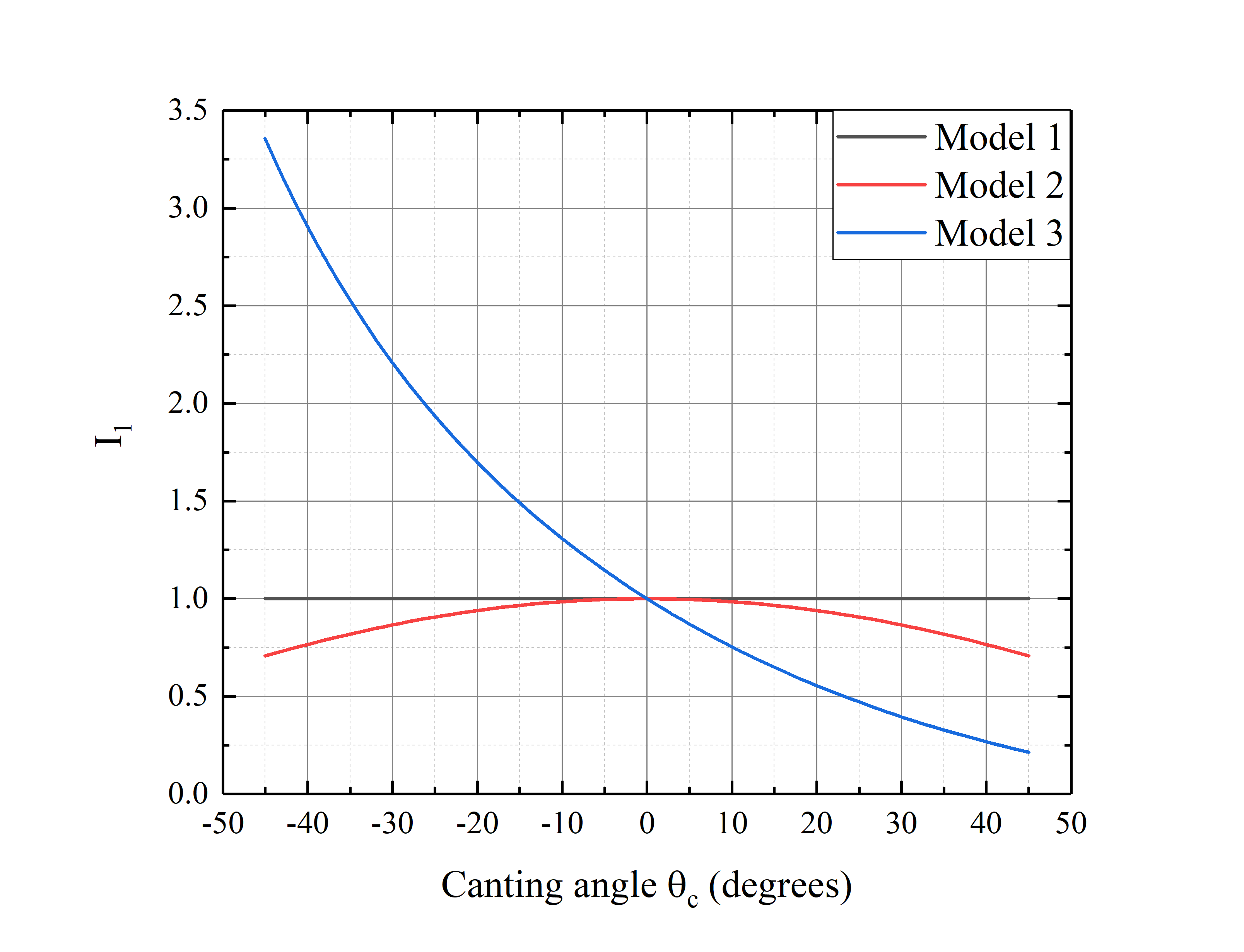}}
		\subfloat[$I_2$.]{\includegraphics[trim=1cm 1cm 1cm 2cm, clip=true,scale=0.32]{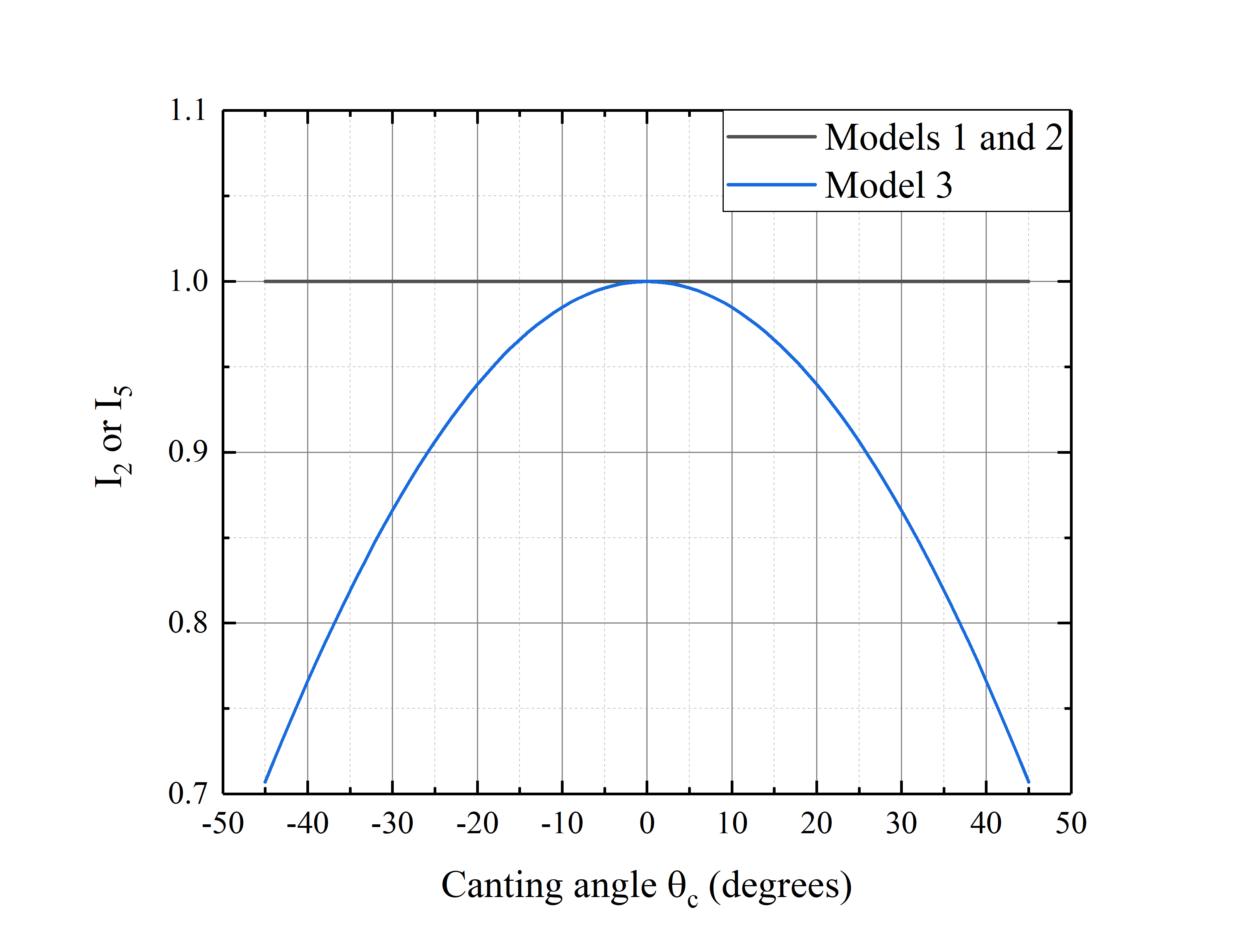}}\\
		\subfloat[$I_3$.]{\includegraphics[trim=1cm 1cm 1cm 2cm, clip=true,scale=0.32]{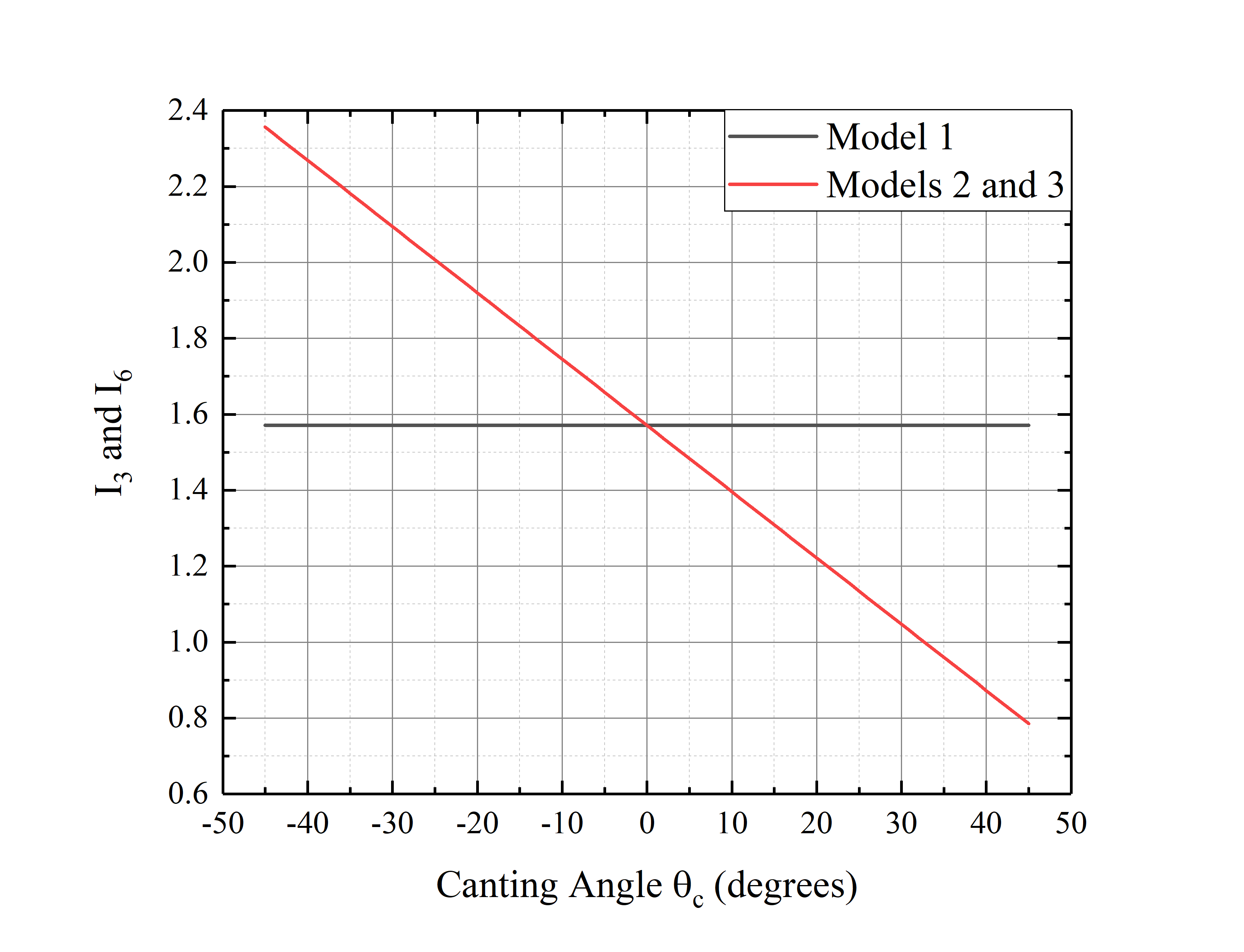}}
		\subfloat[$I_4$.]{\includegraphics[trim=1cm 1cm 1cm 2cm, clip=true,scale=0.32]{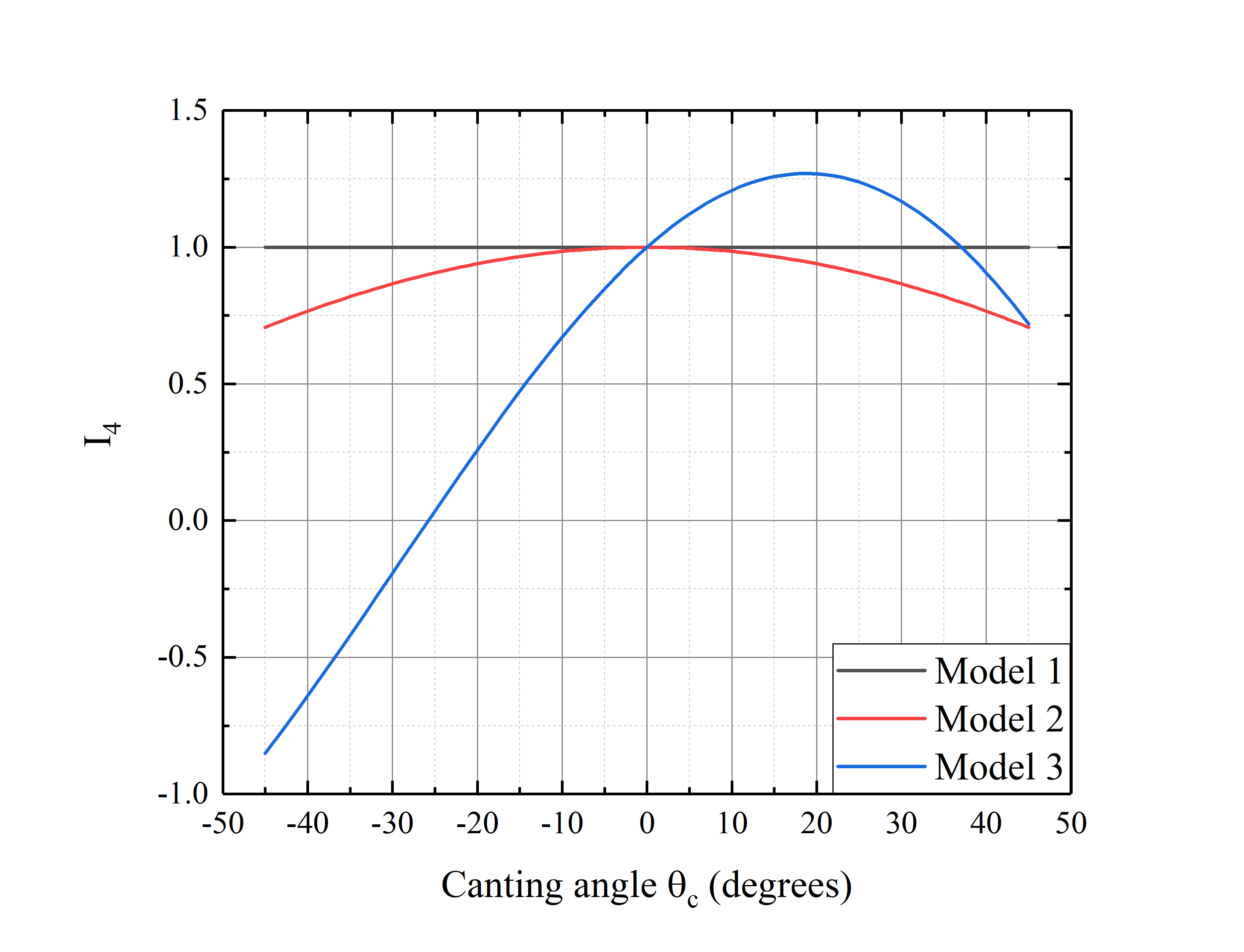}}\\
		\subfloat[$I_7$.]{\includegraphics[trim=1cm 1cm 1cm 2cm, clip=true,scale=0.32]{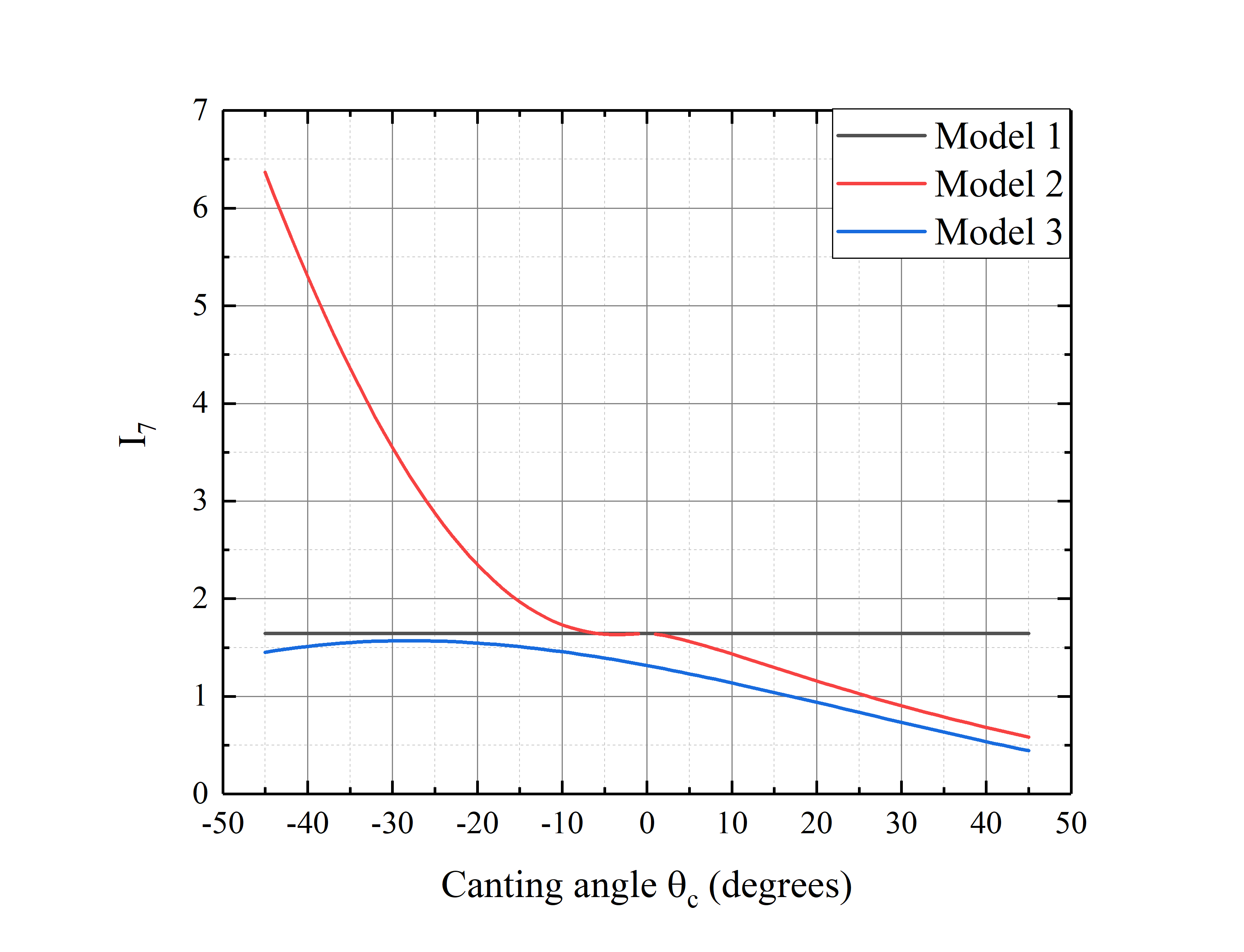}}
		
	\end{center}
	\caption{Variation of the $I_i$ parameters of the collective coordinate models with canting angles. The equations corresponding to each parameter may be found in Table \ref{integ_parameters}.}
	\label{prefactors}
\end{figure}

The mathematical form of the $I_i$s reveal the differences between the model families. Figure \ref{prefactors}  visually depicts the differences between these constants. While both model families 2 and 3 reproduce the constants for model 1 in the limit $\theta_c = 0$ (except for $I_7$ which is different depending on the ansatz), there is considerable difference between the models for the behavior of constants $I_1$, $I_4$, $I_5$ and $I_6$ for non-zero canting angles. Not only the constants for model 2 and 3 do not predict the same value for the same amount of canting, but also their behavior as a function of the canting angle is considerably different for negative canting angles. These differences are key to the different behavior predicted by the different models. 

It should also be noted that with ansatz 2, a few terms (which generated complex number solutions) were neglected in the integration of some of these factors. Specifically, parameter $I_4$ had the additional term $-  \frac{sin^2\theta_c}{2} ln\left(- sin\theta_c cos\theta_c \right)$, and parameter $I_6$ had the additional term $+ \frac{tan\theta_c}{2} ln (- sin\theta_c cos\theta_c)$. These terms will likely affect the accuracy of the models, specially when canting is not playing a role in the dynamics.

\begin{table}
	\centering
	\caption{Summary of model parameters derived from integration for the three different models. Model 1 is the model based on the Bloch profile without canting, model 2 is also based on the Bloch profile but takes into account canting in the domains through the canting angle in the domains ($\theta_c$), and model 3 is based on an inherently canted ansatz. Details of the different parameters and their derivation are provided in the supplementary materials. }
	\begin{tabular}{|c|c|c|c|}
		\hline
		& Model 1  & Model 2 & Model 3\\
		\hline
		$I_1$ & 1 & $\cos\theta_c$ & $1 - \left(\frac{\pi}{2} - \theta_c \right) \tan\theta_c $\\
		\hline
		$I_2$ & 1 & 1 & $\cos\theta_c$ \\
		\hline
		$I_3$ & $\frac{\pi}{2}$ & $\frac{\pi}{2} - \theta_c$ & $\frac{\pi}{2} - \theta_c$\\
		\hline
		$I_4$ &1 & $\cos\theta_c$& $\cos^2\theta_c  + \left(\frac{\pi}{2} - \theta_c\right) \sin\theta_c \cos\theta_c $\\
		\hline
		$I_5$ &1 & $\cos\theta_c$& $\cos\theta_c$\\
		\hline
		$I_6$ & $ \frac{\pi}{2}$ & $\frac{\pi}{2} - \theta_c$ & $\frac{\pi}{2} - \theta_c  $  \\
		\hline
		$I_7$ & $\frac{\pi^2}{6}$ & Equation \ref{I_7_model 2} & Equation \ref{I_7_model 3} \\
		\hline
	\end{tabular}
	\label{integ_parameters}
\end{table}

\subsection{Validity of the CCMs}

\begin{figure}
 	\centering
 	\subfloat[Results from model 3.]{\includegraphics[trim = 2cm 1cm 2cm 2cm , clip=true, scale=0.32]{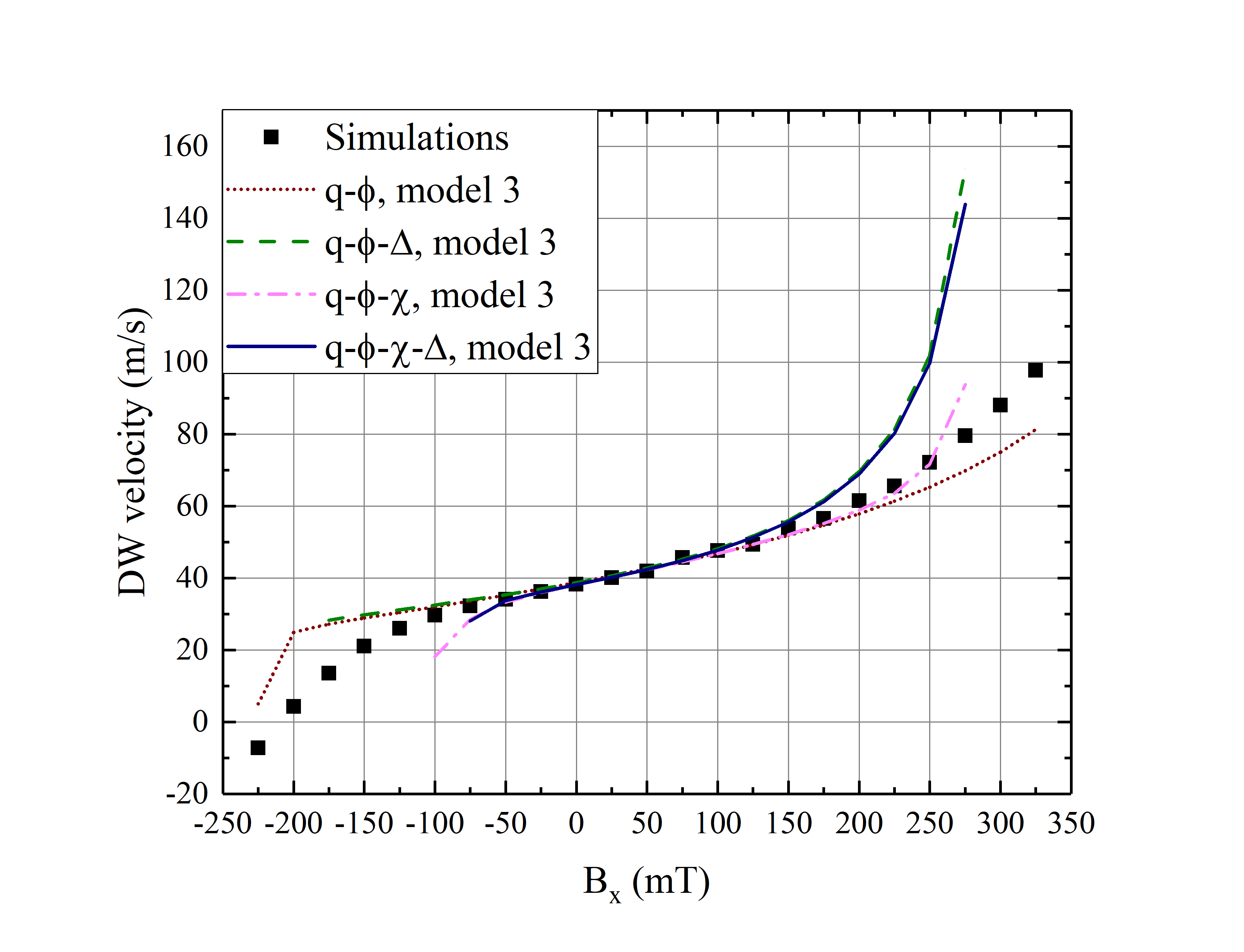}}
 	\subfloat[Results from most accurate models.]{\includegraphics[trim = 2cm 1cm 2cm 2cm , clip=true, scale=0.32]{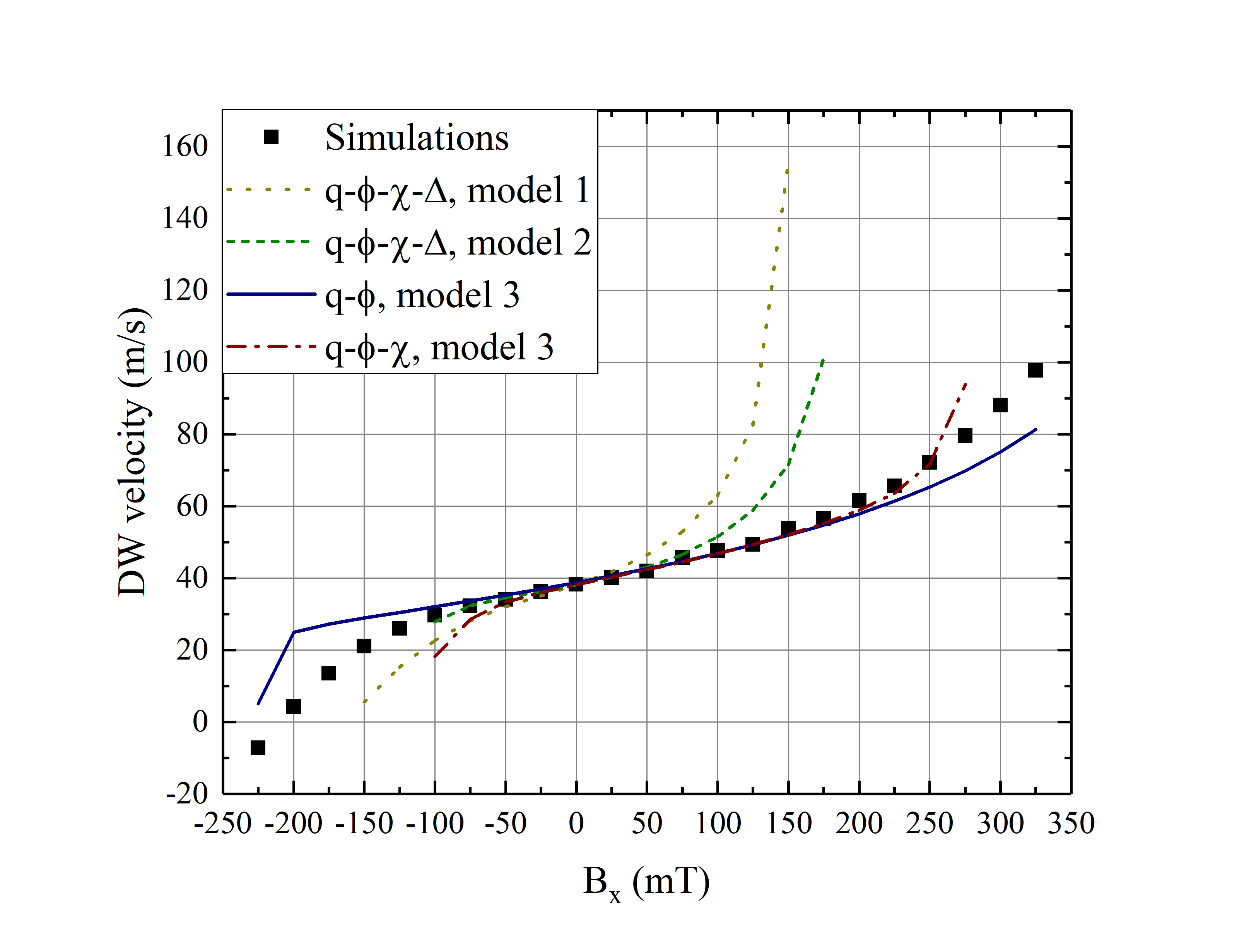}}\\
 	\subfloat[Results from model 3.]{\includegraphics[trim = 2cm 1cm 2cm 2cm , clip=true, scale=0.32]{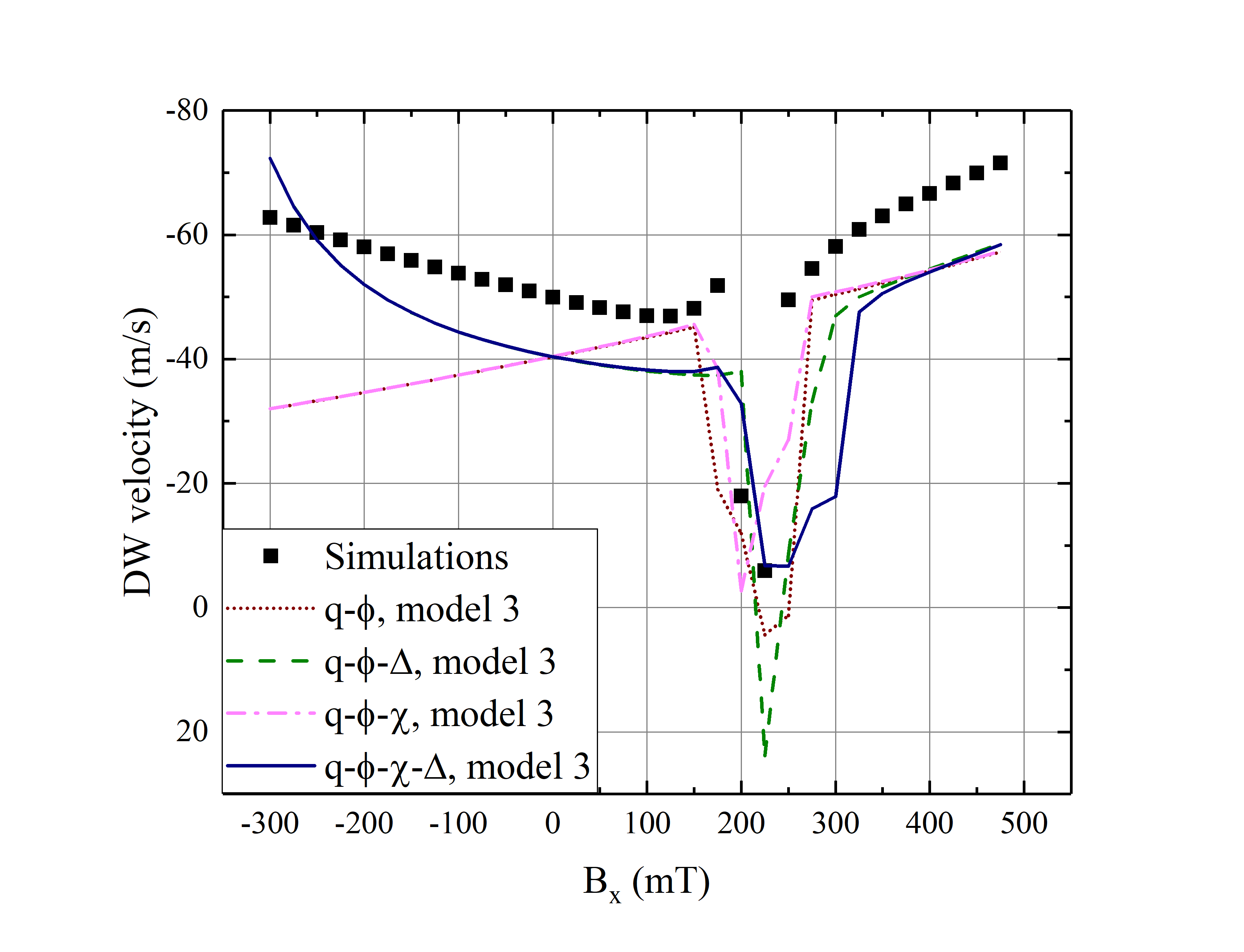}}
 	\subfloat[Results from most accurate models.]{\includegraphics[trim = 2cm 1cm 2cm 2cm , clip=true, scale=0.32]{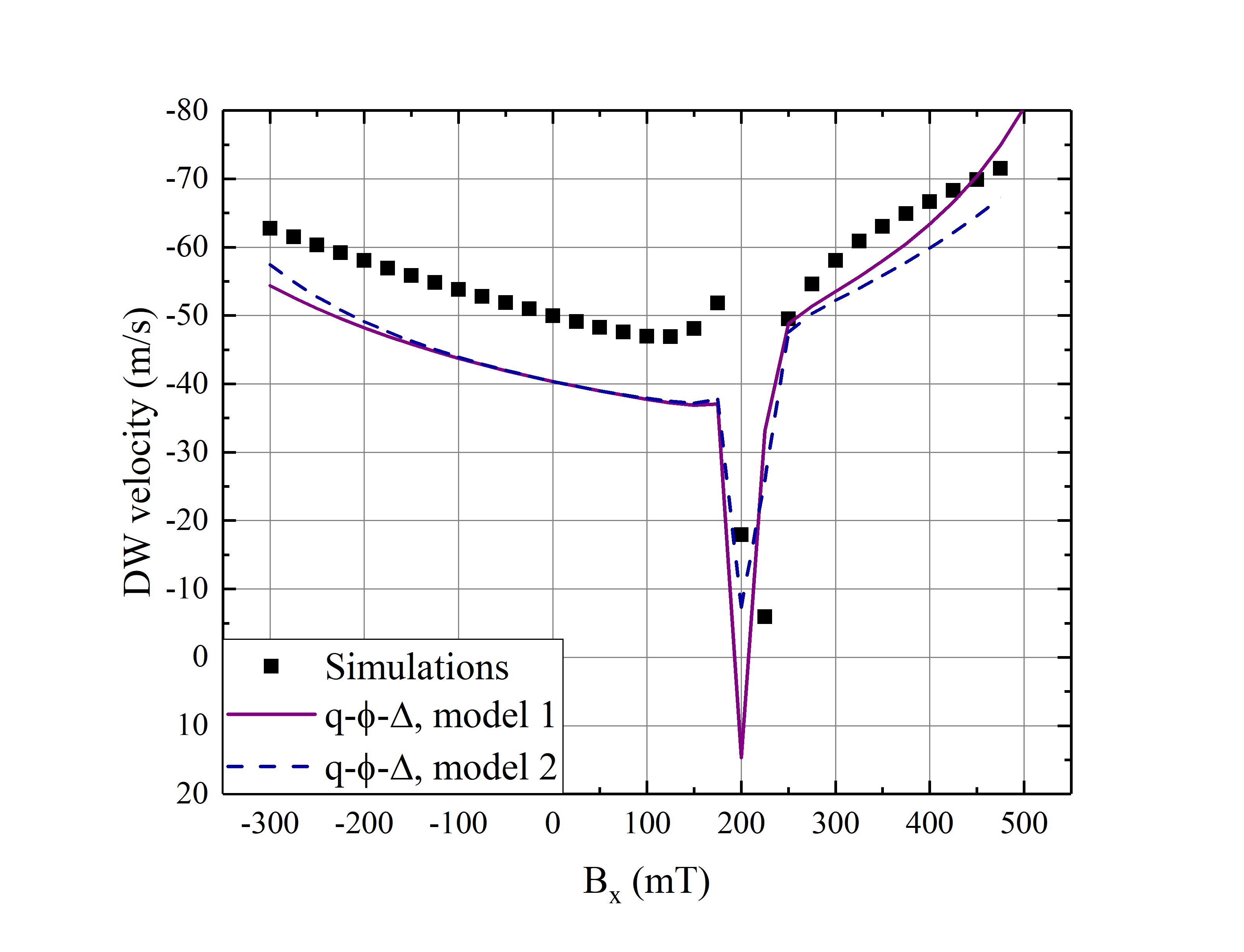}}
 	\caption{Comparison of the instantaneous DW velocity predictions in different material systems from micromagnetics and collective coordinate models. (a) and (b) show DW motion in $Pt/CoFe/MgO$ sample, with an applied current of  $J_x = 0.1 \, TA/m^2$, while (c) and (d) shows DW motion in $Pt/Co/Ni/Co/MgO/Pt$ under an applied field of $B_z = 10 \, mT$. (a) and (b) show that in $Pt/CoFe/ MgO$ with its high DMI and wider DWs, model 3 with two coordinates outperforms all models in terms of accuracy, highlighting the important role of canting in DW motion. (c) and (d) show that in $Pt/Co/Ni/Co/MgO/Pt$ with its lower DMI (low DW tilting), and narrower DWs, the DW width is an important parameter. Due to the high anisotropy of the system, canting also plays a minor role in this case and models without canting can predict DW motion correctly. We also see the prediction of Walker Breakdown in this case, both by the micromagnetic simulations and collective coordinate models. }
 	\label{ansatz2_compare_vel}
\end{figure}

To assess the accuracy of the collective coordinate models, we initially applied them to a case of current-driven DW motion in $Pt/CoFe/MgO$ under a current density of $J_x = 0.1 TA/m^2$ (the same case studies in reference \cite{NAS-17}), and field-driven DW motion in $Pt/Co/Ni/Co/MgO/Pt$ under an applied field of $B_z = 10 mT$, as outlined in Figure \ref{ansatz2_compare_vel}. 

In the $Pt/CoFe/MgO$ sample, models with inherent canting (model 3) show superior capability in replicating the micromagnetic results, and require a lower number of degrees of freedom for accurate predictions (only $q$ and $\phi$) as highlighted in Figure \ref{ansatz2_compare_vel}. b. This shows the importance of including canting in the domains when studying samples with lower anisotropy under in-plane fields. As the most accurate models with canting is the $q-\phi$ form of model 3, the profile used to approximate the DW seems to be more important than adding additional collective coordinates. We further investigate the accuracy of the models in the next section. 

In the $Pt/Co/Ni/Co/MgO/Pt$ sample (which has stronger anisotropy and lower DMI which lead to smaller canting and tilting), according to Figure \ref{ansatz2_compare_vel}. (c) we see that the DW tilting has minimal effect on accuracy of the models, while DW width plays an important role in these narrower DWs. We verified this in other models as well, observing that in this case models without canting are better suited to reproduce micromagnetic results (Figure \ref{ansatz2_compare_vel}. d) likely due to the smaller value of the canting and the fact that ansatz three had some neglected terms. We also observe that the models are able to predict the Walker Breakdown initiation and cessation properly, and showcase the right qualitative trends. 

In summary, it seems that narrower DWs are better modeled by CCMs that include DW width while for wider DWs this parameter plays a minor role. In addition, in systems with high anisotropy canting effects can be neglected. In the next section, we provide a more detailed analysis of different cases of DW motion under in-plane field using the models verified here.

\section{Results and Discussion}
Micromagnetic simulations were performed on the two nanowires outlined in Table \ref{materials} with DWs driven by fields or Slonczewski-like spin-orbit torques under the applications of longitudinal ($B_x$) and transverse ($B_y$) magnetic fields. To interpret the results of the micromagnetic simulations, we used the four time dependent collective coordinates identified earlier. In this section, we present the results of these studies.

It is well-known from micromagnetic studies that the motion of the DW reaches steady state conditions after a transient time, which we also verified for our system in micromagnetic simulations. In our simulations, steady state conditions were reached after about $2.5 ns$ in most cases, with $\dot{\phi} \ sim \dot{\chi} \sim \dot{\Delta} \sim 0$. While we found the evolution of the micromagnetic model to not exactly match the CCMs, a steady state condition was identified in the CCMs as well. In CCMs without the tilting of the DW, a steady state condition was observed with $\dot{\phi} \sim \dot{\Delta} \sim 0$, while in tilted models we found only $\dot{\Delta} \sim 0$ (although in many cases $\dot{\phi} \sim \dot{\chi} < 1$). In steady state conditions, the collective coordinate models may be simplified to better understand the critical points which can be identified in the micromagnetic simulations. In this section we use a $^*$ is used to denote steady state values of the collective coordinate. 

In the next subsections, we will present more detailed results of specific cases studied. We showcase which CCMs were able to better predict the micromagnetic results in each case, and use these models to highlight features or critical points in the dynamics of the DW. By better predicting the micromagnetic results, we mean doing so with the lowest error over a wider range of fields. Note that the range of in-plane field values over which different collective coordinate models can be solved with a convergent solution is different for different materials and drive-conditions; as such, in some cases no solution is plotted for a particular range of in-plane fields.

\subsection{General Observations}

We identified several general features in the simulations. First, as outlined in our previous work \cite{NAS-17}, one notes that domains under large in-plane fields can no longer be assumed to be fully perpendicularly magnetized, but clearly show some canting of the magnetization into the plane of the sample. This effect was much smaller in $Pt/Co/Ni/Co/MgO/Pt$ compared to $Pt/CoFe/MgO$, due mainly to the difference in the uniaxial magnetic anisotropy of the two samples. 

Second, in the $Pt/Co/Ni/Co/MgO/Pt$ sample we observed limited tilting of the DW (only up to 10 degrees in many cases) which likely is due to the much lower DMI of this sample compared to the $Pt/CoFe/MgO$ sample. As a result of this observation, we expect the $\chi$ coordinate to play a small role in modeling this system.

Third, we found that DW shape and rigidity (lack of elasticity) are affected by in-plane fields. Depending on the combination of drive interaction and in-plane fields, the DW might have a rigid line shape, or a curved shape (either S-shaped or an arch of a circle). With large in-plane fields (longitudinal and transverse), the DW might lose its rigidity, and instead extend elastically through the system. In $Pt/CoFe/MgO$ system, we found both longitudinal and transverse field where the DW starts to elongate instead of moving rigidly. This field was dependent on the material properties, and also the driving interaction applied to the system. In the $Pt/Co/Ni/Co/MgO/Pt$ system, these effects were not observed, likely due to the high uniaxial anisotropy of the system which helps maintain the DW shape. However, in this material the DW shape was disrupted due to other events which will be discussed later. 



\subsection{Domain Wall Motion Under Longitudinal in-plane Fields}

\subsubsection{Field-Driven Case}
\begin{figure}
	\centering
	\subfloat[DW velocity.]{\includegraphics[trim=1cm 1cm 1cm 2cm, clip=true,scale=0.32]{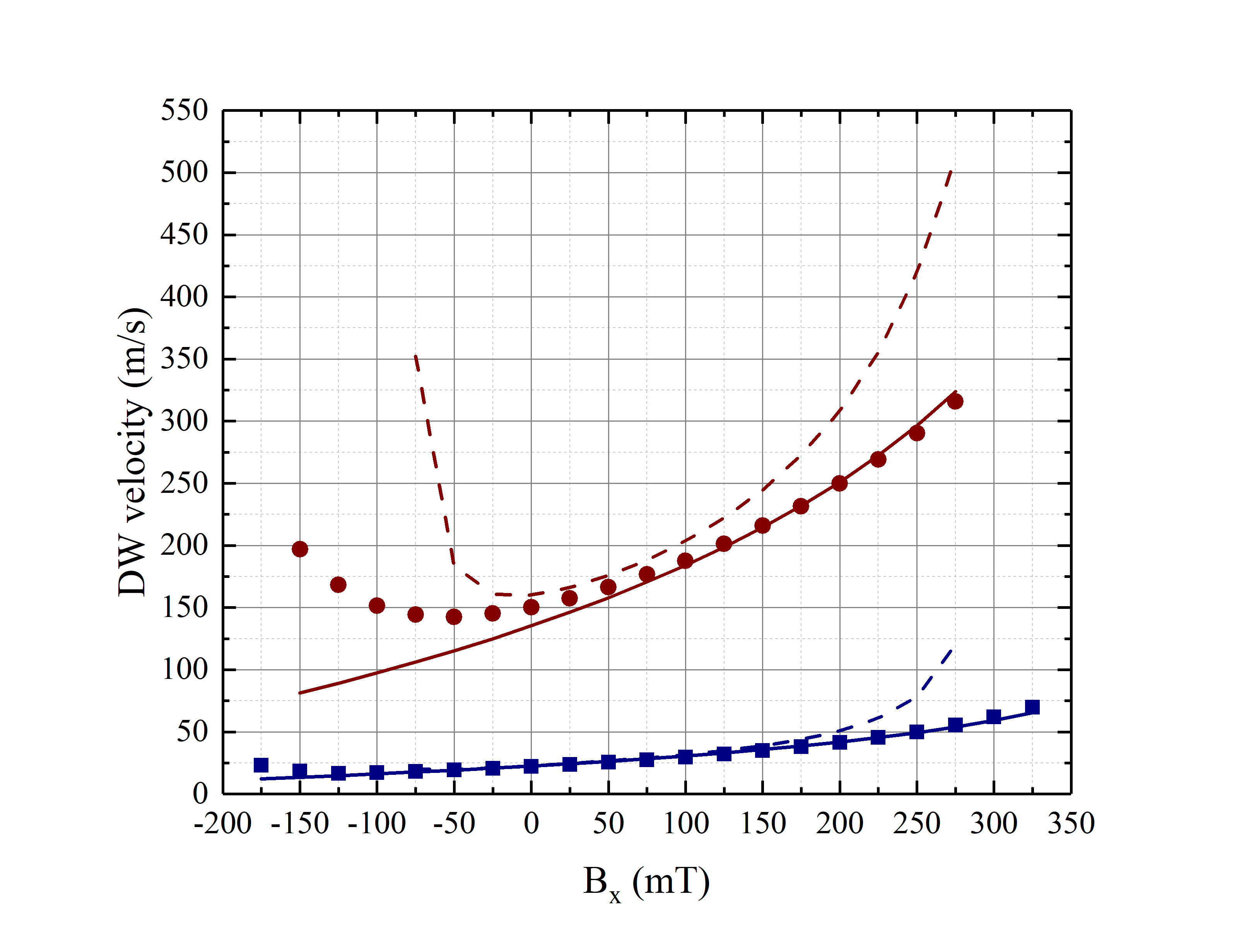}}
	\subfloat[DW width parameter ($\Delta$).]{\includegraphics[trim=1cm 1cm 1cm 2cm, clip=true,scale=0.32]{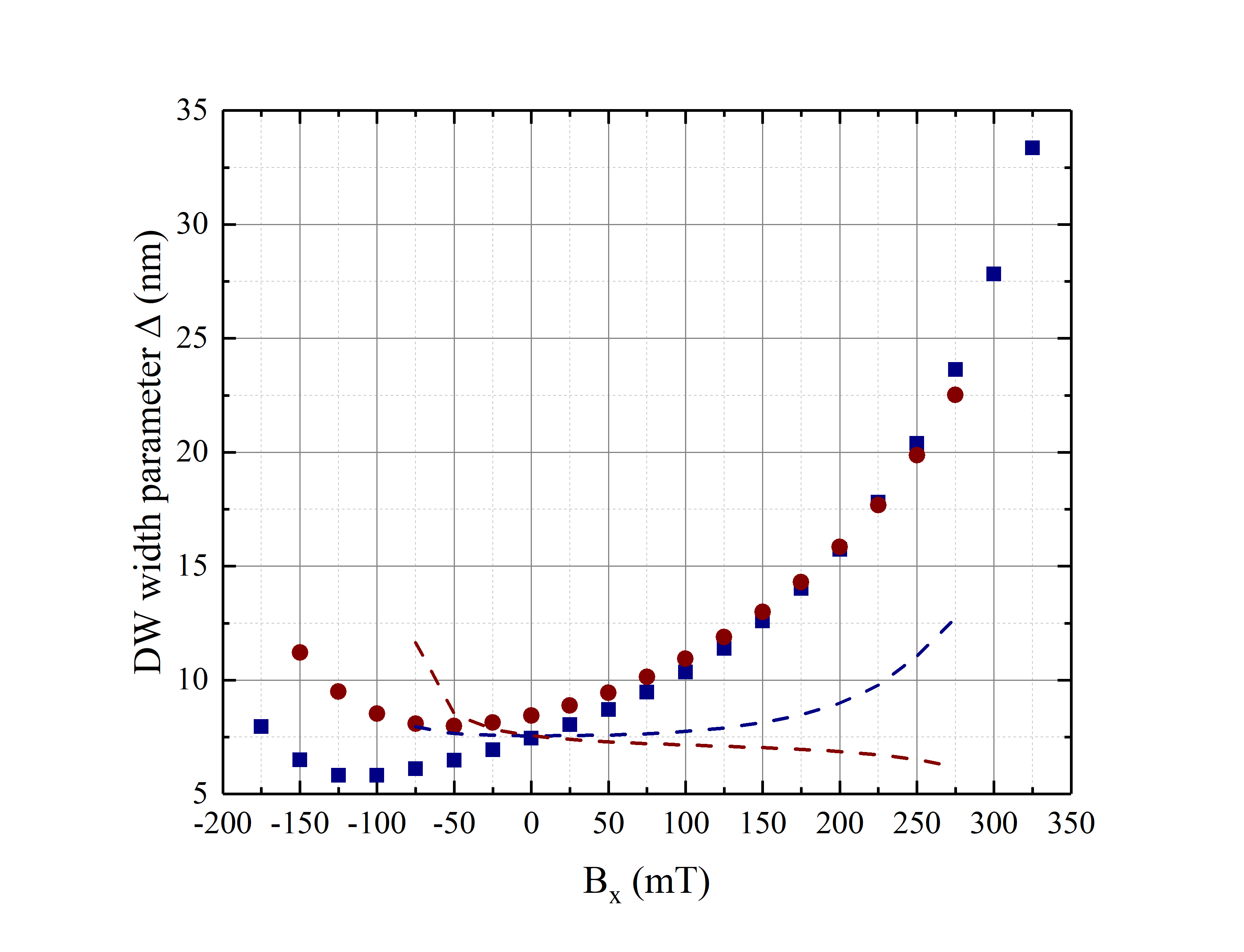}}\\
	\subfloat[$\phi-\chi$]{\includegraphics[trim=1cm 1cm 1cm 2cm, clip=true,scale=0.32]{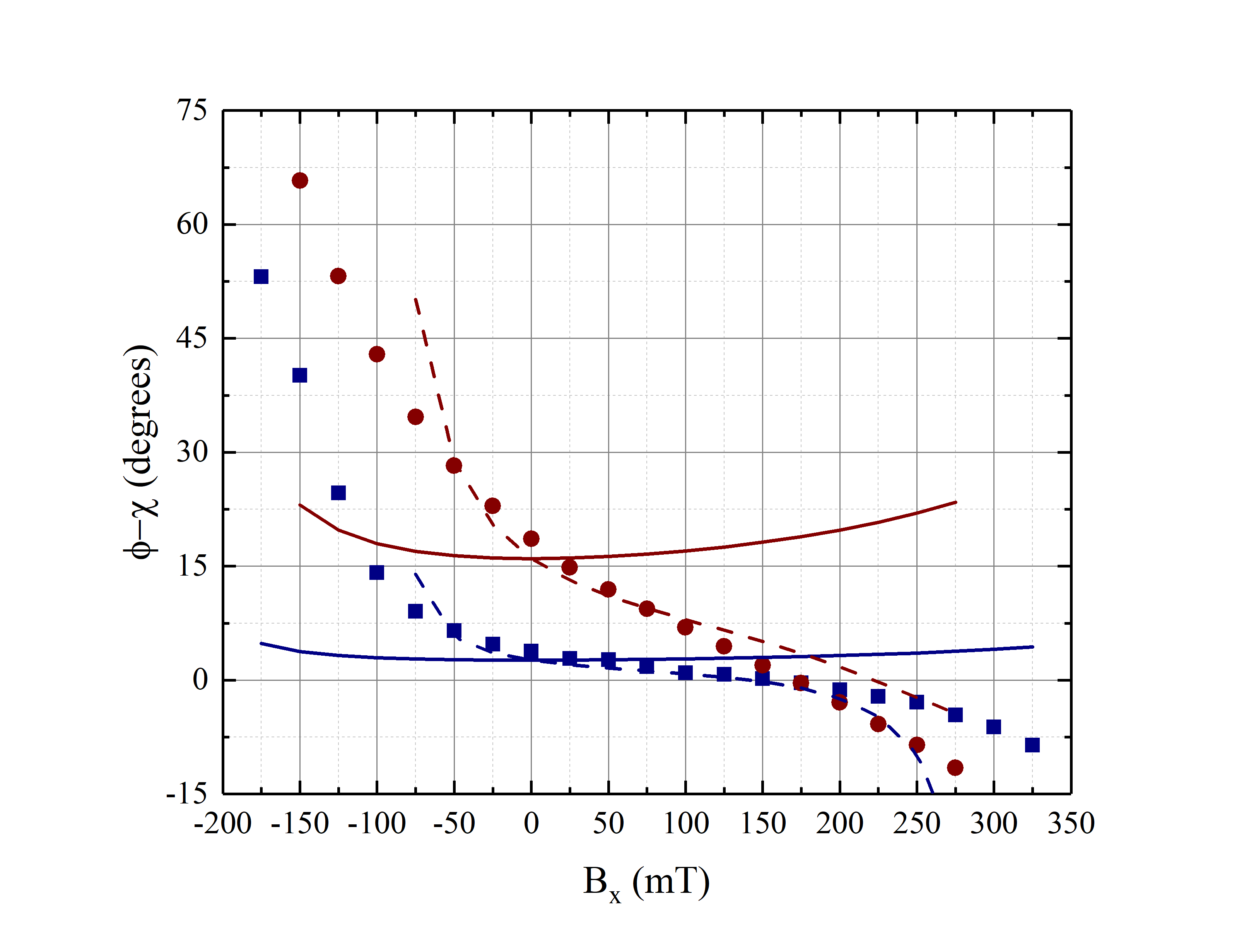}}
	\subfloat[DW magnetization angle ($\phi$).]{\includegraphics[trim=1cm 1cm 1cm 2cm, clip=true,scale=0.32]{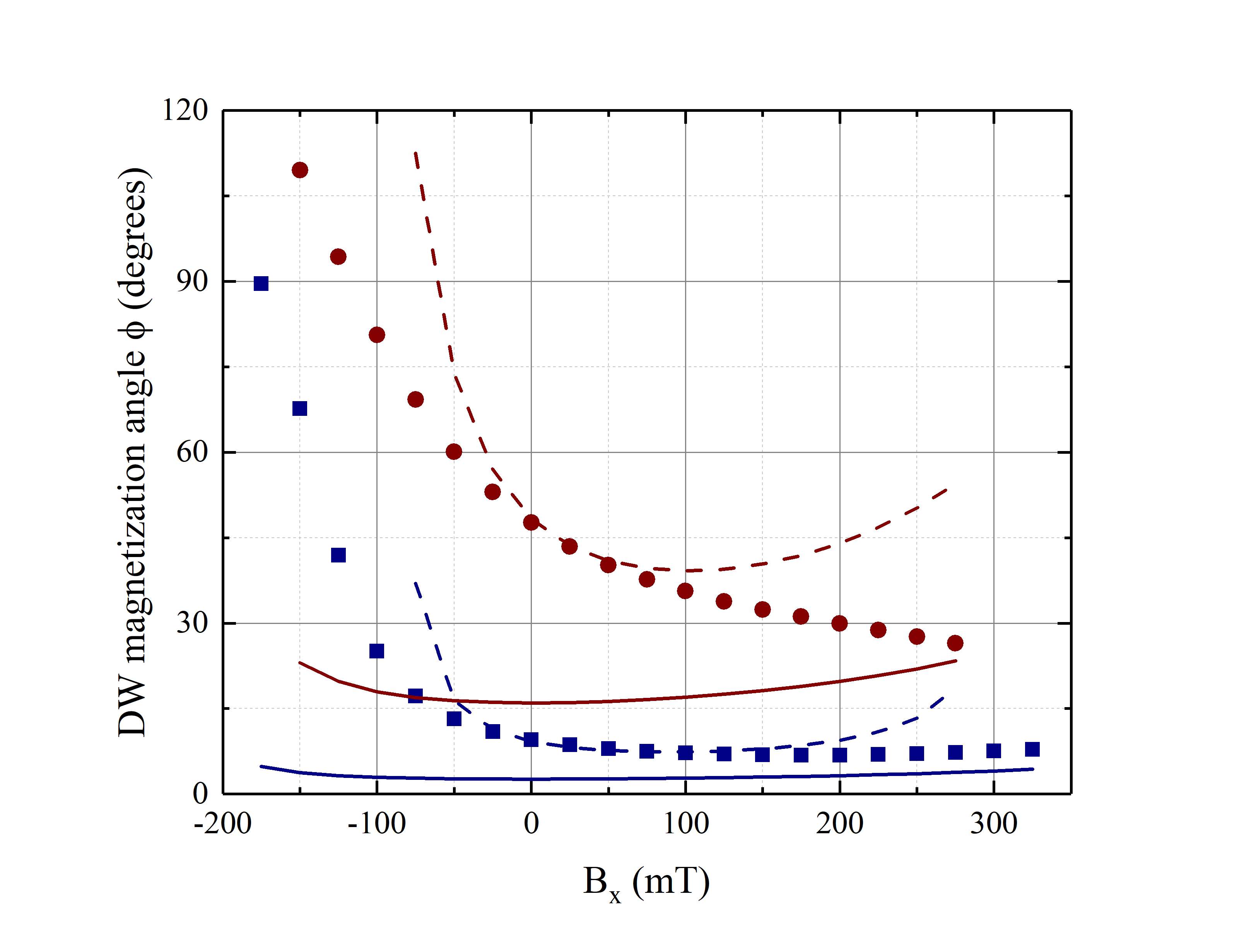}}\\
	\subfloat[DW tilting angle ($\chi$).]{\includegraphics[trim=1cm 1cm 1cm 2cm, clip=true,scale=0.32]{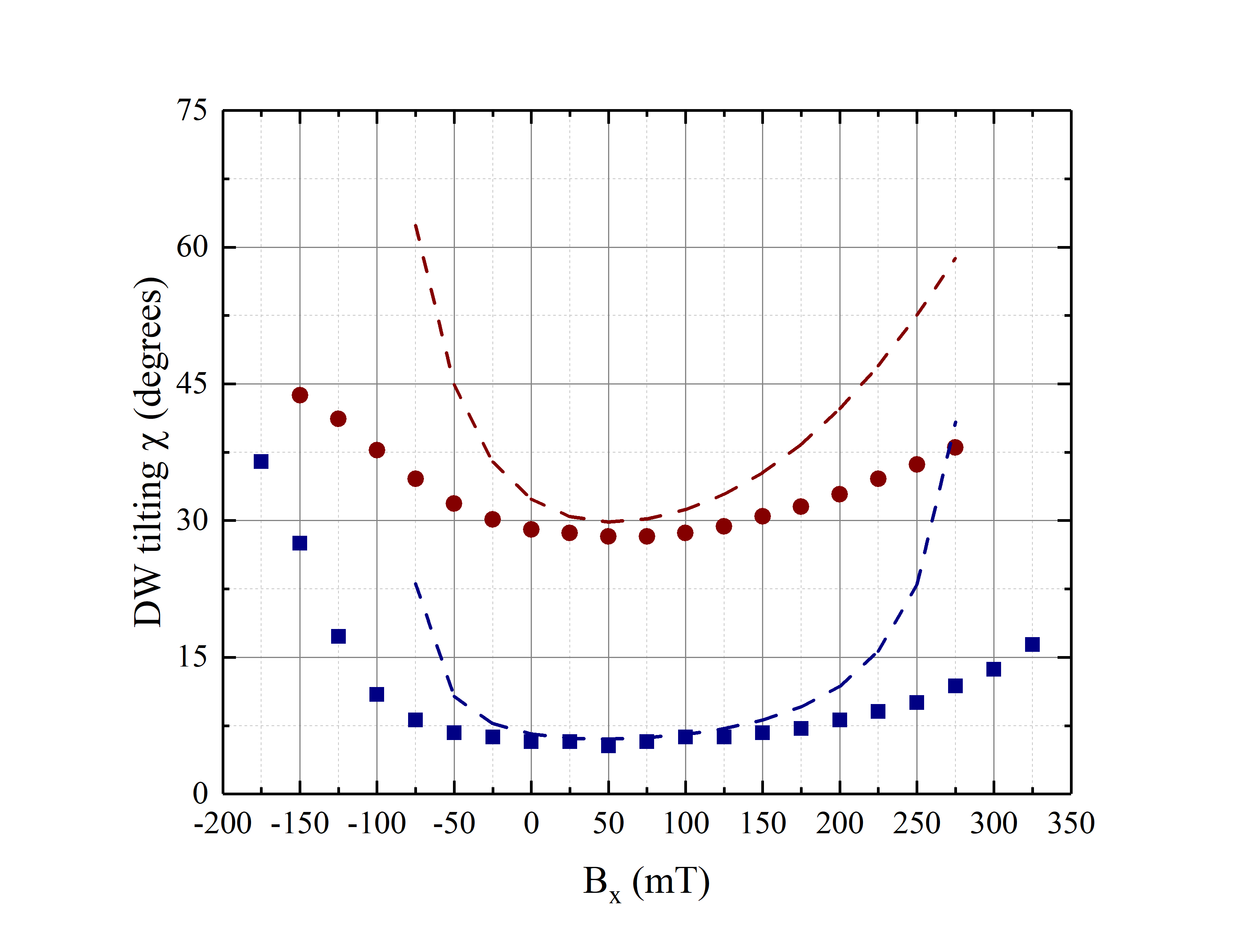}}
	\subfloat{\includegraphics[scale=0.4]{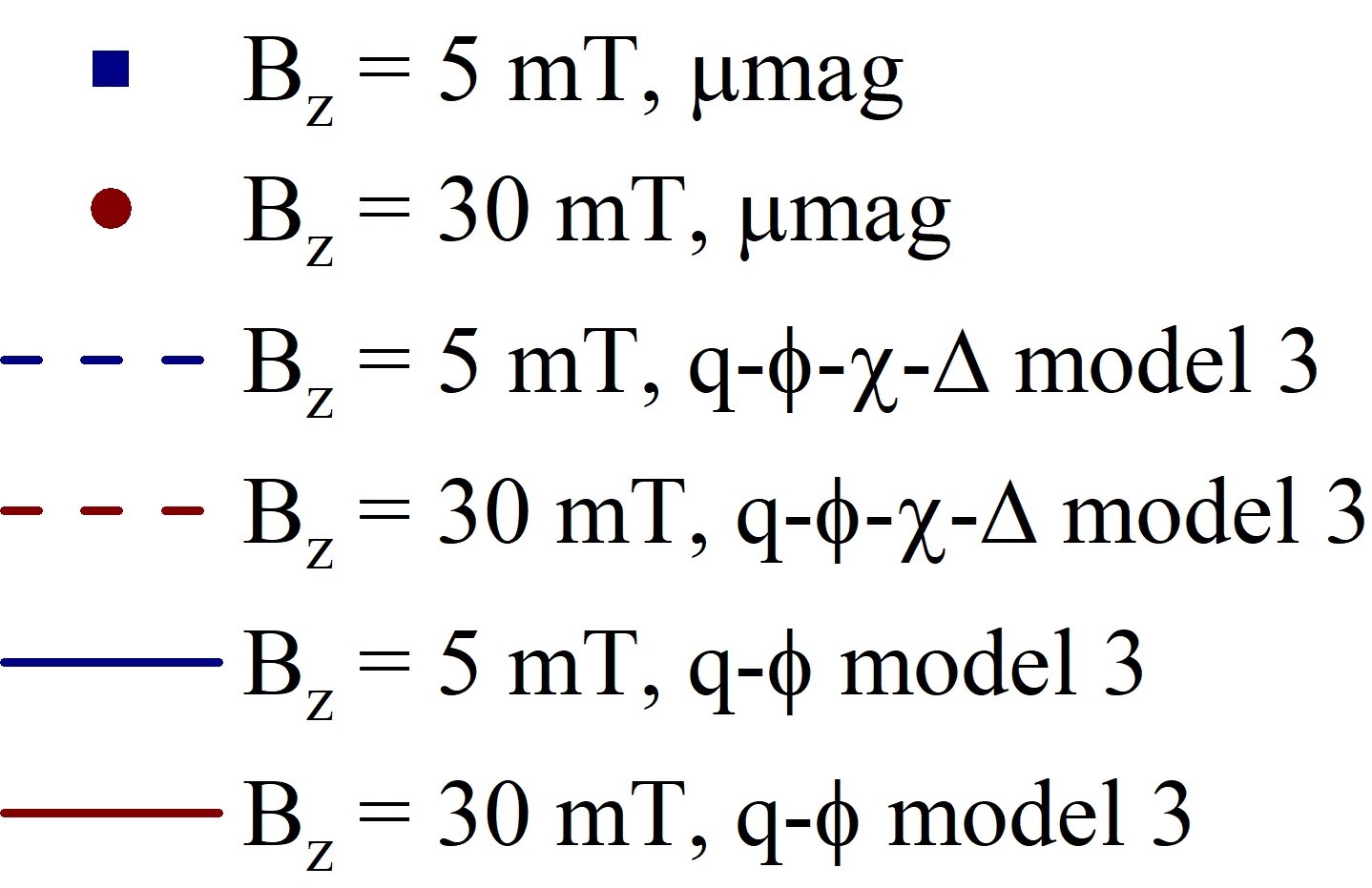}}
	\caption{Instantaneous steady state DW characteristics for field-driven DW motion in $Pt/CoFe/MgO$ with different out of plane and longitudinal fields applied. Only the collective coordinate models with highest accuracy in predicting the DW velocity are shown.}
	\label{fields_Bx_micromag}
\end{figure}

\begin{figure}
	\centering
	\subfloat[DW average speed.]{\includegraphics[trim=1cm 1cm 1cm 2cm, clip=true,scale=0.32]{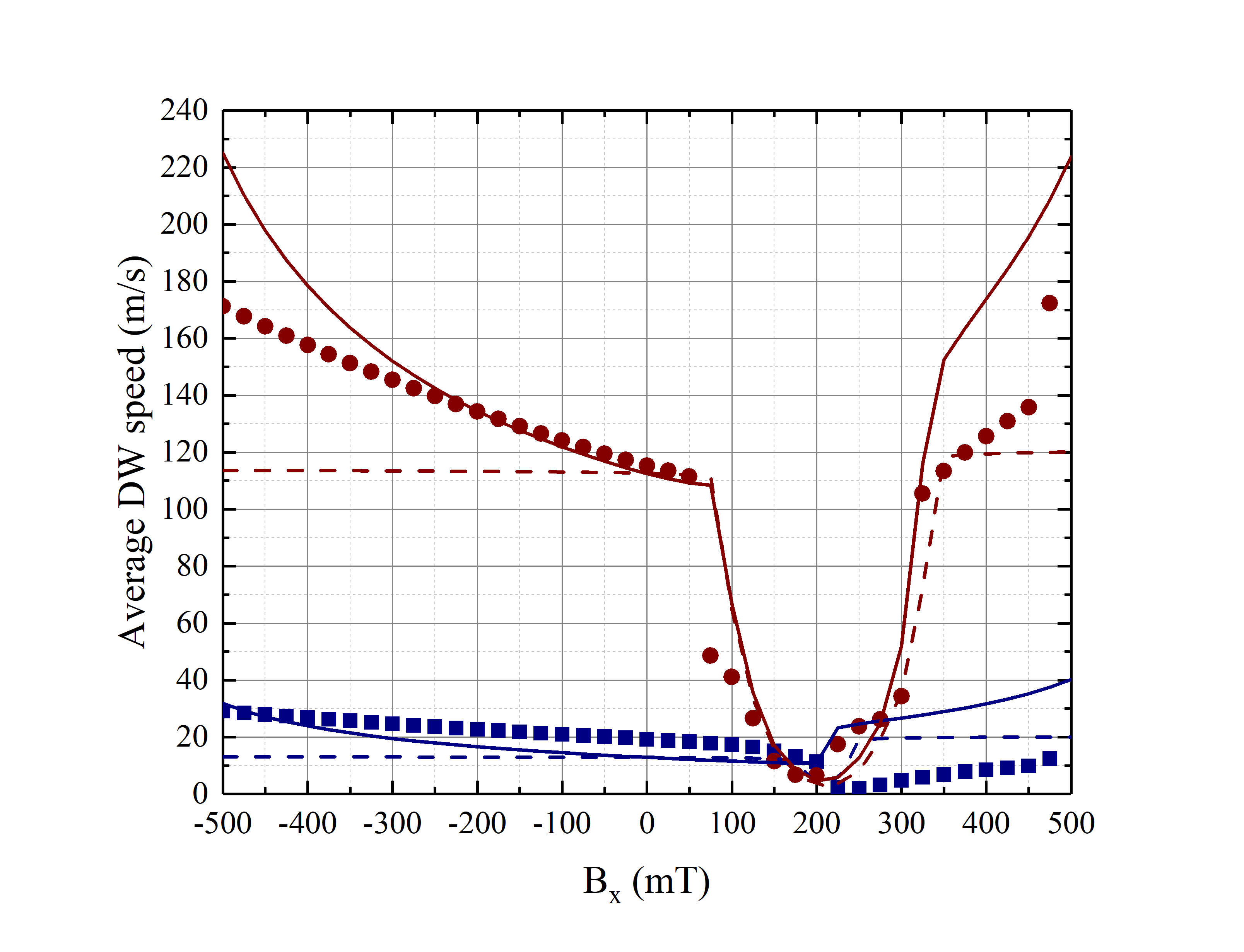}}
	\subfloat[Instantaneous DW width parameter ($\Delta$).]{\includegraphics[trim=1cm 1cm 1cm 2cm, clip=true,scale=0.32]{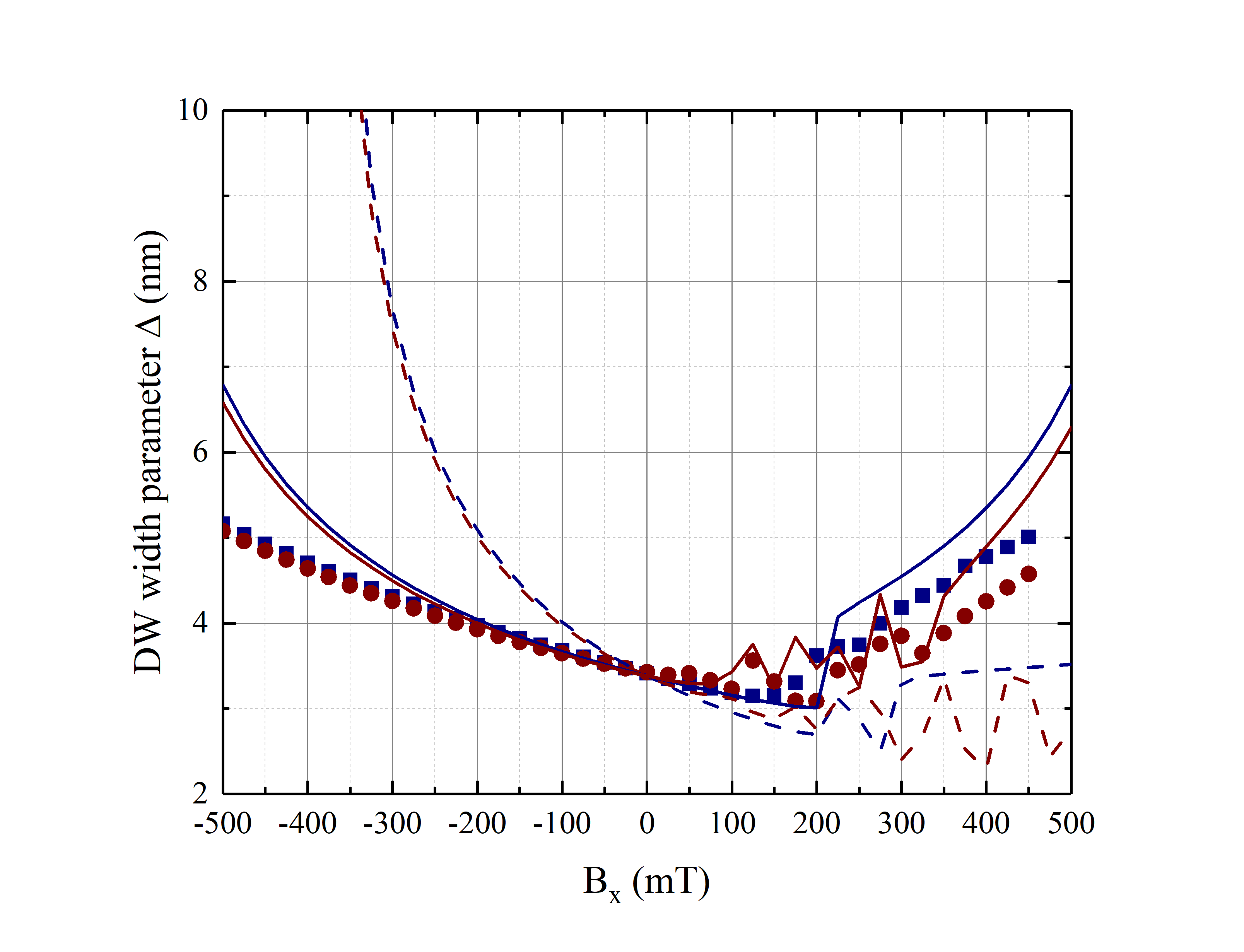}}\\
	\subfloat[Instantaneous $\phi-\chi$]{\includegraphics[trim=1cm 1cm 1cm 2cm, clip=true,scale=0.32]{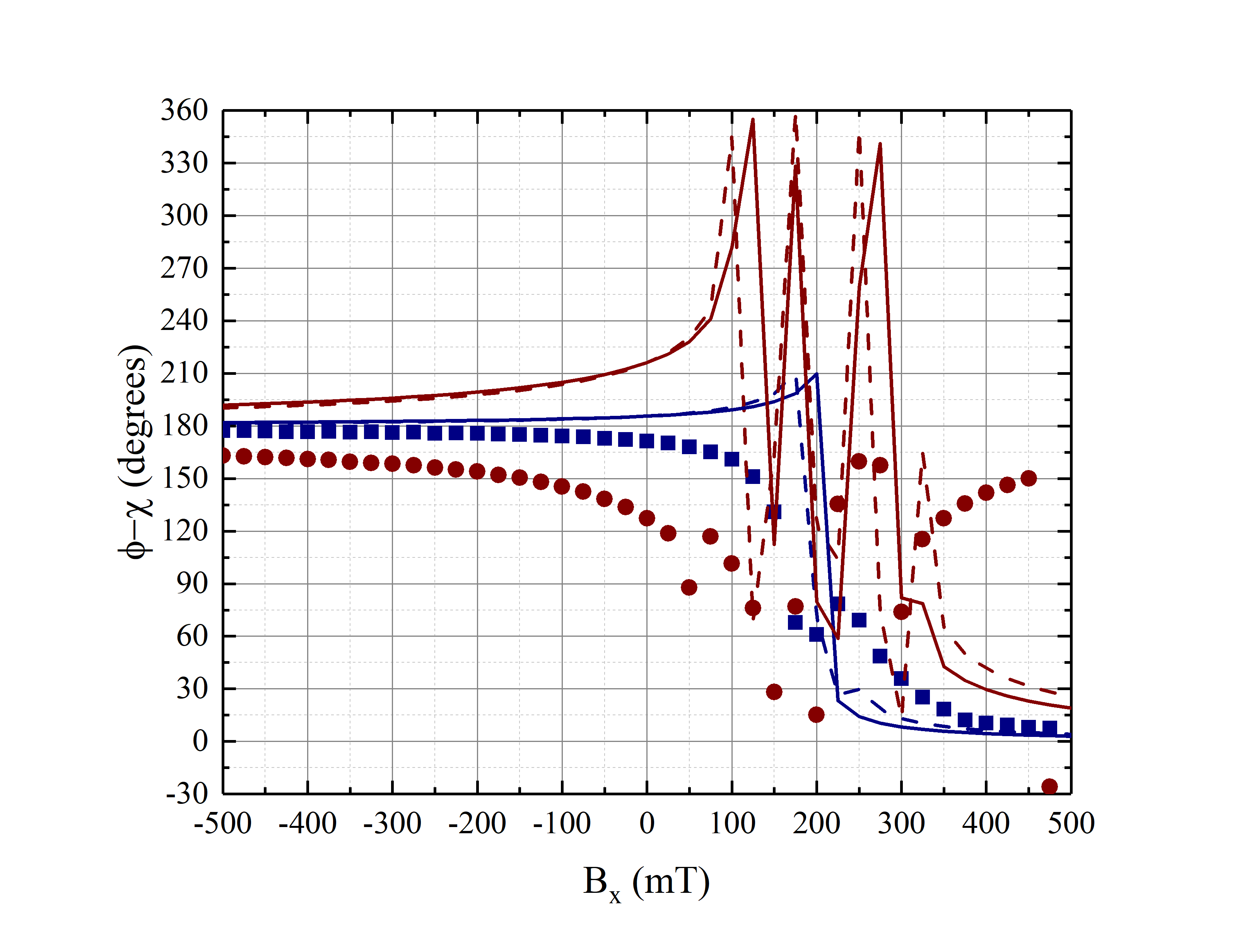}}
	\subfloat[Instantaneous DW magnetization angle ($\phi$).]{\includegraphics[trim=1cm 1cm 1cm 2cm, clip=true,scale=0.32]{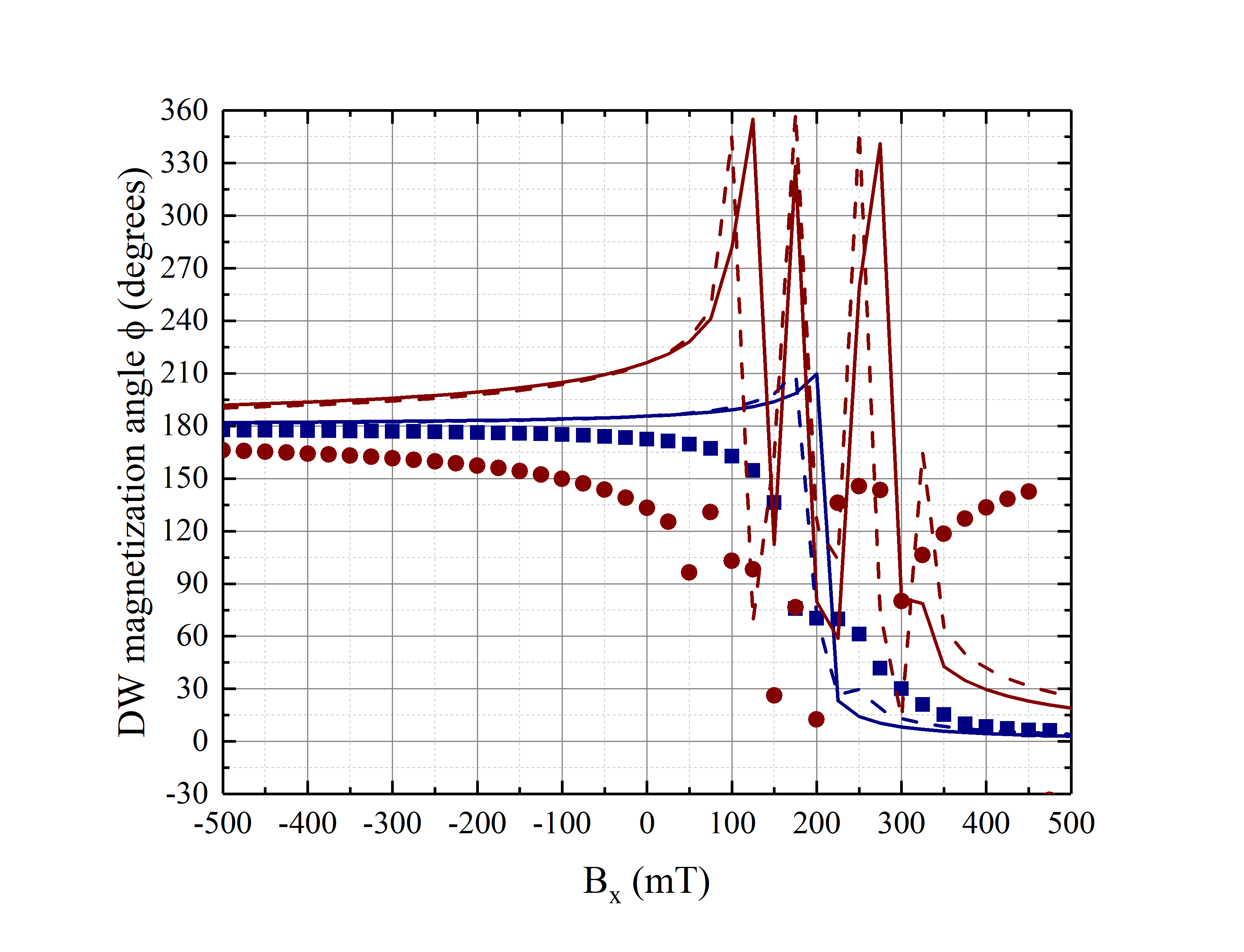}}\\
	\subfloat[Instantaneous DW tilting angle ($\chi$).]{\includegraphics[trim=1cm 1cm 1cm 2cm, clip=true,scale=0.32]{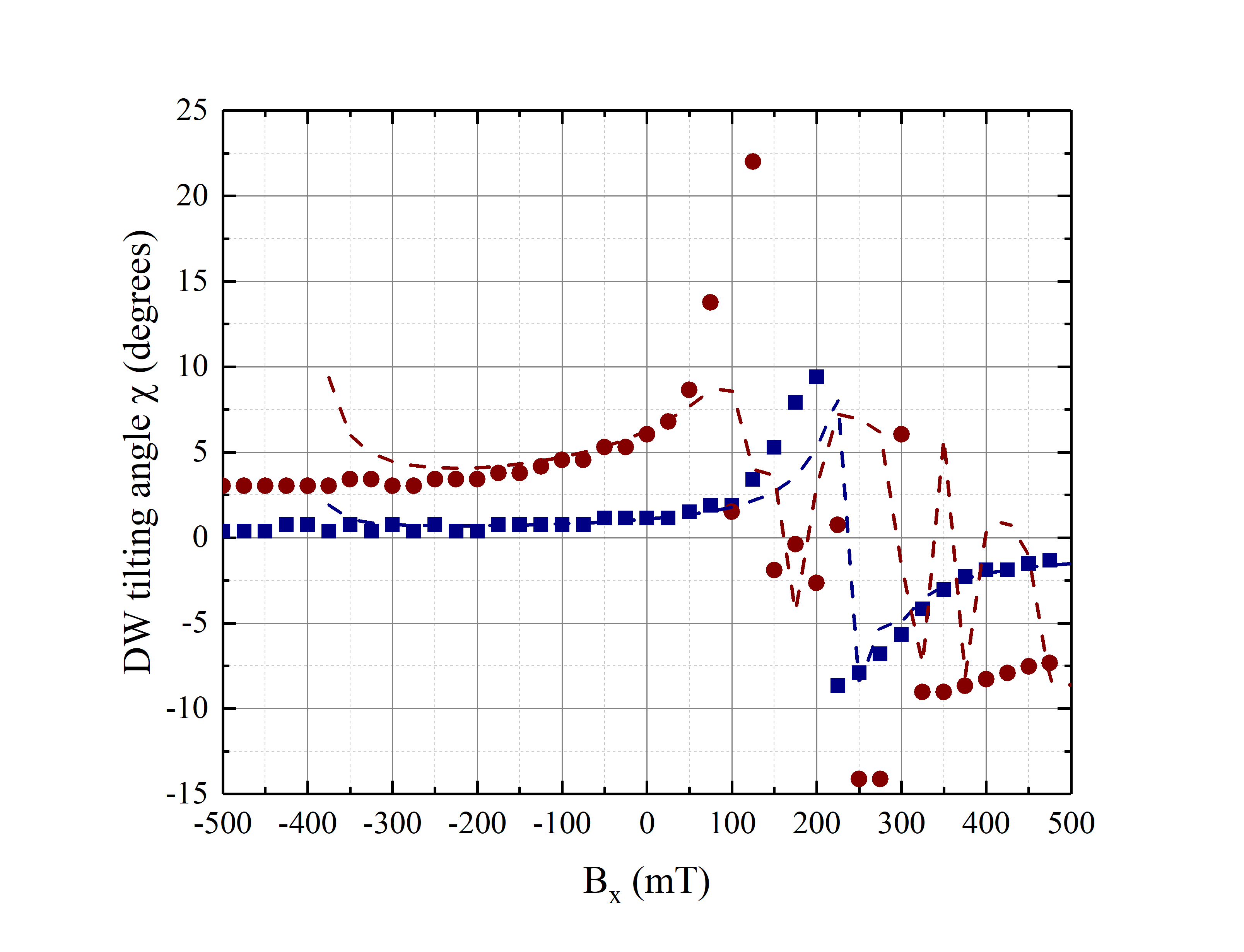}}
	\subfloat{\includegraphics[scale=0.4]{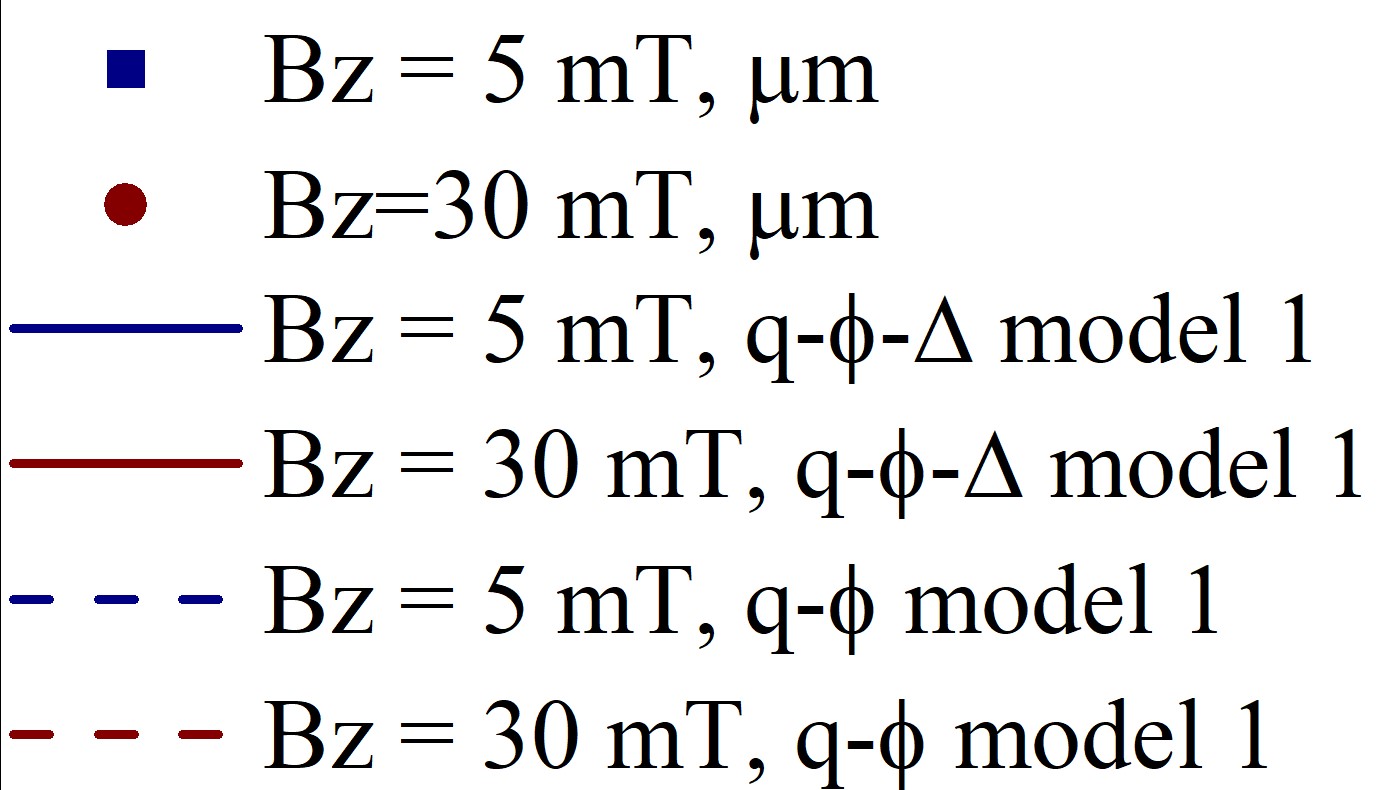}}
	\caption{DW characteristics for field-driven DW motion in $Pt/Co/Ni/Co/MgO/Pt$ with different out of plane and longitudinal fields applied. Only the collective coordinate models with highest accuracy in predicting the DW velocity are shown. We decided to show the average DW speed in place of the instantaneous velocity of the DW, due to the walker breakdown behavior; this behavior is observed in panels (b)-(e) for $50 \, mT \, < \, B_x \, < \, 300 \, mT$.}
	\label{fields_Bx_micromag_PtCoNiCoMgOPt}
\end{figure}

The two samples were studied under drive fields of $B_z = 5, 30 mT$ and longitudinal in-plane fields. The results of these micromagnetic simulations are presented in Figures \ref{fields_Bx_micromag} and \ref{fields_Bx_micromag_PtCoNiCoMgOPt}, and compared to the most accurate collective coordinate models.

Comparing the variation of velocity for the two samples, as depicted in Figure \ref{fields_Bx_micromag}.a and \ref{fields_Bx_micromag_PtCoNiCoMgOPt}.a, we see that in both cases the general trend with the drive field is the same; the velocity and nonlinearity of the curve increases with increasing drive field ($B_z$), while changing $B_x$ tunes the velocity to an extend (with the curve having a minimum with respect to the longitudinal field). The DW velocity predictions are quantitatively in agreement the behavior observed in experiments \cite{VAN-15a, TON-15}. 

\begin{figure}[H]
	\centering
	\subfloat[$B_z = 10 \, mT$, $B_x = 225 \, mT$.]{\includegraphics[trim = 1100 0 1100 0 , clip=true, scale=0.2]{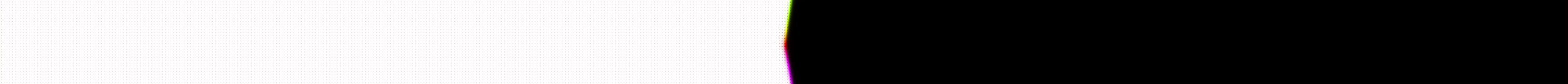}}
	\hspace{0.3 in}
	\subfloat[$B_z = 30 \, mT$, $B_x = 225 \, mT$.]{\includegraphics[trim = 1100 0 1100 0 , clip=true, scale=0.2]{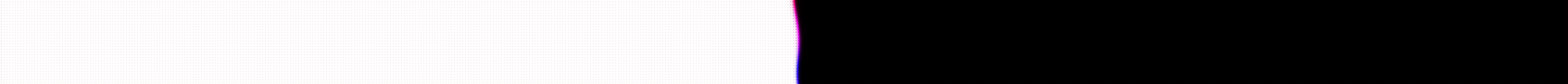}}
	\hspace{0.3 in}
	\subfloat[$B_z = 30 \, mT$, $B_x = 100 \, mT$.]{\includegraphics[trim = 1100 0 1100 0 , clip=true, scale=0.2]{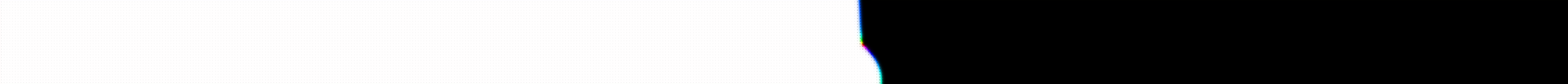}}
	\caption{Snapshots of the moving DW under different conditions in the $Pt/Co/Ni/Co/MgO/Pt$ sample. The rapid change of magnetization along the DW owing to the Walker Breakdown can be observed.}
	\label{snapshots_WB_fields}
\end{figure}

However, the $Pt/Co/Ni/Co/MgO/Pt$ sample shows the additional effect of a sudden drop in DW velocity for a range of in-plane fields applied. This Walker Breakdown (WB) like behavior \cite{SCH-74} was verified by looking at the snapshots of the DW motion (depicted in Figure \ref{snapshots_WB_fields}), where we can see local precession of the magnetization and formation of vertical Bloch lines arising from the edge that affect the DW structure \cite{YOS-16}. This behavior could be attributed to the higher anisotropy of this material, which reduces the local field needed to reach WB. Note that this behavior is local; the wall does not oscillate back and forth as a single object, but the overall effect of the local precession of magnetization over time is equivalent to the DW moving back and forth rigidly, which is why the collective coordinate model can replicate this effect to an extent. 

In terms of the CCMs, we found models without canting to better reproduce the results for $Pt/Co/Ni/Co/MgO/Pt$ (where canting and DW tilting are small). However, in this material the DW width parameter $\Delta$ was important in predicting the DW behavior properly. In $Pt/CoFe/MgO$ with its higher canting and tilting of the DW, we found that model 3 (with inherent canting) is better suited in predicting the DW behavior with the two coordinate $q-\phi$ model.

Looking closely at Figures \ref{fields_Bx_micromag}.b and \ref{fields_Bx_micromag_PtCoNiCoMgOPt}.b, we find a serious flaw in model 3; this model seems to not be able to predict the DW width correctly, which in turn can affect its outputs. As such, when a two coordinate form of this model is used, it is able to better predict the DW motion. This also shows why this model is not suitable for the $Pt/Co/Ni/Co/MgO/Pt$ sample where lack of canting and tilting mean $\Delta$ is one of the main parameters affecting the DW. Overall, this observation suggests that the $q-\phi$ form of model 3 is the most suitable for studying DW motion in these systems.

A major difference between the two cases can be observed in the DW's tilting behavior; while $Pt/CoFe/MgO$ DWs always maintain a positive tilting, in the case of $Pt/Co/Ni/Co/MgO/Pt$ negative tilting can be observed which is likely due to the lower DMI strengths and the higher applied fields used (see Figures \ref{fields_Bx_micromag}.e and \ref{fields_Bx_micromag_PtCoNiCoMgOPt}.e).

Another notable feature of the DW behavior could be seen in Figure \ref{fields_Bx_micromag}.c where at a specific field $\phi-\chi \sim 0$ independent of the drive field, while in Figure \ref{fields_Bx_micromag_PtCoNiCoMgOPt}.c a point could be observed for which $\phi-\chi \sim \frac{\pi}{2}$. We identify these points as critical in-plane fields which will be discussed in details in later sections.

\pdfcomment{For the cases with angles there is too much data. Should I remove some of the models?}

\subsubsection{Current-Driven Case}

\begin{figure}
	\centering
	\subfloat[DW velocity.]{\includegraphics[trim=1cm 1cm 1cm 2cm, clip=true,scale=0.32]{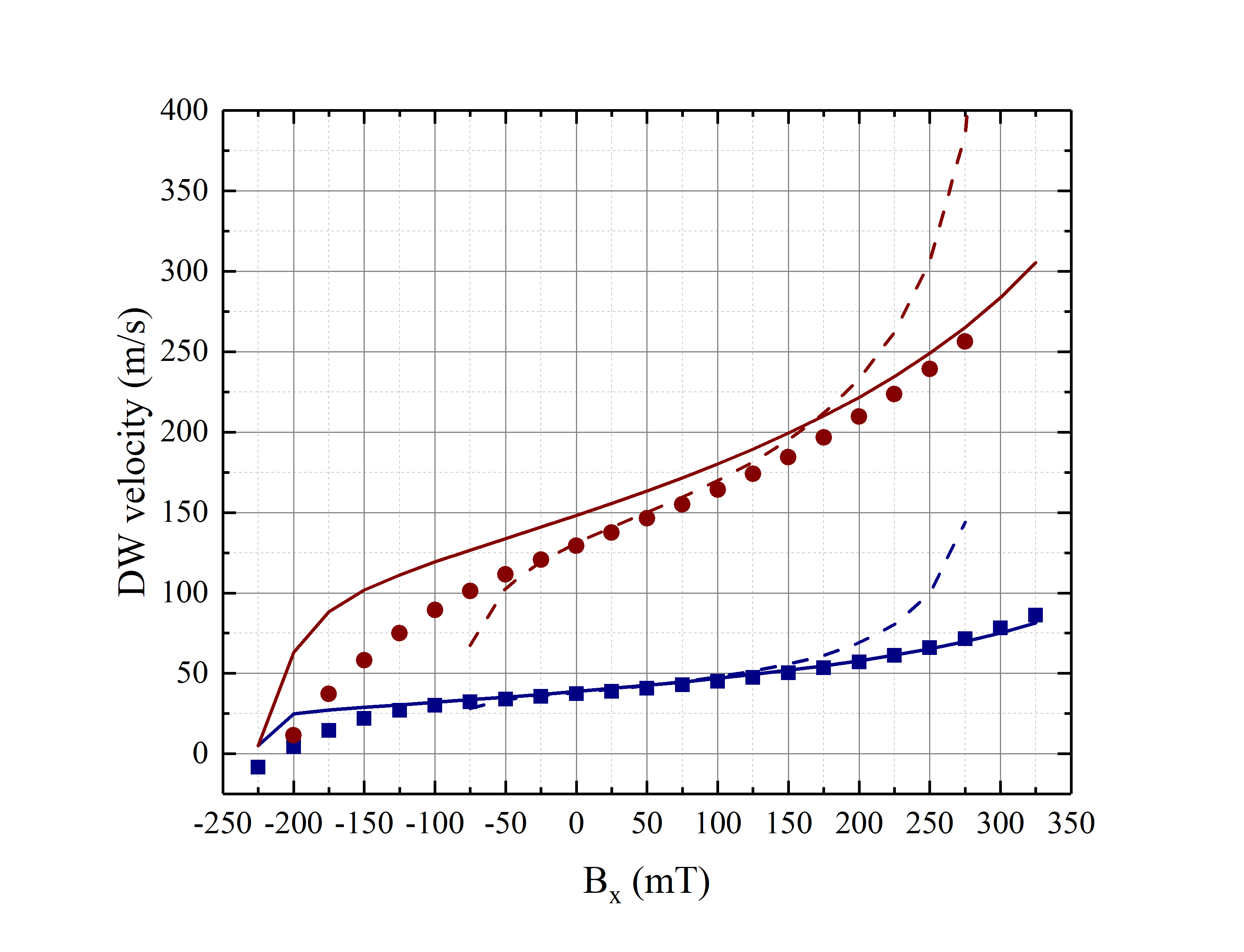}}
	\subfloat[DW width parameter ($\Delta$).]{\includegraphics[trim=1cm 1cm 1cm 2cm, clip=true,scale=0.32]{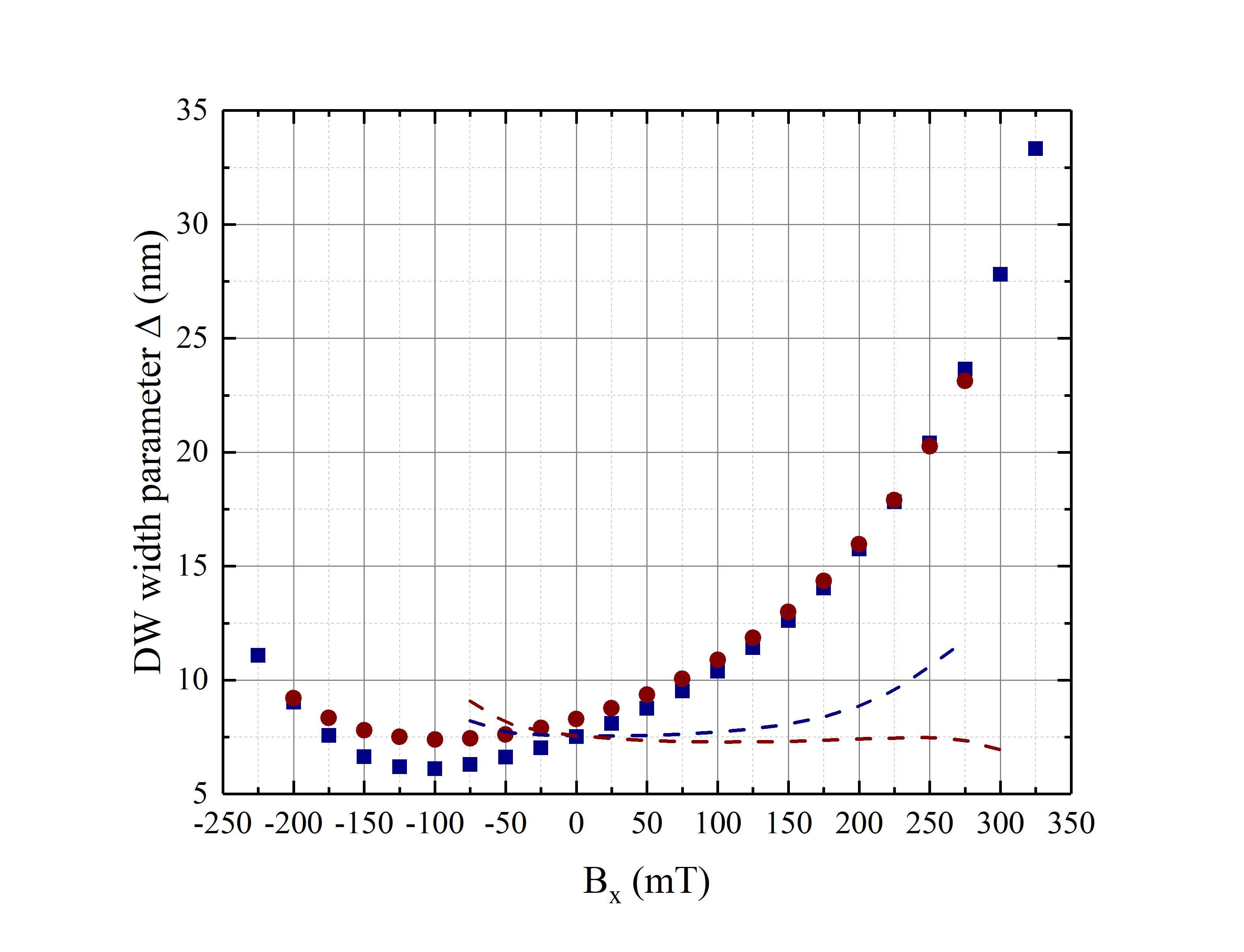}}\\
	\subfloat[$\phi-\chi$]{\includegraphics[trim=1cm 1cm 1cm 2cm, clip=true,scale=0.32]{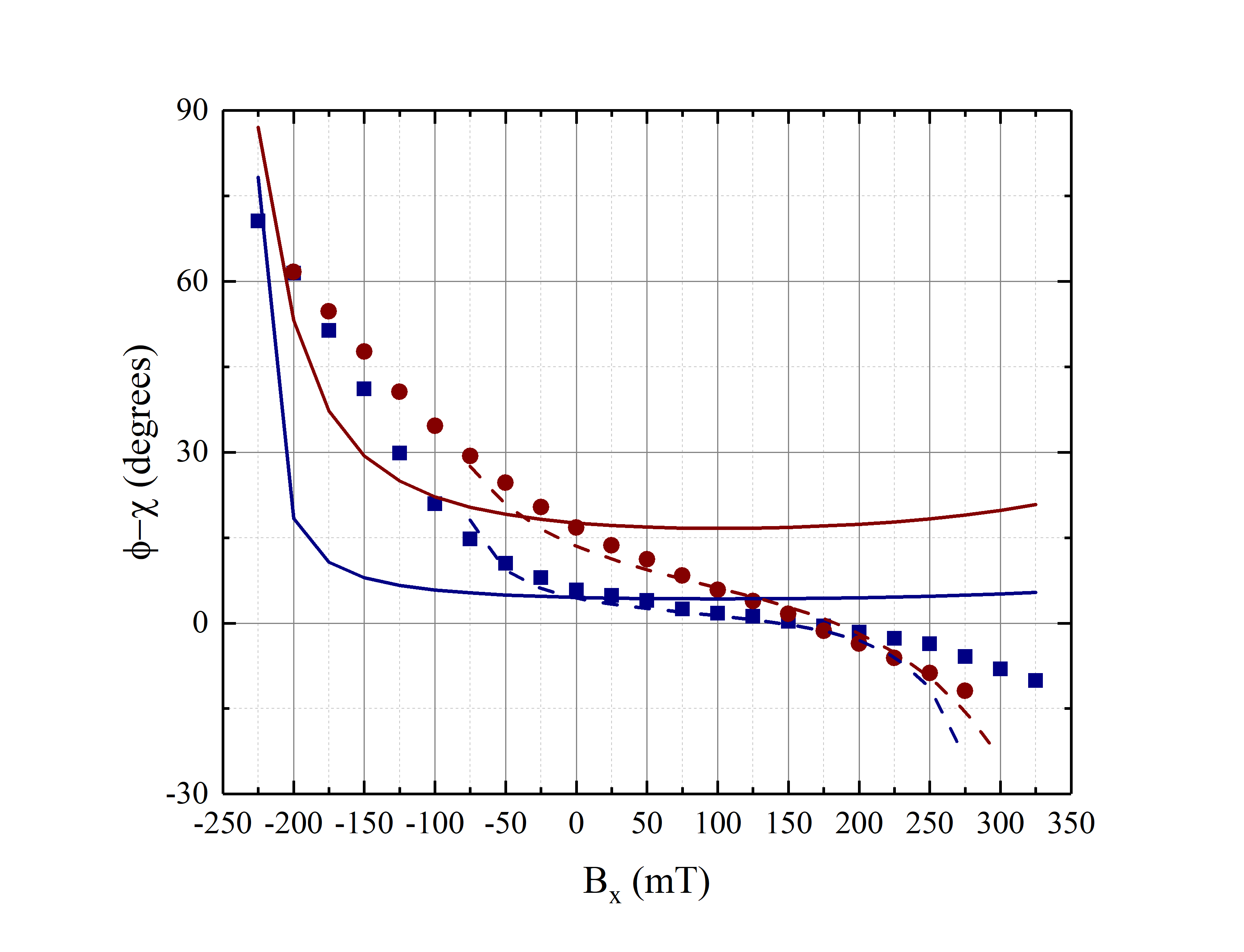}}
	\subfloat[DW magnetization angle ($\phi$).]{\includegraphics[trim=1cm 1cm 1cm 2cm, clip=true,scale=0.32]{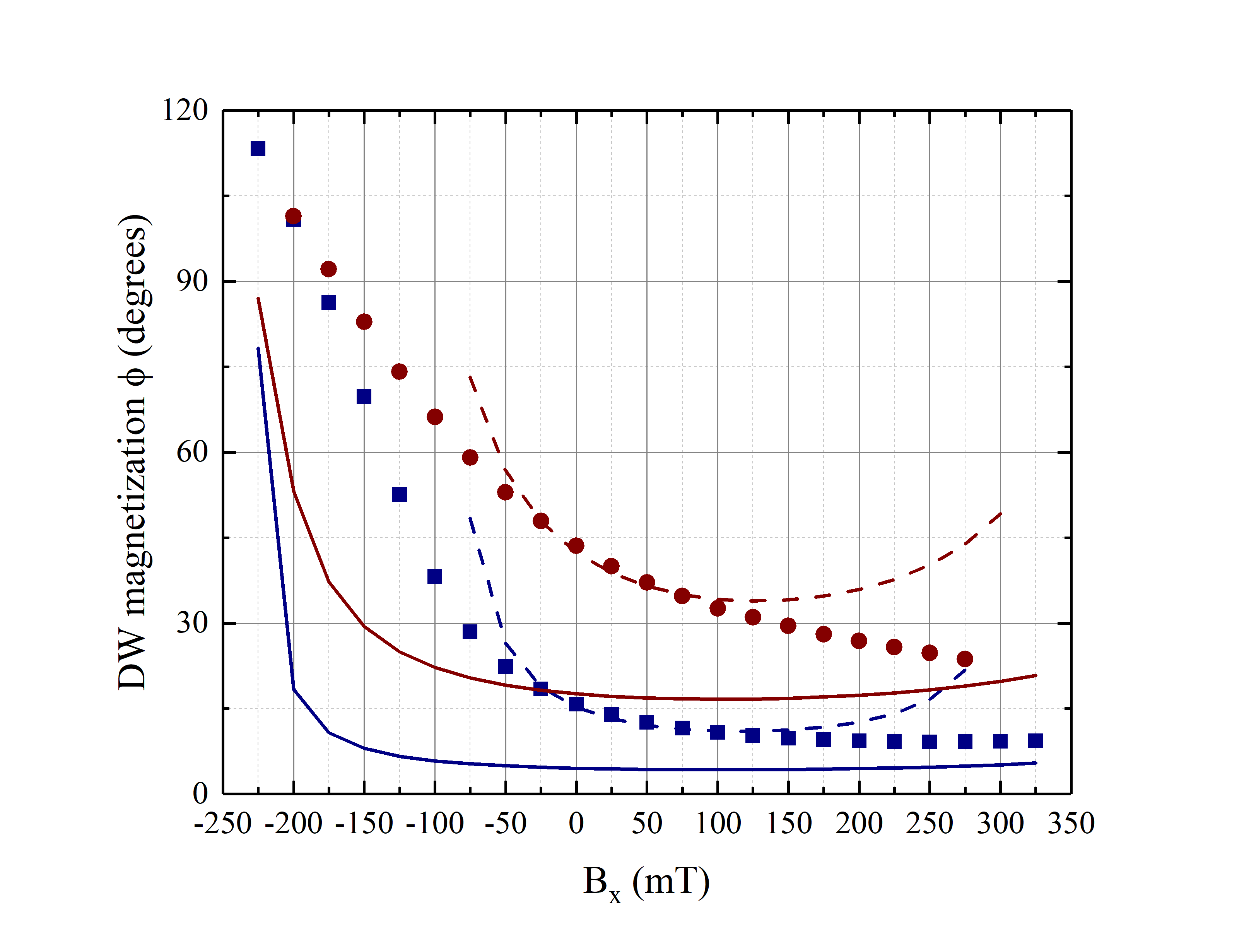}}\\
	\subfloat[DW tilting angle ($\chi$).]{\includegraphics[trim=1cm 1cm 1cm 2cm, clip=true,scale=0.32]{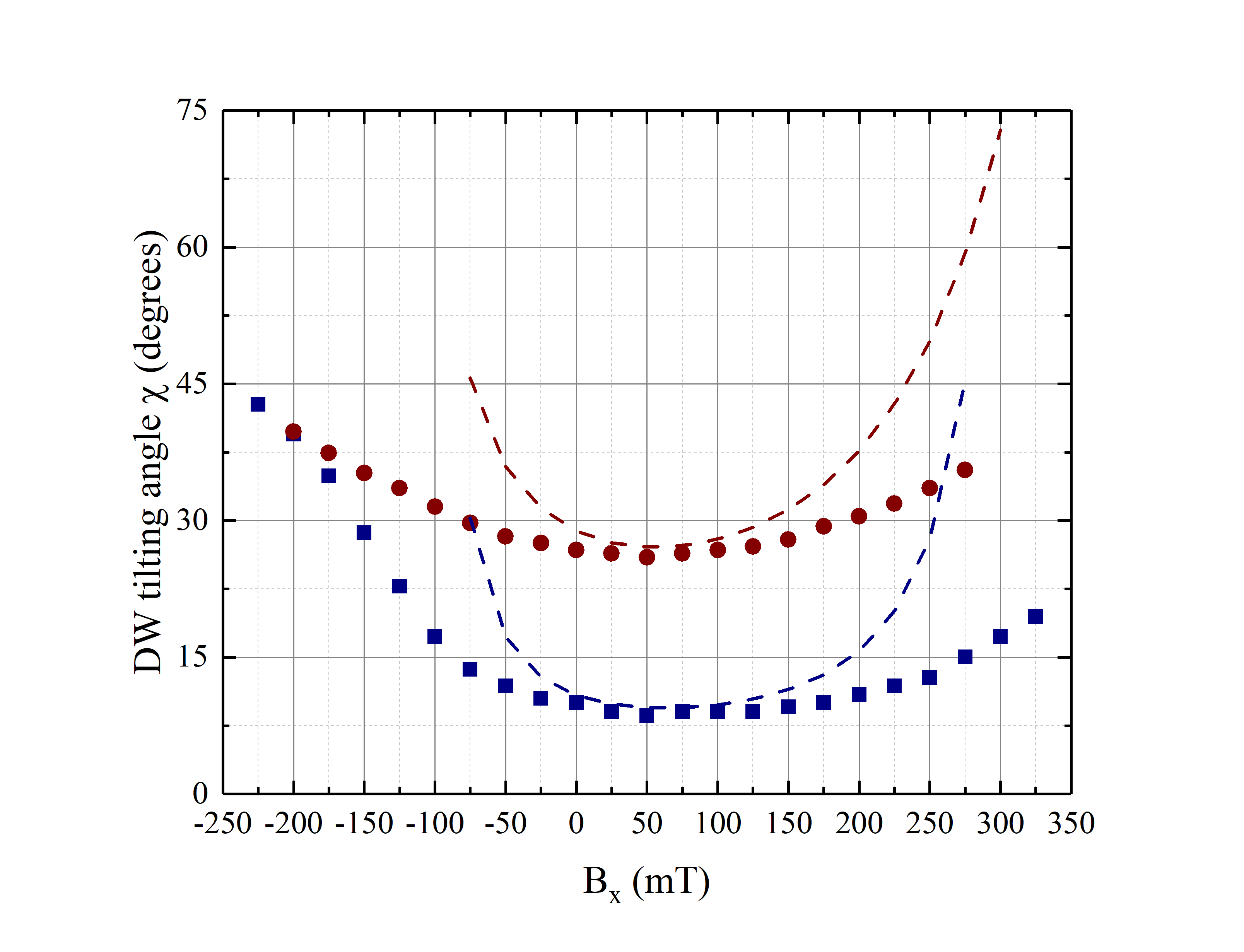}}
	\subfloat{\includegraphics[scale=0.4]{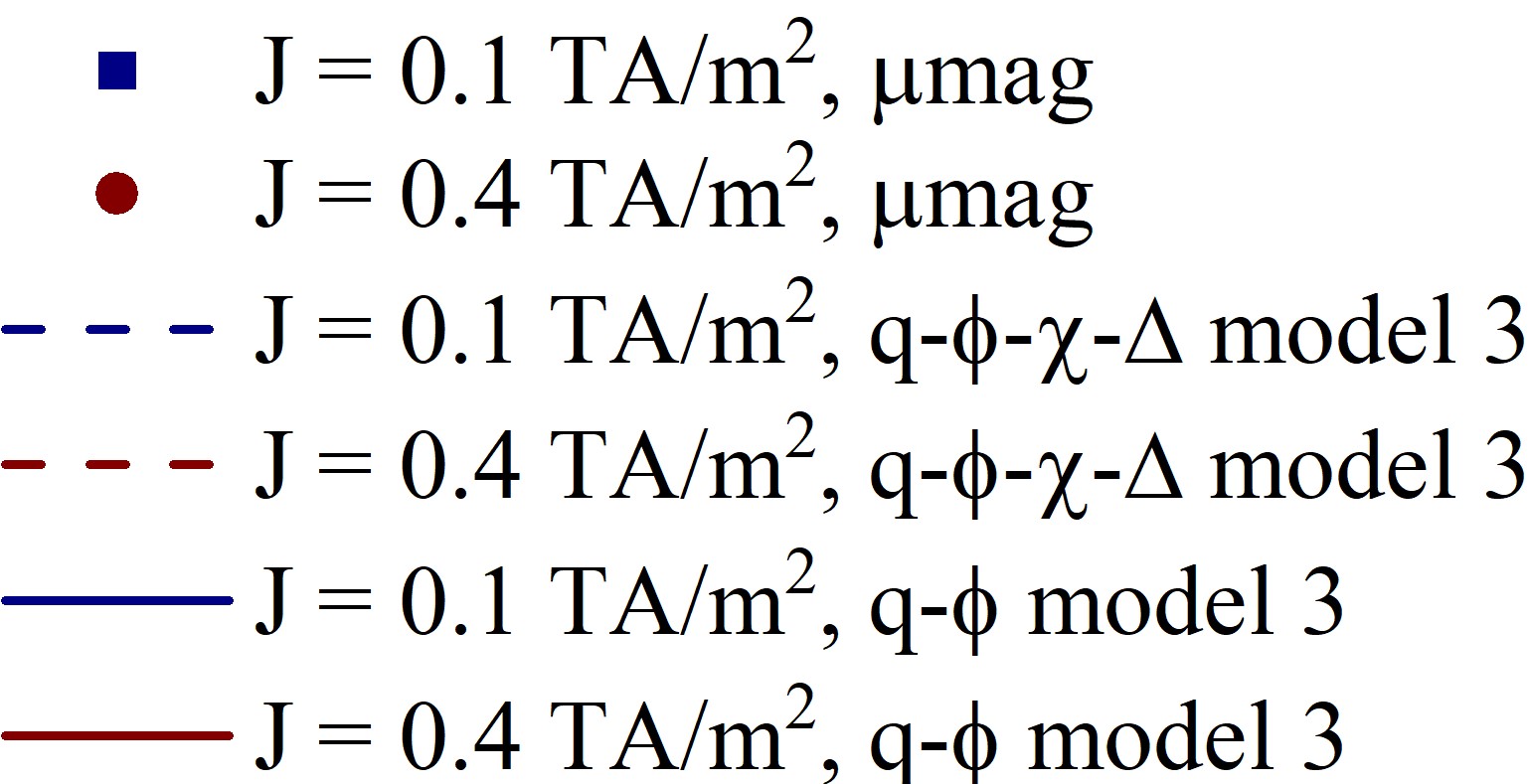}}
	\caption{Instantaneous steady state DW characteristics for SHE-driven DW motion in $Pt/CoFe/MgO$ with different currents and longitudinal fields applied. Only the collective coordinate models with highest accuracy in predicting the DW velocity are shown.}
	\label{current_Bx_micromag}
\end{figure}

\begin{figure}
	\centering
	\subfloat[DW velocity.]{\includegraphics[trim=1cm 1cm 1cm 2cm, clip=true,scale=0.32]{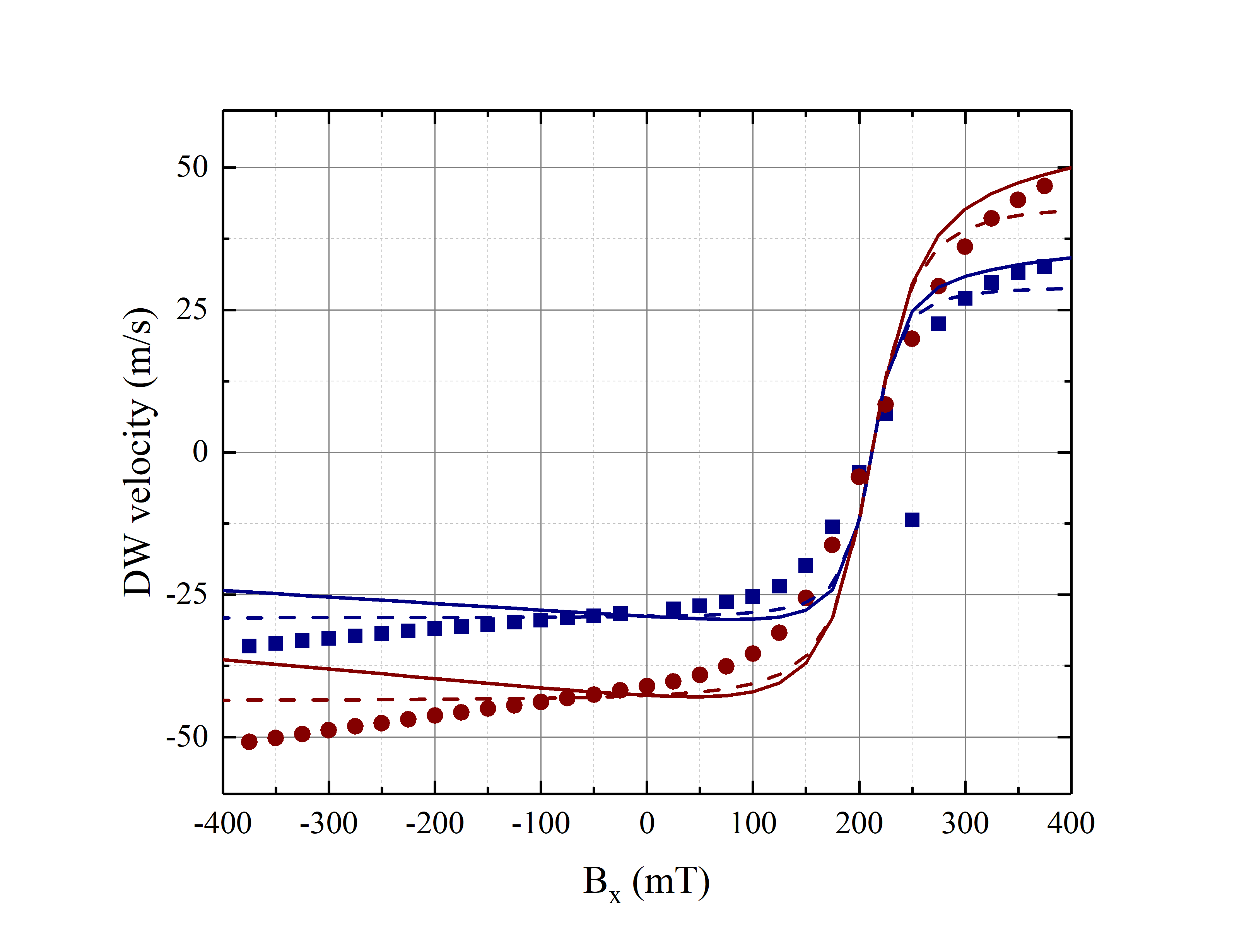}}
	\subfloat[DW width parameter ($\Delta$).]{\includegraphics[trim=1cm 1cm 1cm 2cm, clip=true,scale=0.32]{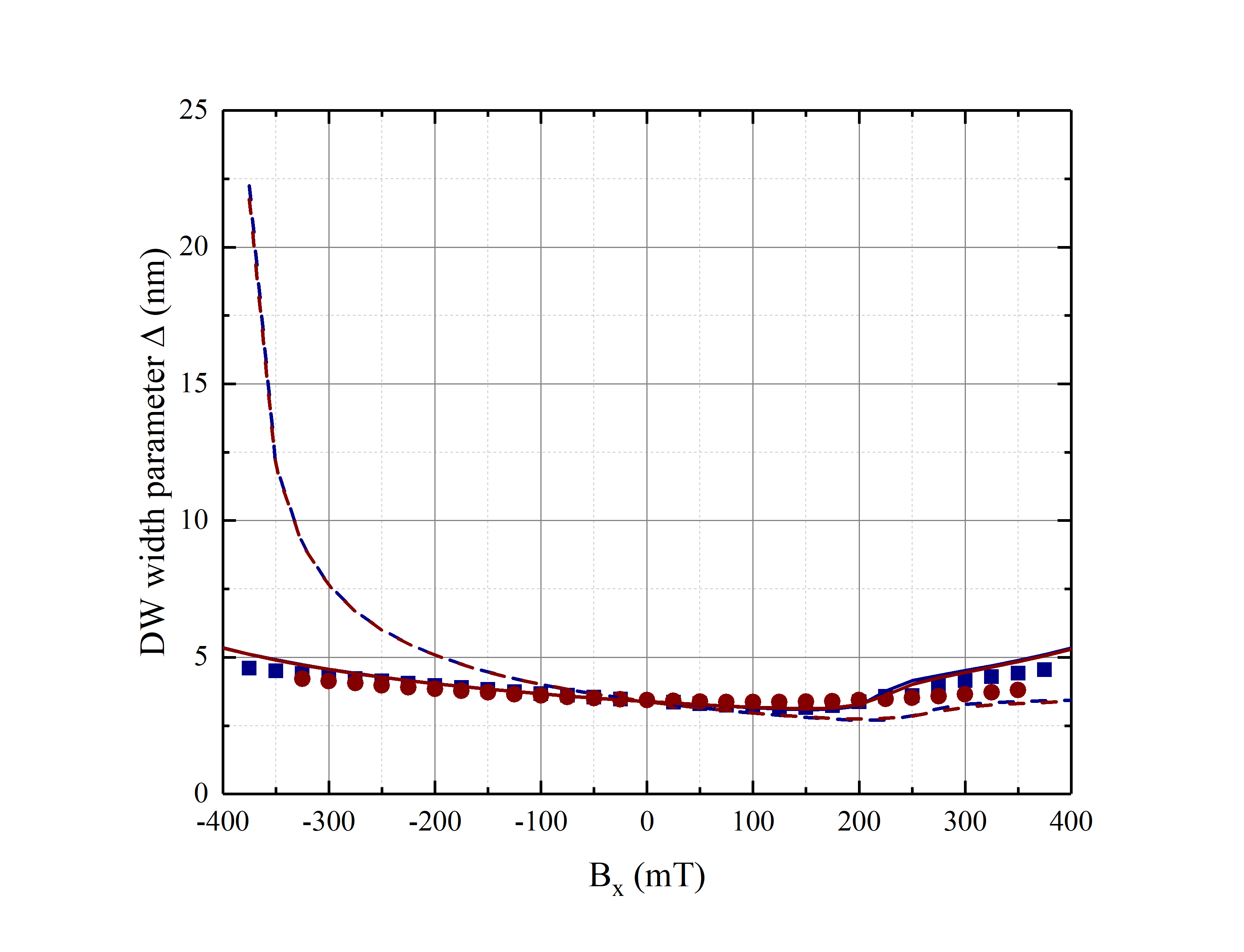}}\\
	\subfloat[$\phi-\chi$]{\includegraphics[trim=1cm 1cm 1cm 2cm, clip=true,scale=0.32]{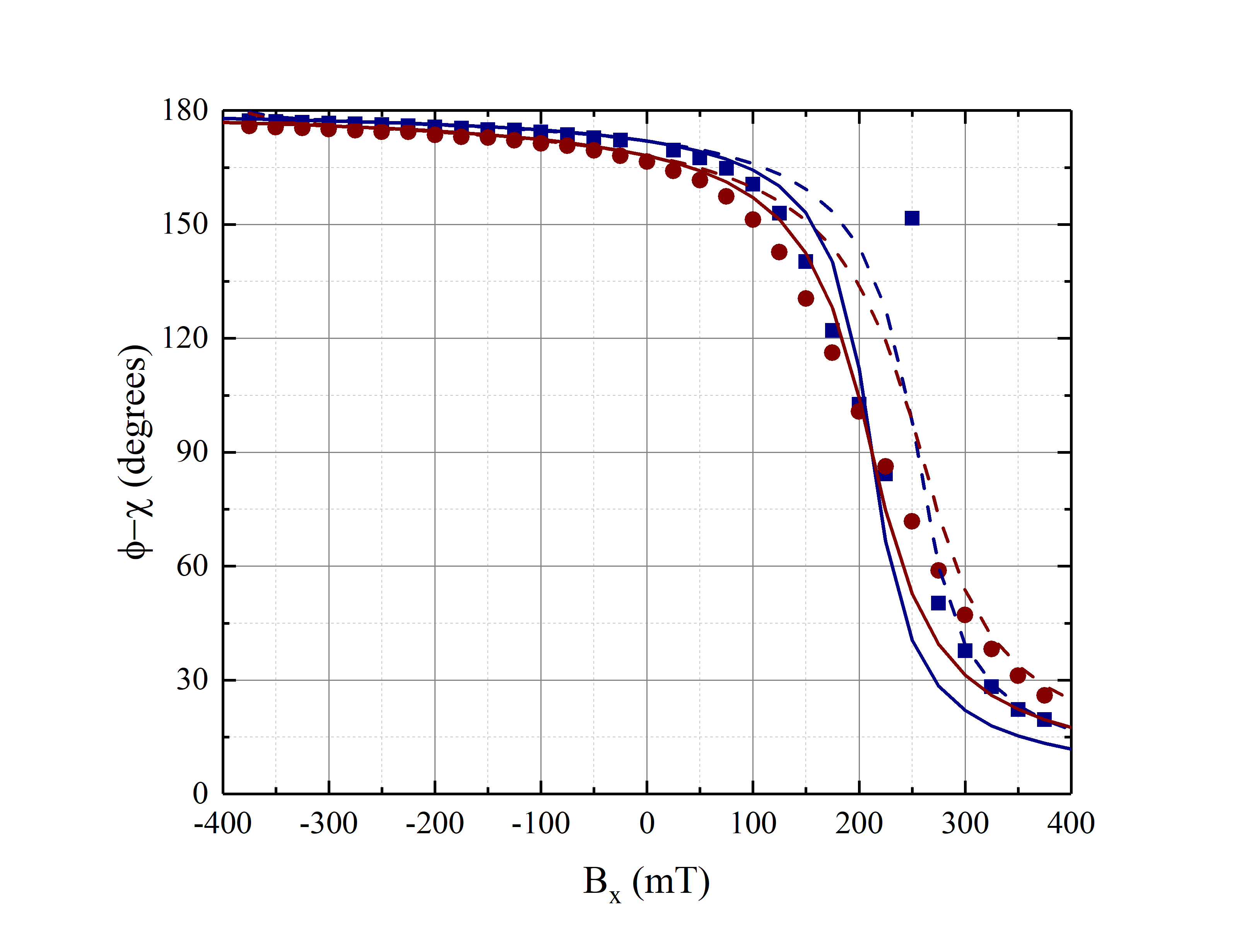}}
	\subfloat[DW magnetization angle ($\phi$).]{\includegraphics[trim=1cm 1cm 1cm 2cm, clip=true,scale=0.32]{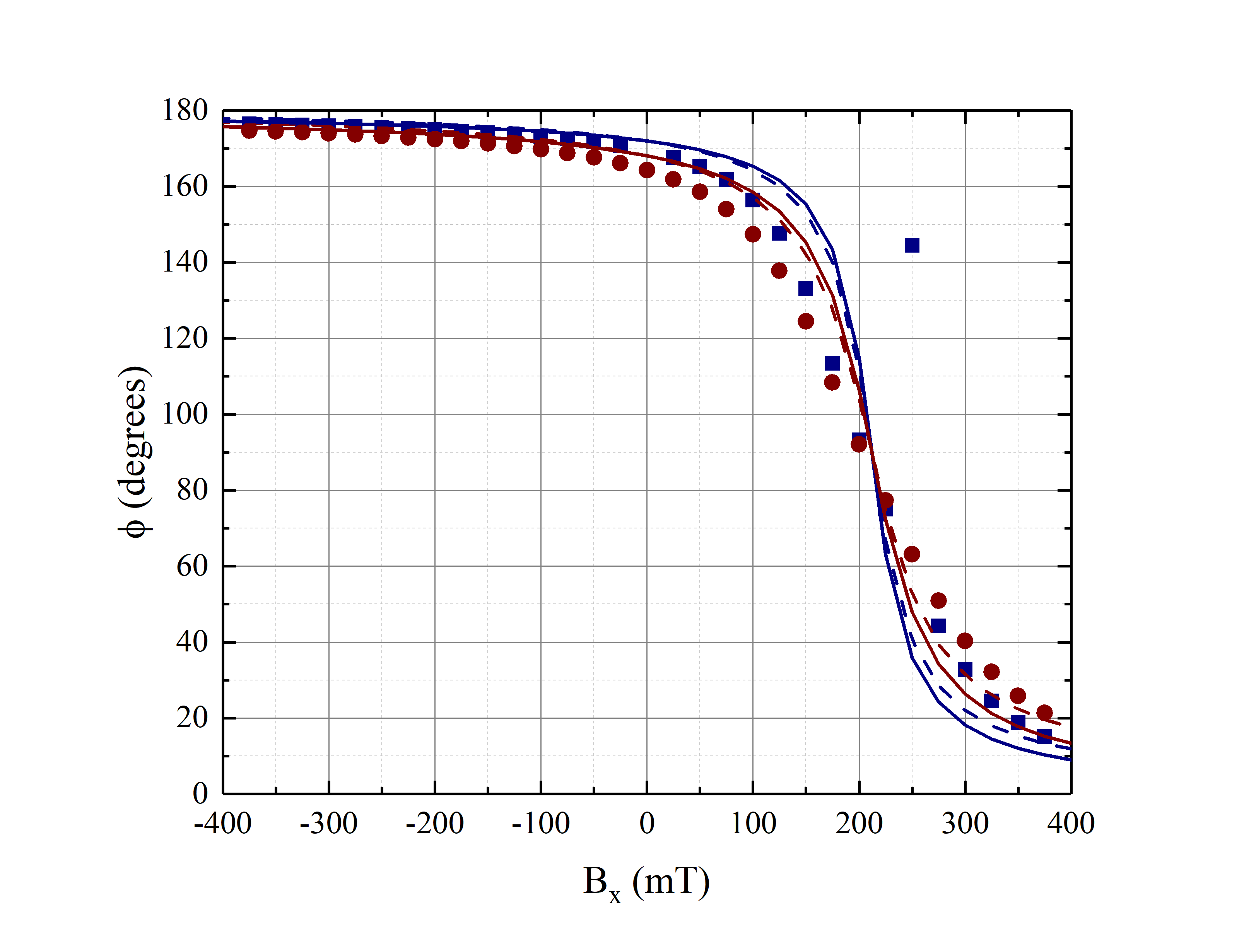}}\\
	\subfloat[DW tilting angle ($\chi$).]{\includegraphics[trim=1cm 1cm 1cm 2cm, clip=true,scale=0.32]{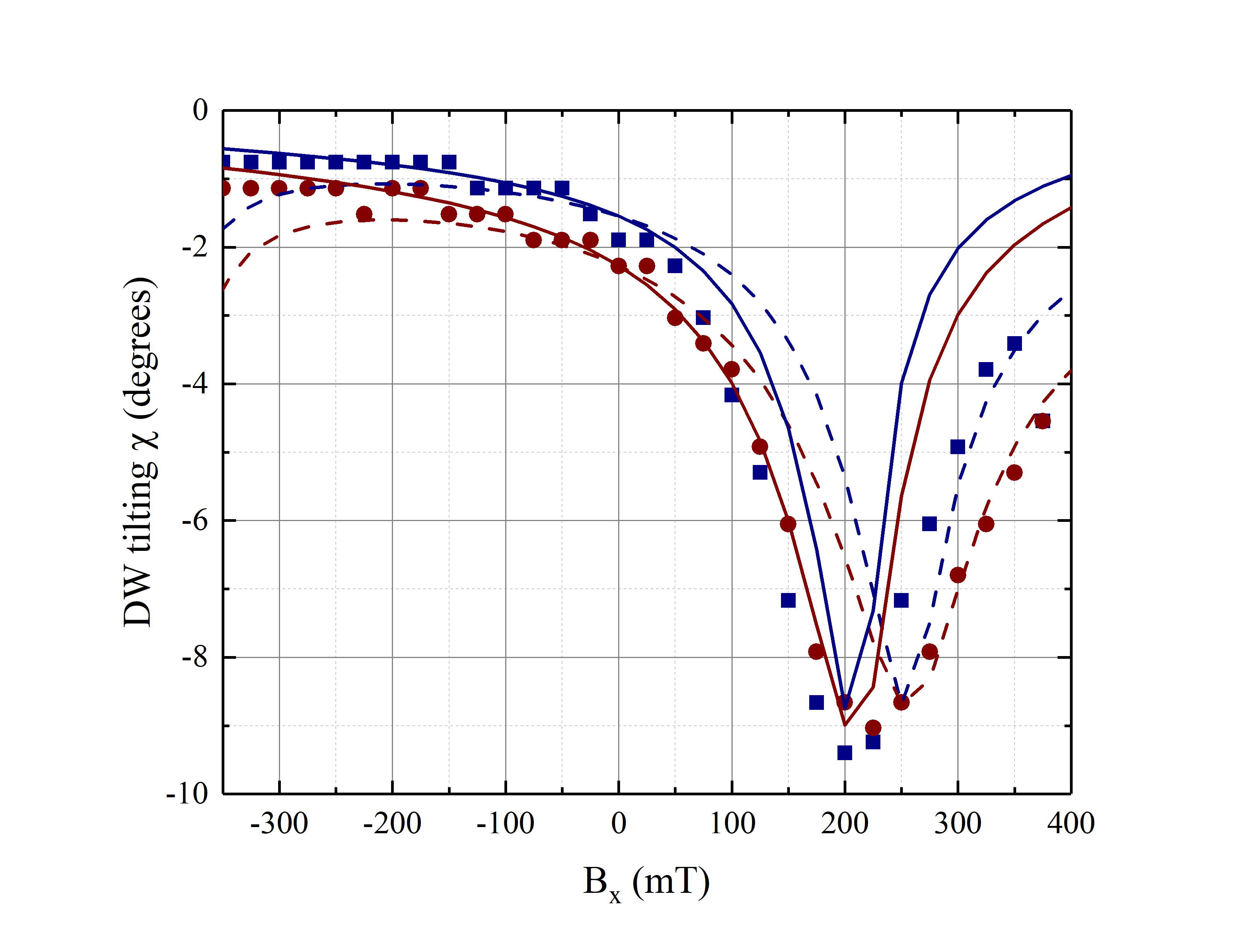}}
	\subfloat{\includegraphics[scale=0.4]{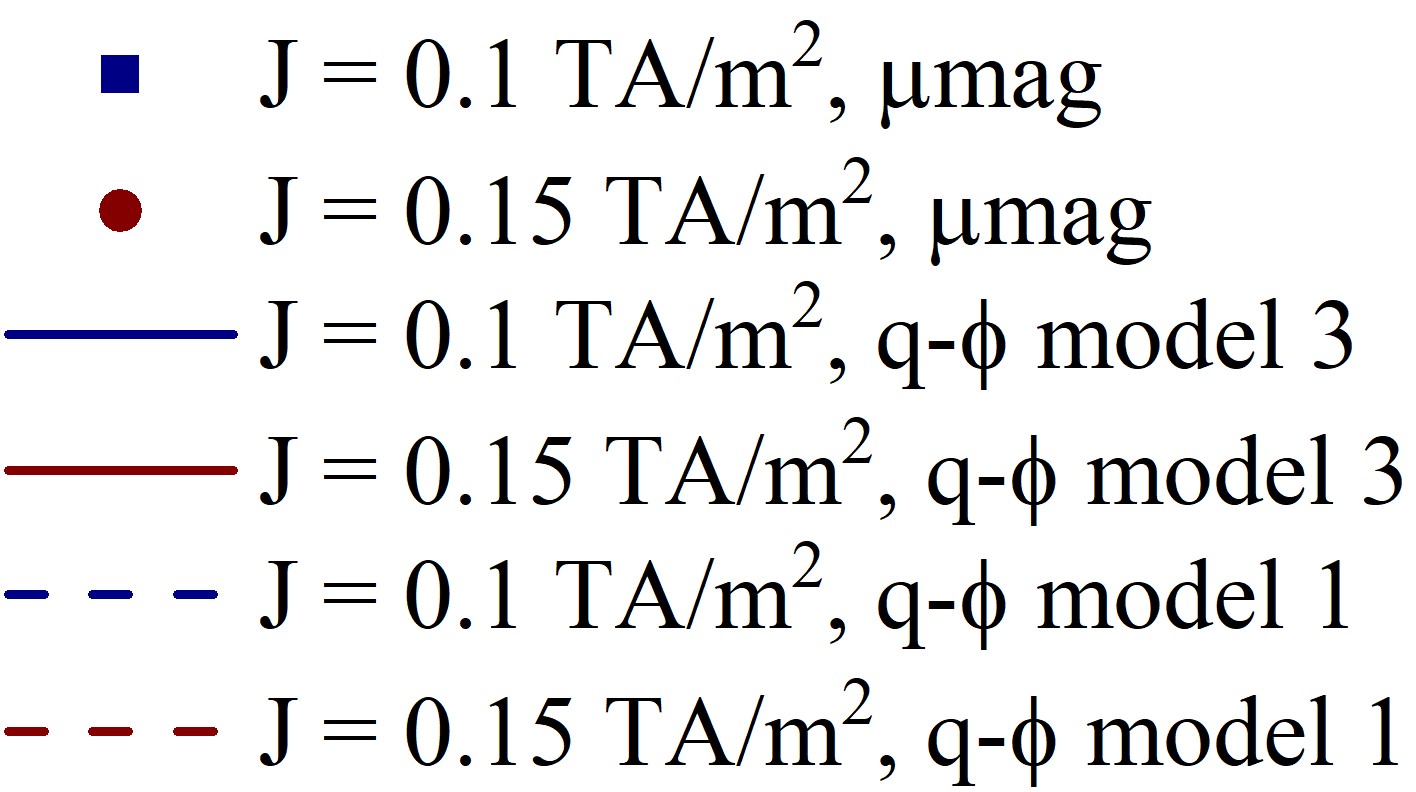}}
	\caption{Instantaneous steady state DW characteristics for SHE-driven DW motion in $Pt/Co/Ni/Co/MgO/Pt$ with different currents and longitudinal fields applied. We found that the best models in these cases were two coordinate models. Canting has a minimal effect on the outputs. We also observed Walker Breakdown in these cases. Only the collective coordinate models with highest accuracy in predicting the DW velocity are shown.}
	\label{current_Bx_micromag_PtCoNiMgOPt}
\end{figure}

Figures \ref{current_Bx_micromag} and \ref{current_Bx_micromag_PtCoNiMgOPt} illustrates the results of micromagnetic simulations for current-driven DW motion under longitudinal fields. 

The trends observed in the velocity of current-driven DW motion (Figures \ref{current_Bx_micromag}.a and \ref{current_Bx_micromag_PtCoNiMgOPt}.a) are in general agreement with published experimental results \cite{EMO-13, EMO-14, RYU-13, RYU-14}. A somewhat linear behavior is observed for low longitudinal fields, which becomes non-linear as the in-plane field increases. The non-linearity in behavior also seems to increase with increasing current. 

We also observe a point where the DW velocity is zero in both cases; the in-plane field at which this happens is another critical point of interest. For the system with lower DMI the nonlinearity in the DW velocity seems to be observable mainly around this point, while in the system with larger DMI this nonlinear behavior is observed over all in-plane fields studied. Interestingly, this in-plane field seems to have an additional feature: the DW will have the same tilting angle $\chi$ for different drive interactions ((Figures \ref{current_Bx_micromag}.e and \ref{current_Bx_micromag_PtCoNiMgOPt}.e)). In the system with the higher anisotropy and lower DMI, we also observe that at this point $\phi-\chi \sim 90 \circ$ (a fully Bloch DW). 

One unexpected result was the presence of Walker Breakdown in our initial current-driven simulations of $Pt/Co/Ni/Co/MgO/Pt$; however, in these cases while a vertical Bloch line is nucleated, it is short-lived and simulating for longer durations shows that this is just a transitory effect. One point with such an effect can be observed in Figure \ref{current_Bx_micromag_PtCoNiMgOPt}.a as an outlier at $B_x = 250 mT$. 

In terms of angles, an in-plane field for which  $\phi-\chi \sim 0$ (Figures \ref{current_Bx_micromag}.c and \ref{current_Bx_micromag_PtCoNiMgOPt}.c), in the $Pt/CoFe/MgO$ sample we also see a case of $\phi=0$ (Figures \ref{current_Bx_micromag}.d) which is absent in the $Pt/Co/Ni/Co/MgO/Pt$ sample. This could be a point of interest.

In terms of the CCMs, we saw results similar to the field-driven case, with models without canting being more suitable for the high anisotropy system and those with canting more suitable for the low anisotropy high DMI system.

\pdfcomment{I could add J=1 TA/m instead of 0.15}

\FloatBarrier
\subsection{Domain Wall Motion Under Transverse Fields}

\begin{figure}
	\subfloat[DW velocity.]{\includegraphics[trim=1cm 1cm 1cm 2cm, clip=true,scale=0.32]{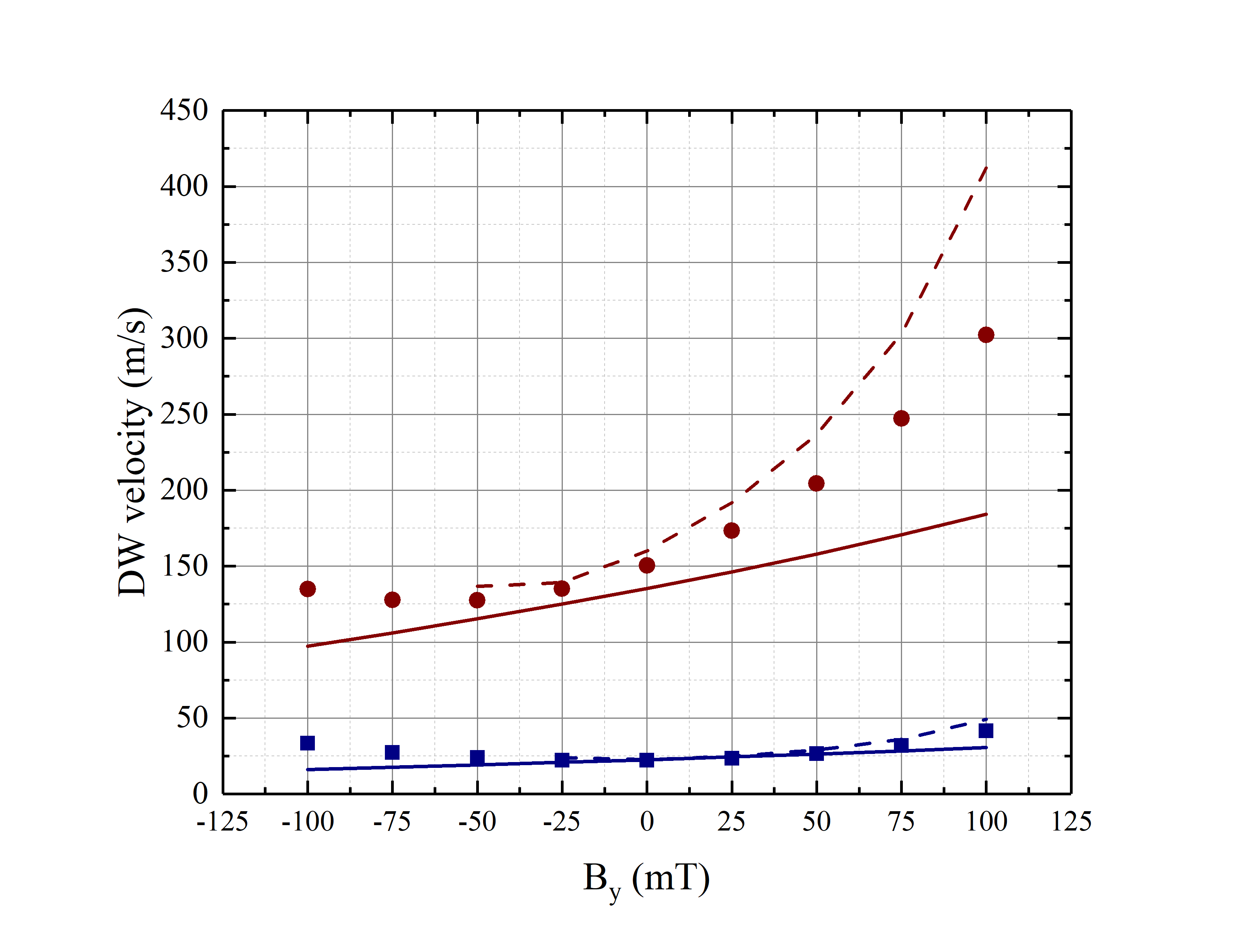}}
	\subfloat[DW width parameter ($\Delta$).]{\includegraphics[trim=1cm 1cm 1cm 2cm, clip=true,scale=0.32]{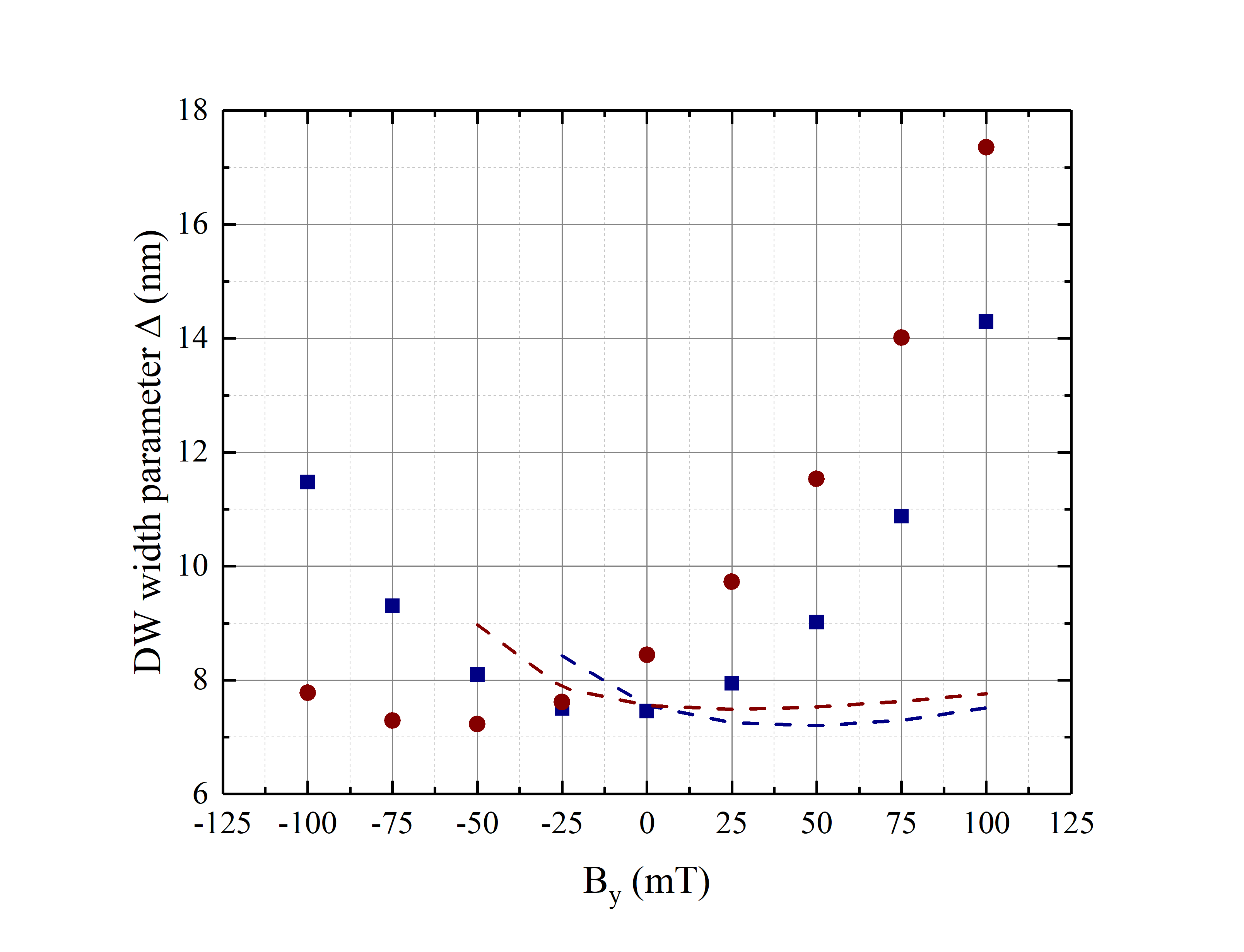}}\\
	\subfloat[$\phi-\chi$]{\includegraphics[trim=1cm 1cm 1cm 2cm, clip=true,scale=0.32]{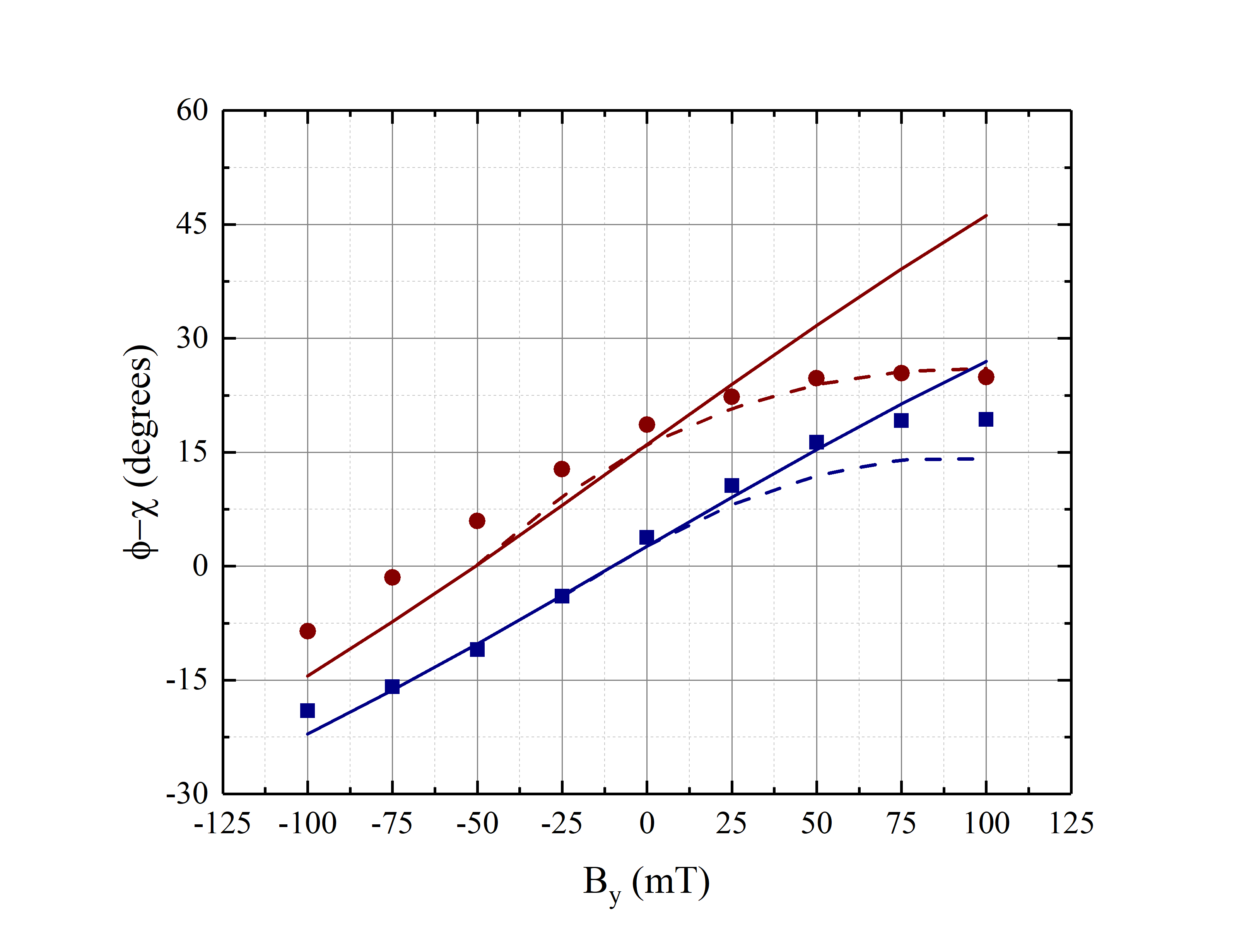}}
	\subfloat[DW magnetization angle ($\phi$).]{\includegraphics[trim=1cm 1cm 1cm 2cm, clip=true,scale=0.32]{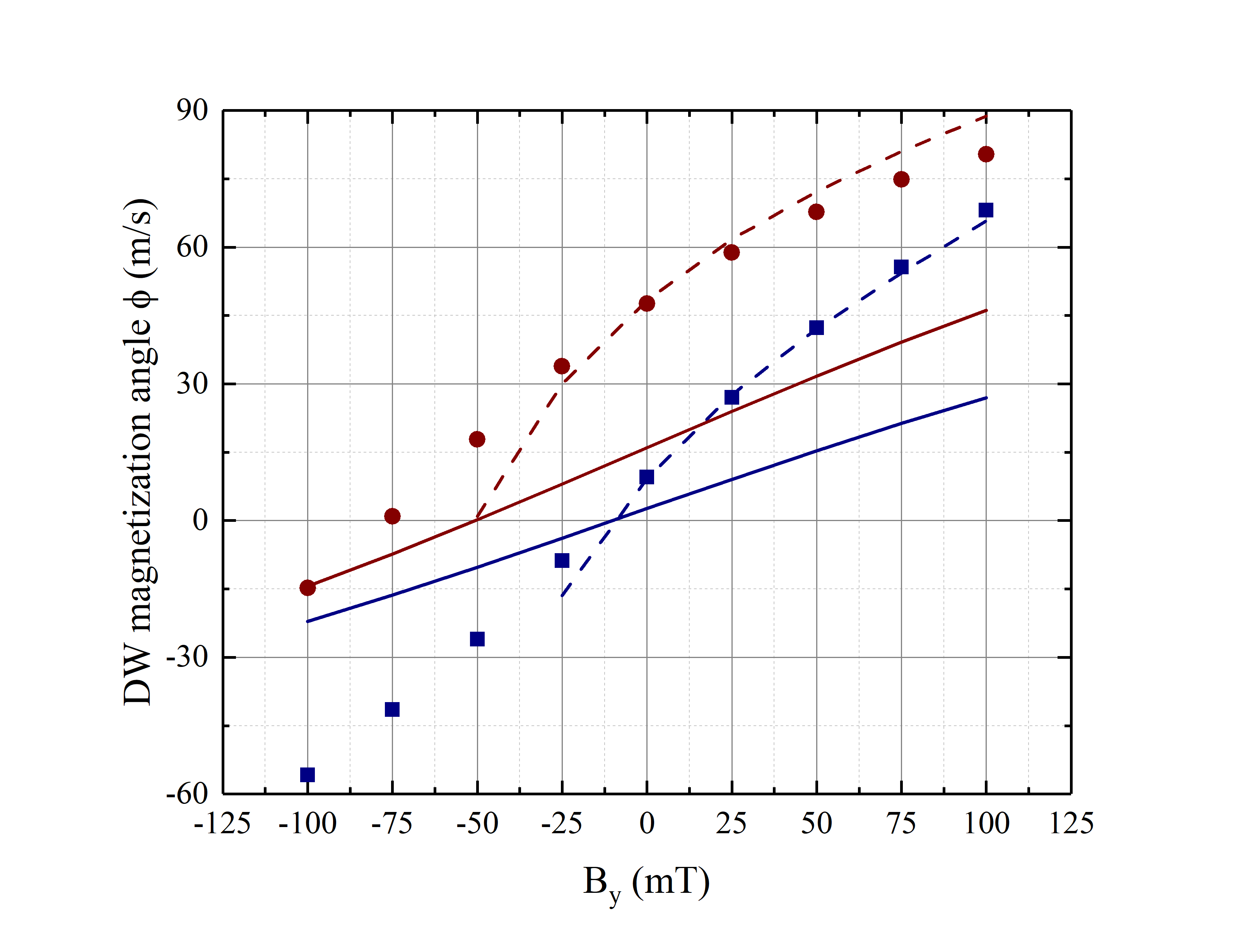}}\\
	\subfloat[DW tilting angle ($\chi$).]{\includegraphics[trim=1cm 1cm 1cm 2cm, clip=true,scale=0.32]{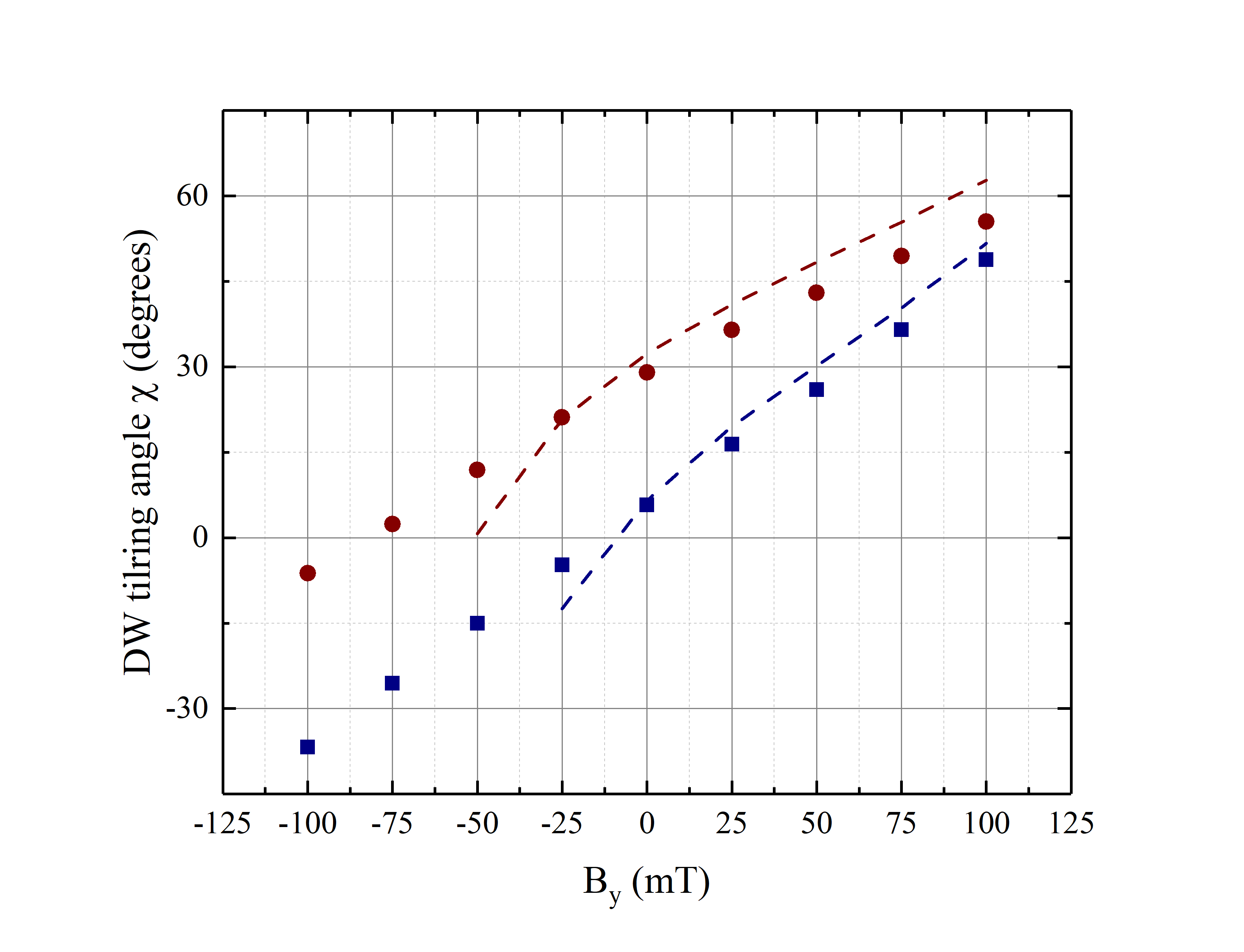}}
	\subfloat{\includegraphics[scale=0.4]{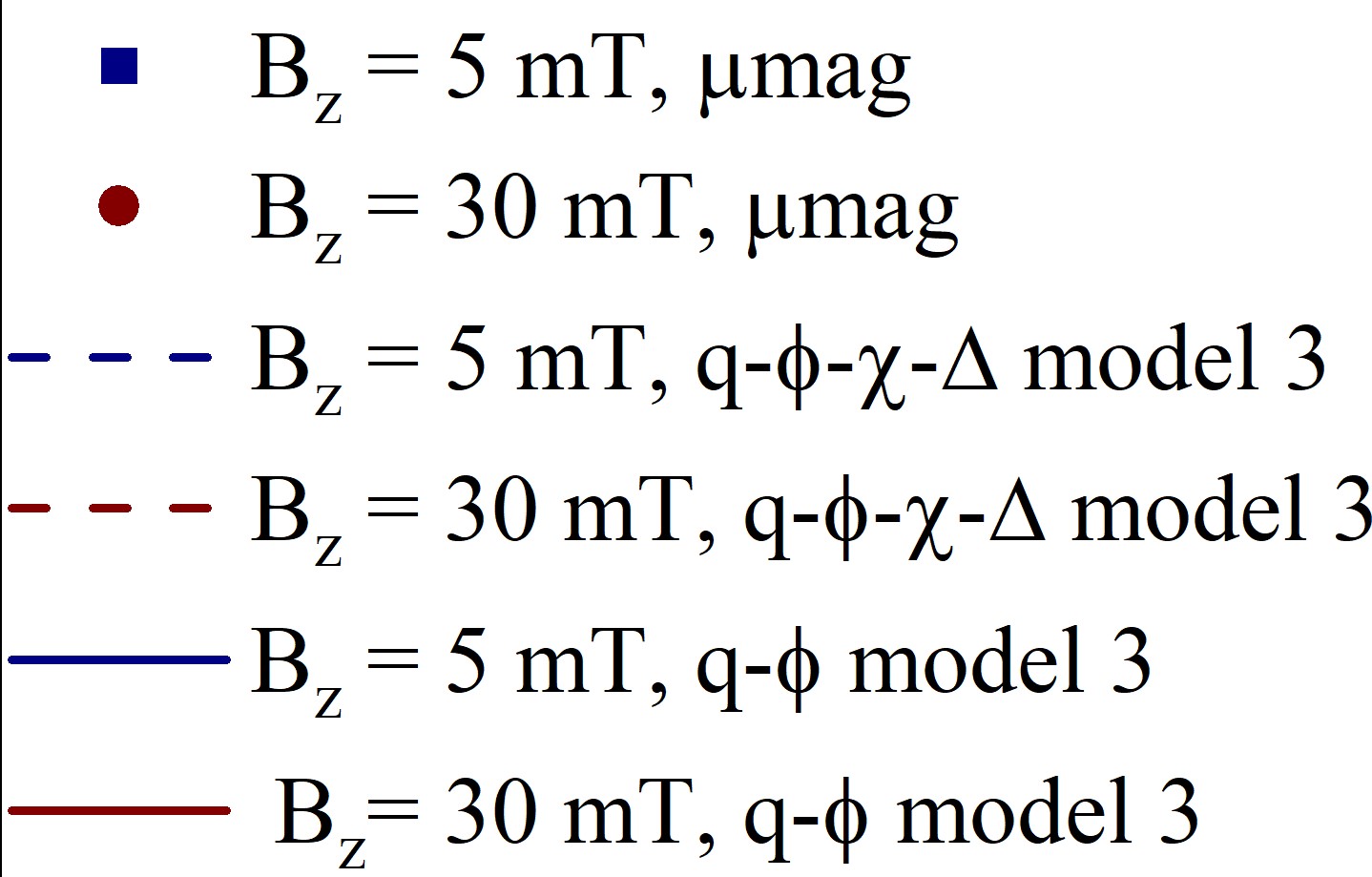}}
	\caption{Instantaneous steady state DW characteristics for field-driven DW motion in $Pt/CoFe/MgO$ with different out of plane and transverse fields applied. Only the collective coordinate models with highest accuracy in predicting the DW velocity are shown.}
	\label{fields_By_micromag}
\end{figure}

\begin{figure}
	\subfloat[DW velocity.]{\includegraphics[trim=1cm 1cm 1cm 2cm, clip=true,scale=0.32]{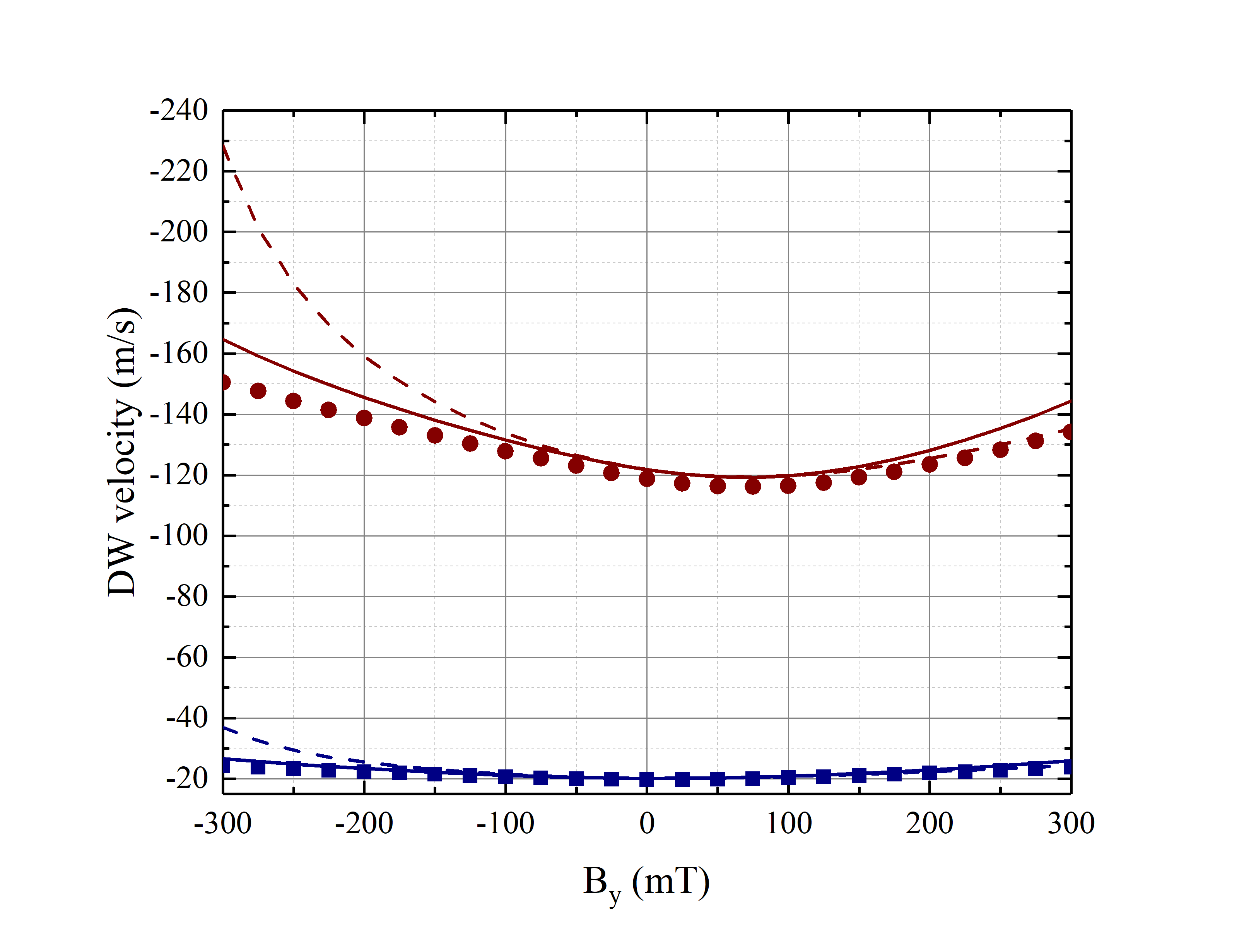}}
	\subfloat[DW width parameter ($\Delta$).]{\includegraphics[trim=1cm 1cm 1cm 2cm, clip=true,scale=0.32]{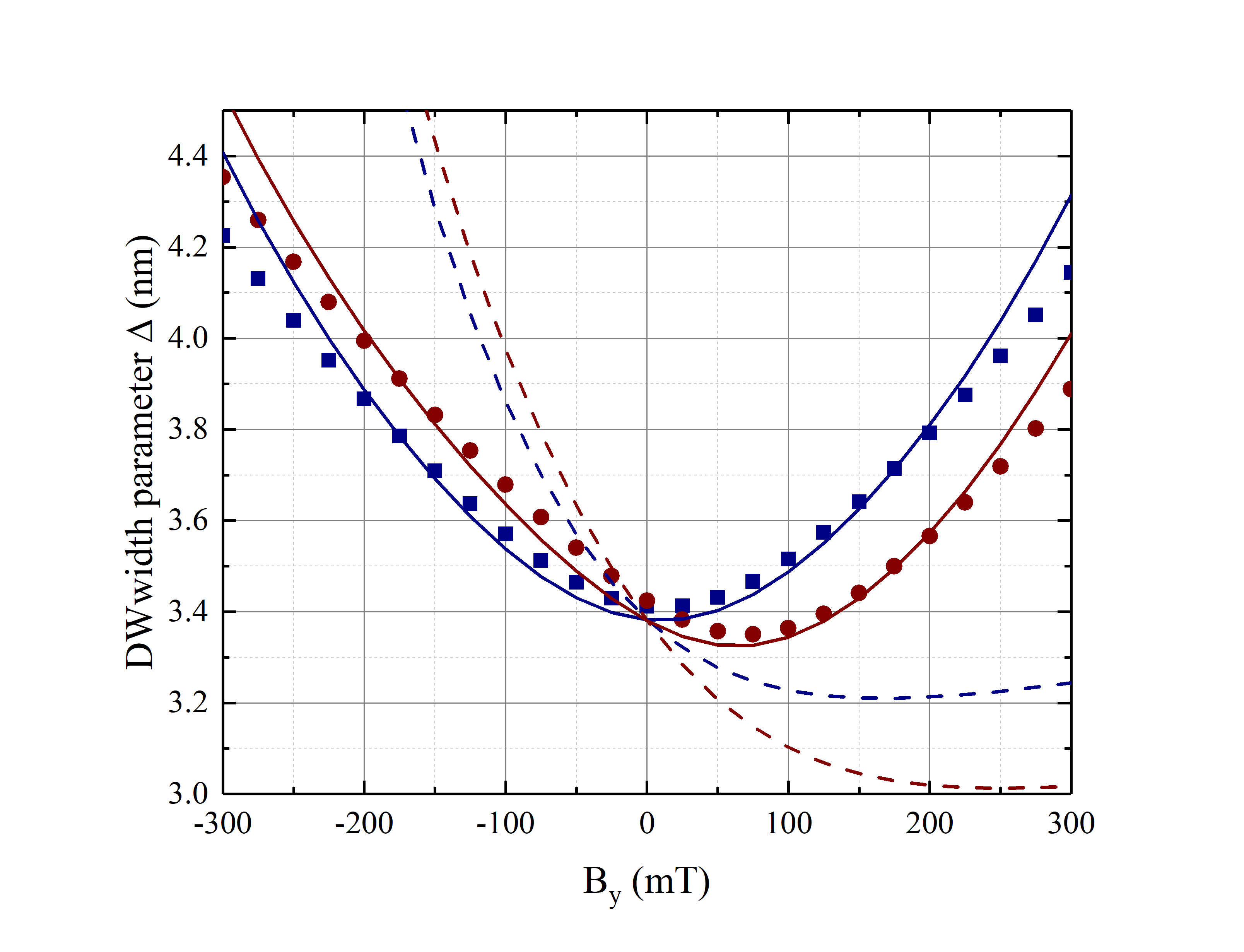}}\\
	\subfloat[$\phi-\chi$]{\includegraphics[trim=1cm 1cm 1cm 2cm, clip=true,scale=0.32]{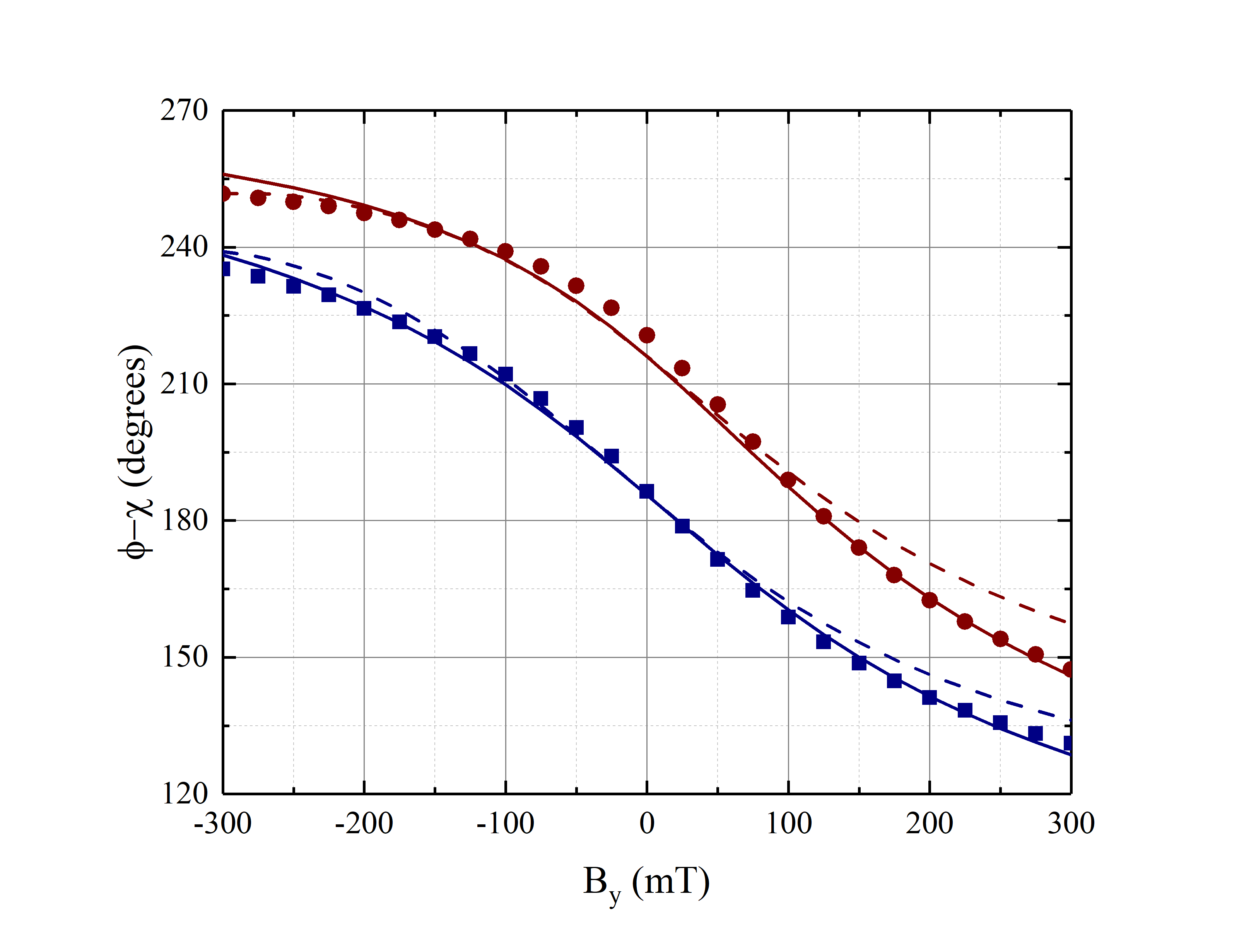}}
	\subfloat[DW magnetization angle ($\phi$).]{\includegraphics[trim=1cm 1cm 1cm 2cm, clip=true,scale=0.32]{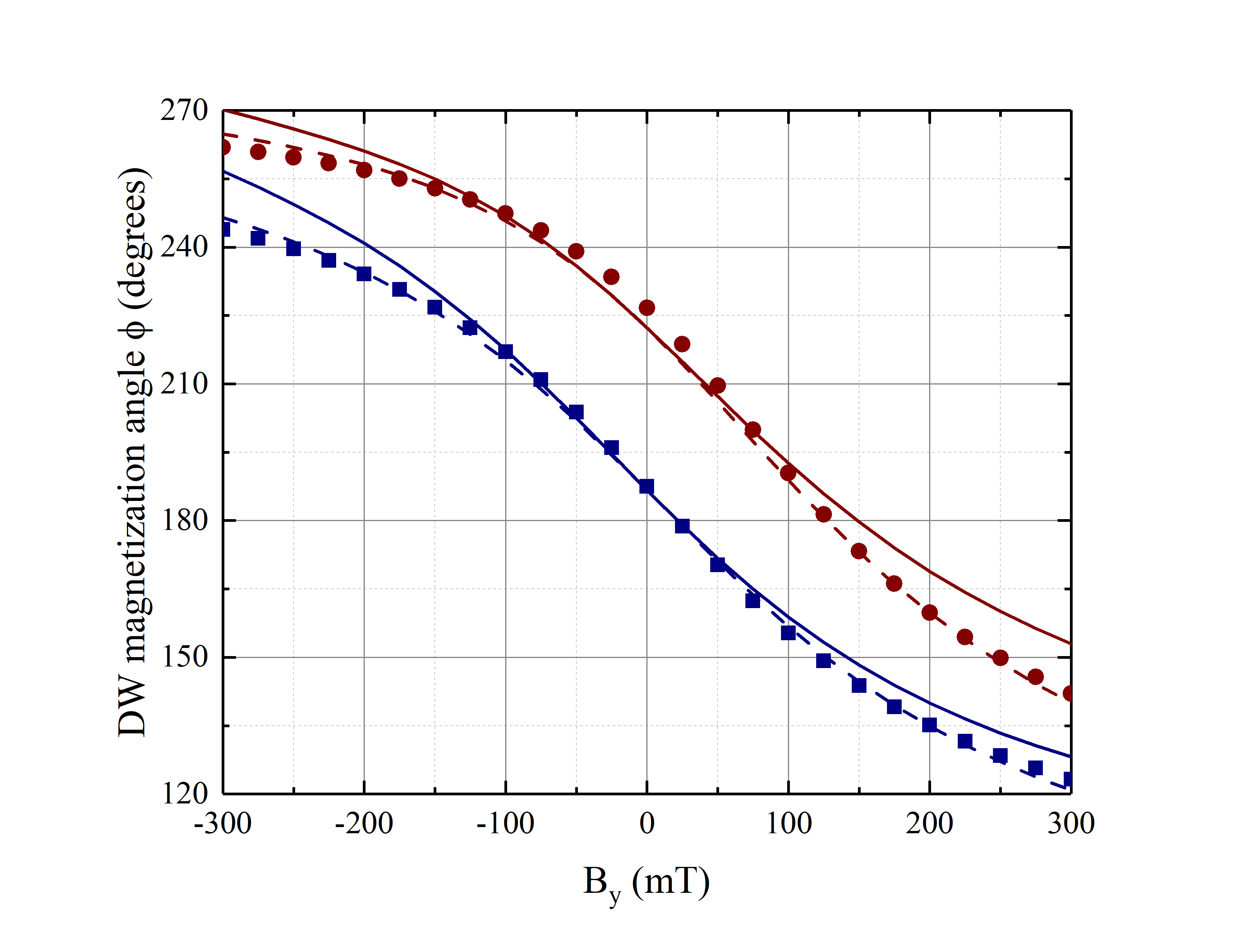}}\\
	\subfloat[DW tilting angle ($\chi$).]{\includegraphics[trim=1cm 1cm 1cm 2cm, clip=true,scale=0.32]{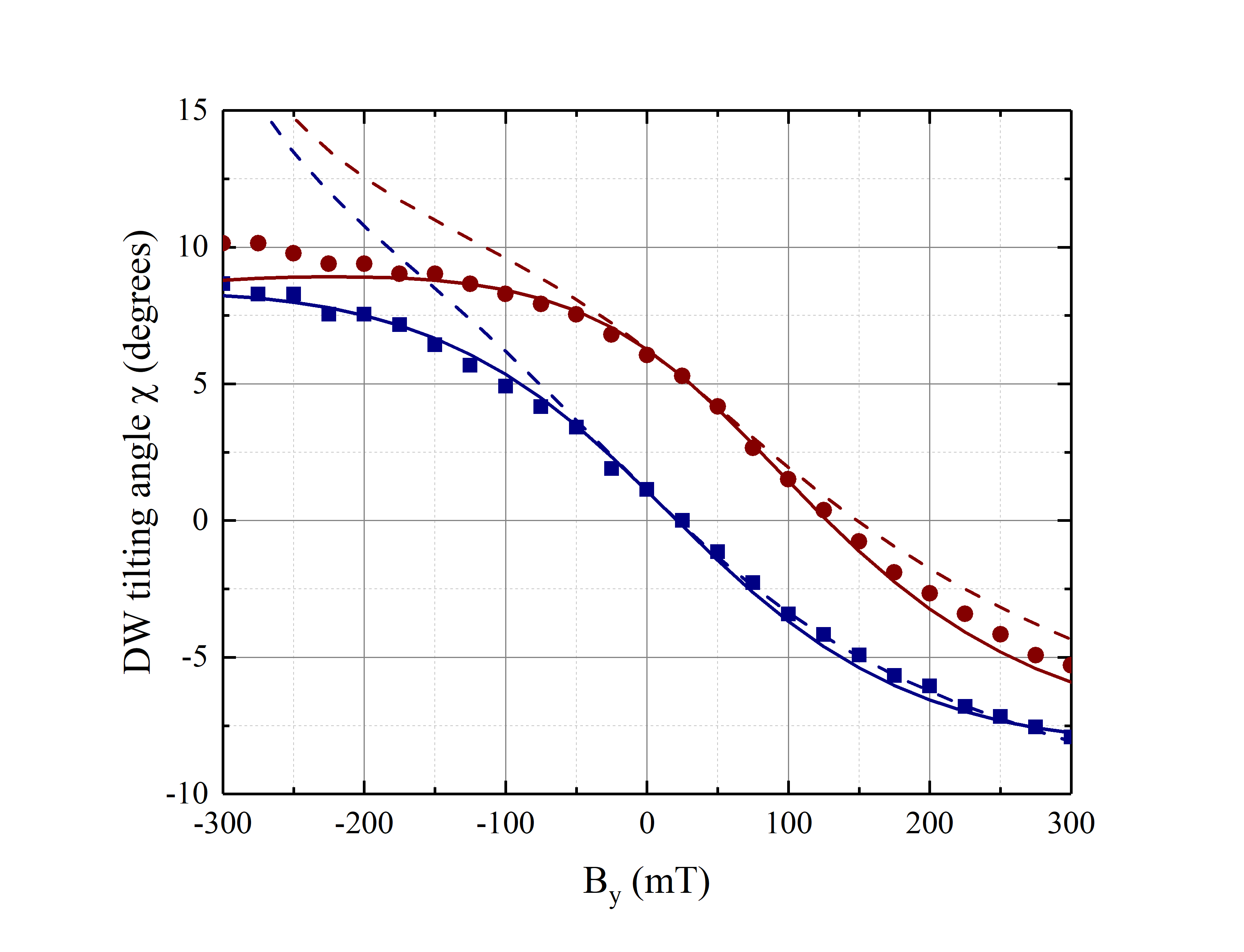}}
	\subfloat{\includegraphics[scale=0.4]{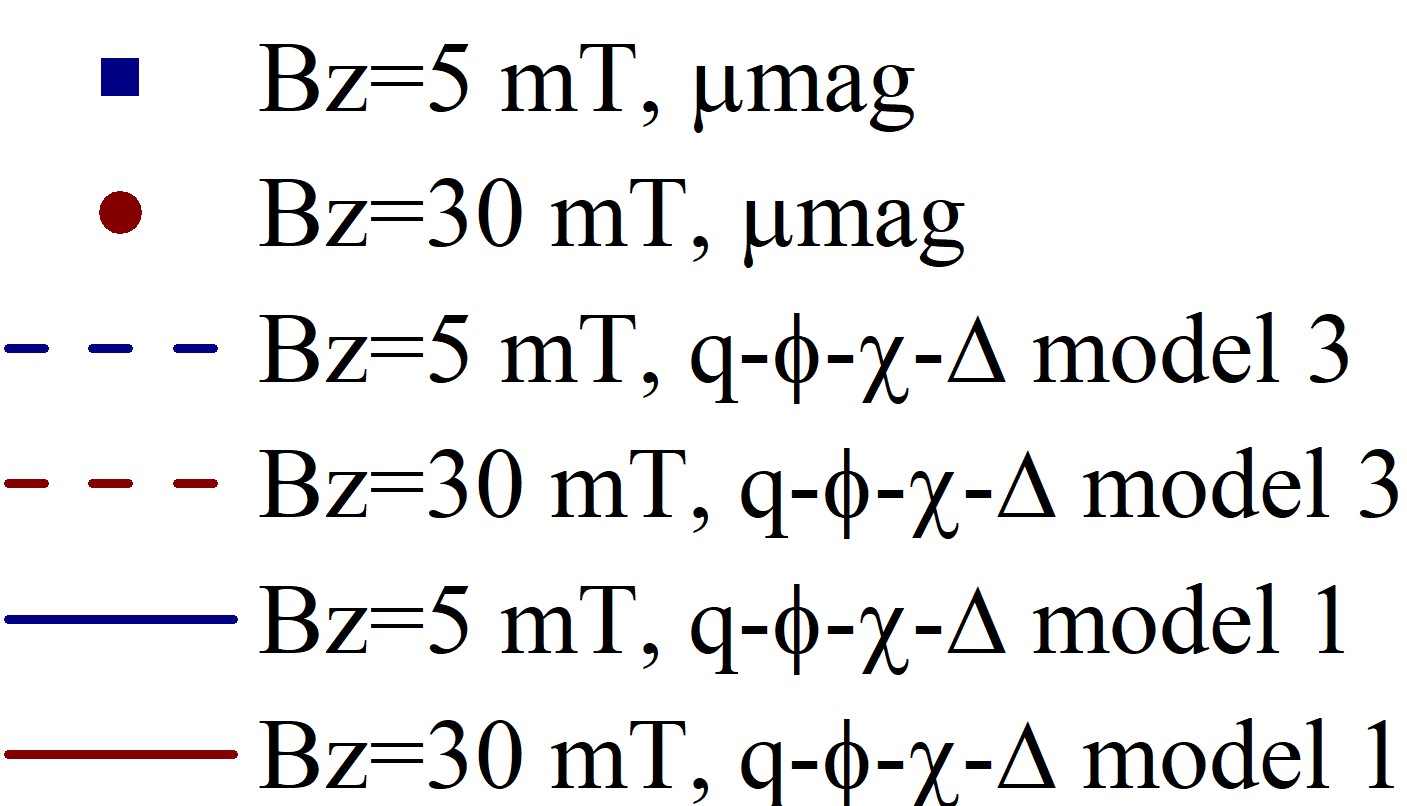}}
	\caption{Instantaneous steady state DW characteristics for field-driven DW motion in $Pt/Co/Ni/Co/MgO/Pt$ with different out of plane and transverse fields applied. Only the collective coordinate models with highest accuracy in predicting the DW velocity are shown.}
	\label{fields_By_micromag_PtCoNiCoMgOPt}
\end{figure}

\begin{figure}
	\subfloat[DW velocity.]{\includegraphics[trim=1cm 1cm 1cm 2cm, clip=true,scale=0.32]{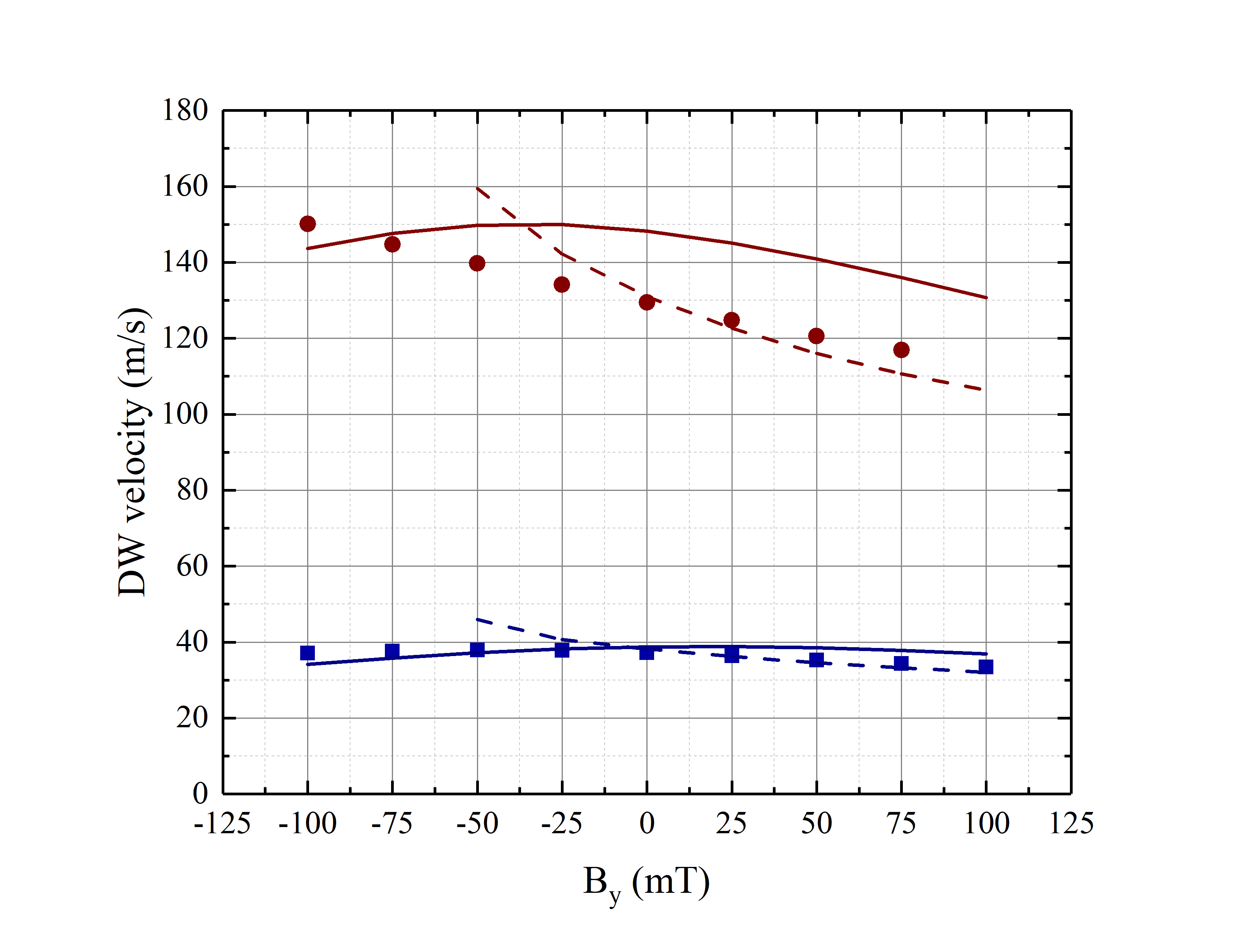}}
	\subfloat[DW width parameter ($\Delta$).]{\includegraphics[trim=1cm 1cm 1cm 2cm, clip=true,scale=0.32]{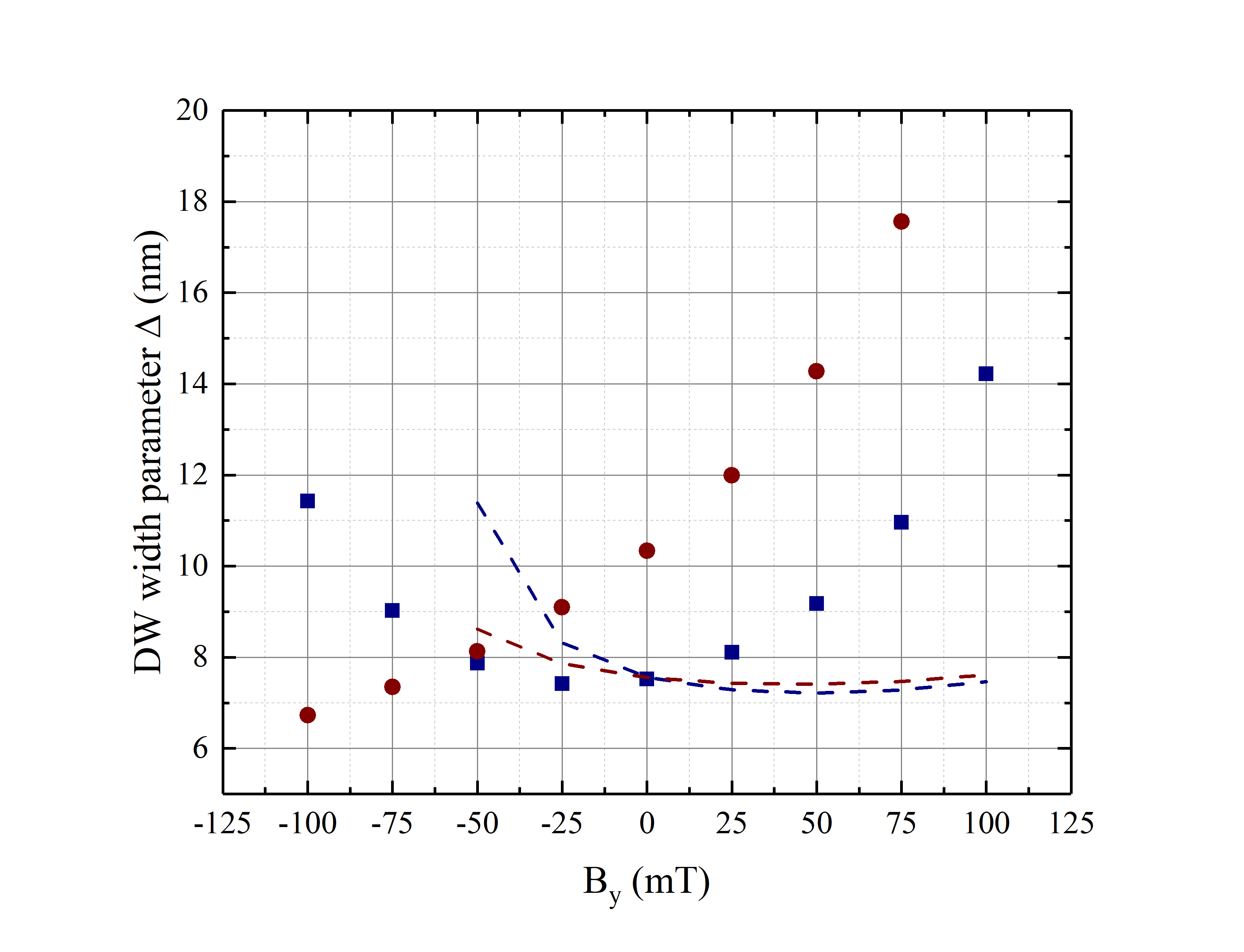}}\\
	\subfloat[$\phi-\chi$]{\includegraphics[trim=1cm 1cm 1cm 2cm, clip=true,scale=0.32]{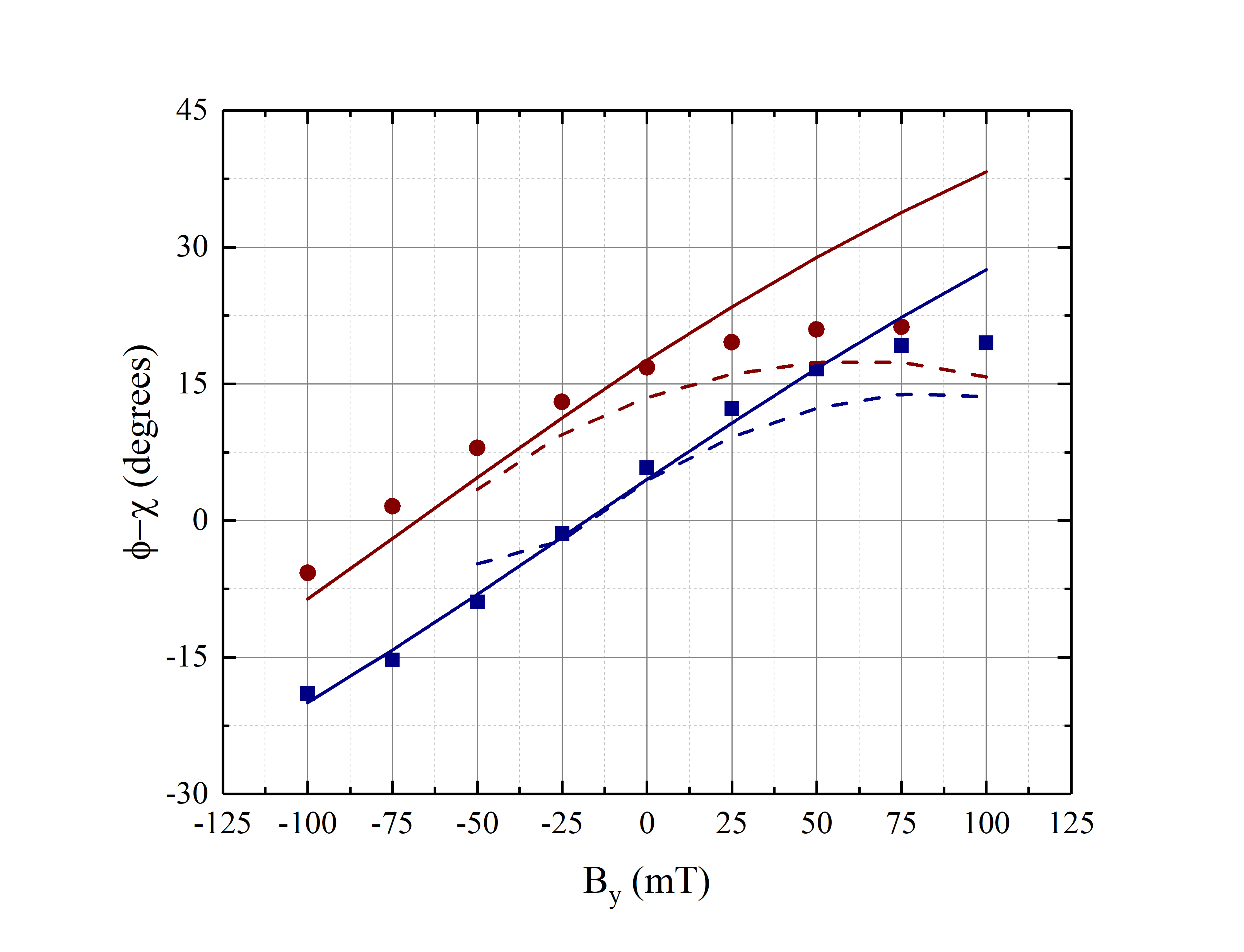}}
	\subfloat[DW magnetization angle ($\phi$).]{\includegraphics[trim=1cm 1cm 1cm 2cm, clip=true,scale=0.32]{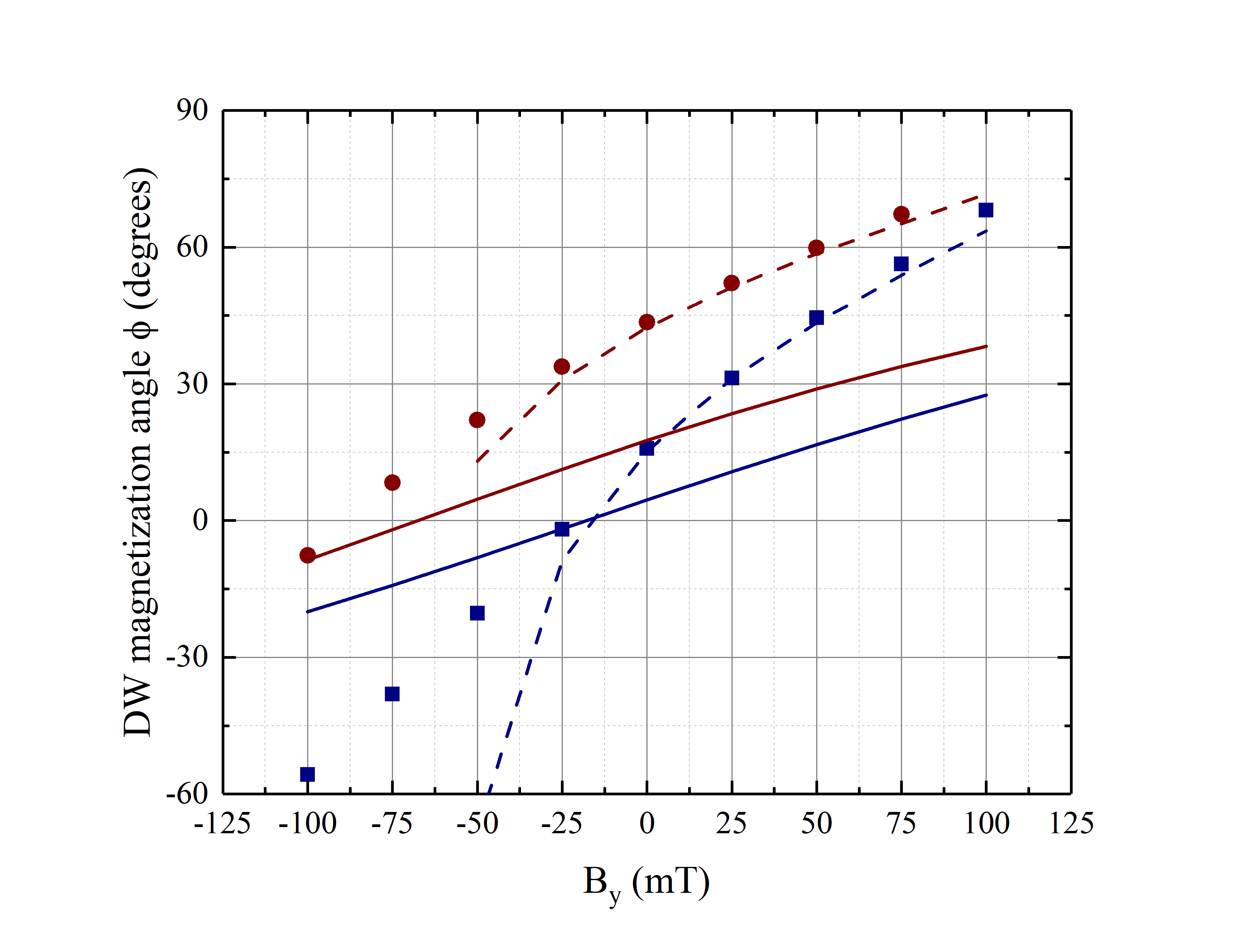}}\\
	\subfloat[DW tilting angle ($\chi$).]{\includegraphics[trim=1cm 1cm 1cm 2cm, clip=true,scale=0.32]{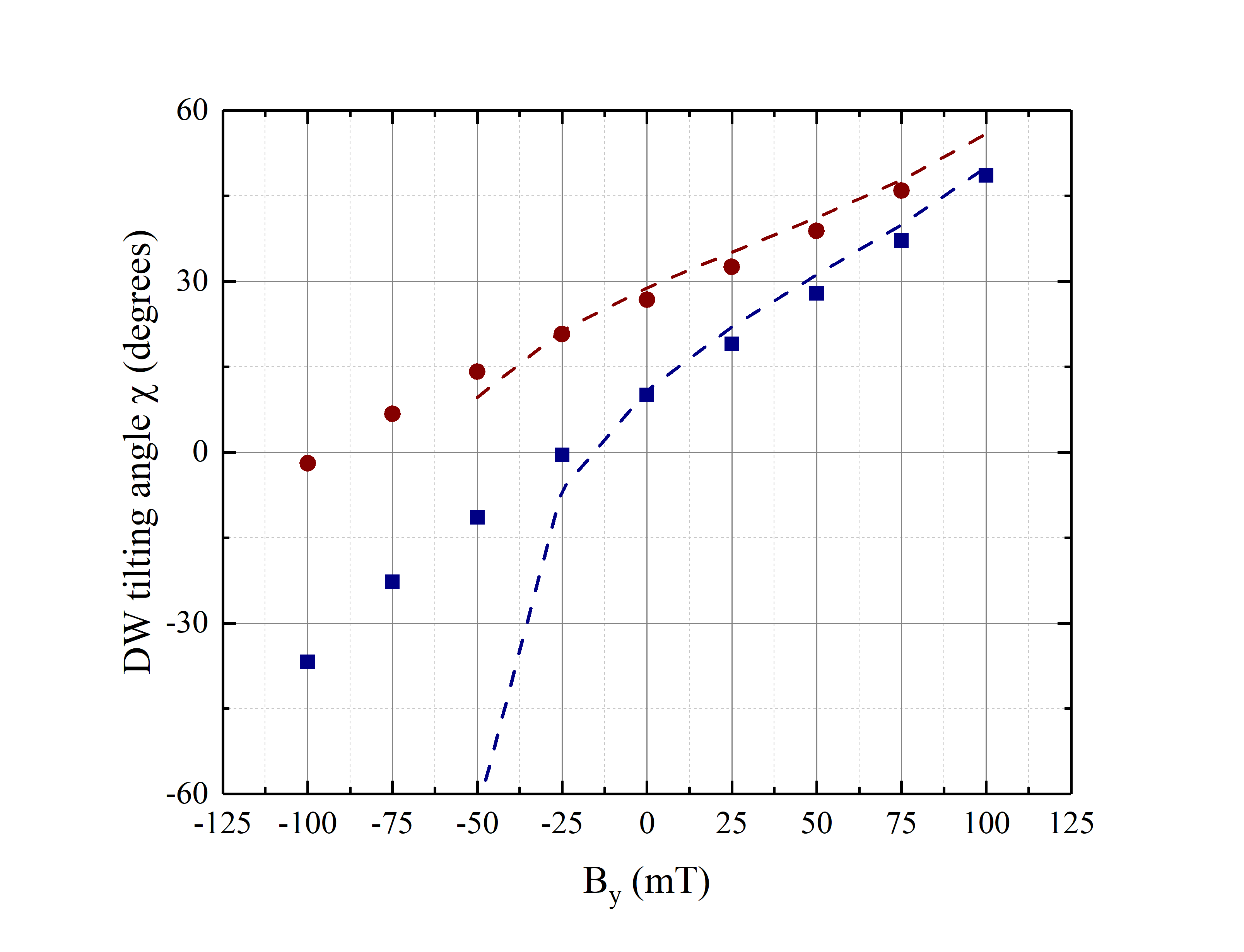}}
	\subfloat{\includegraphics[scale=0.4]{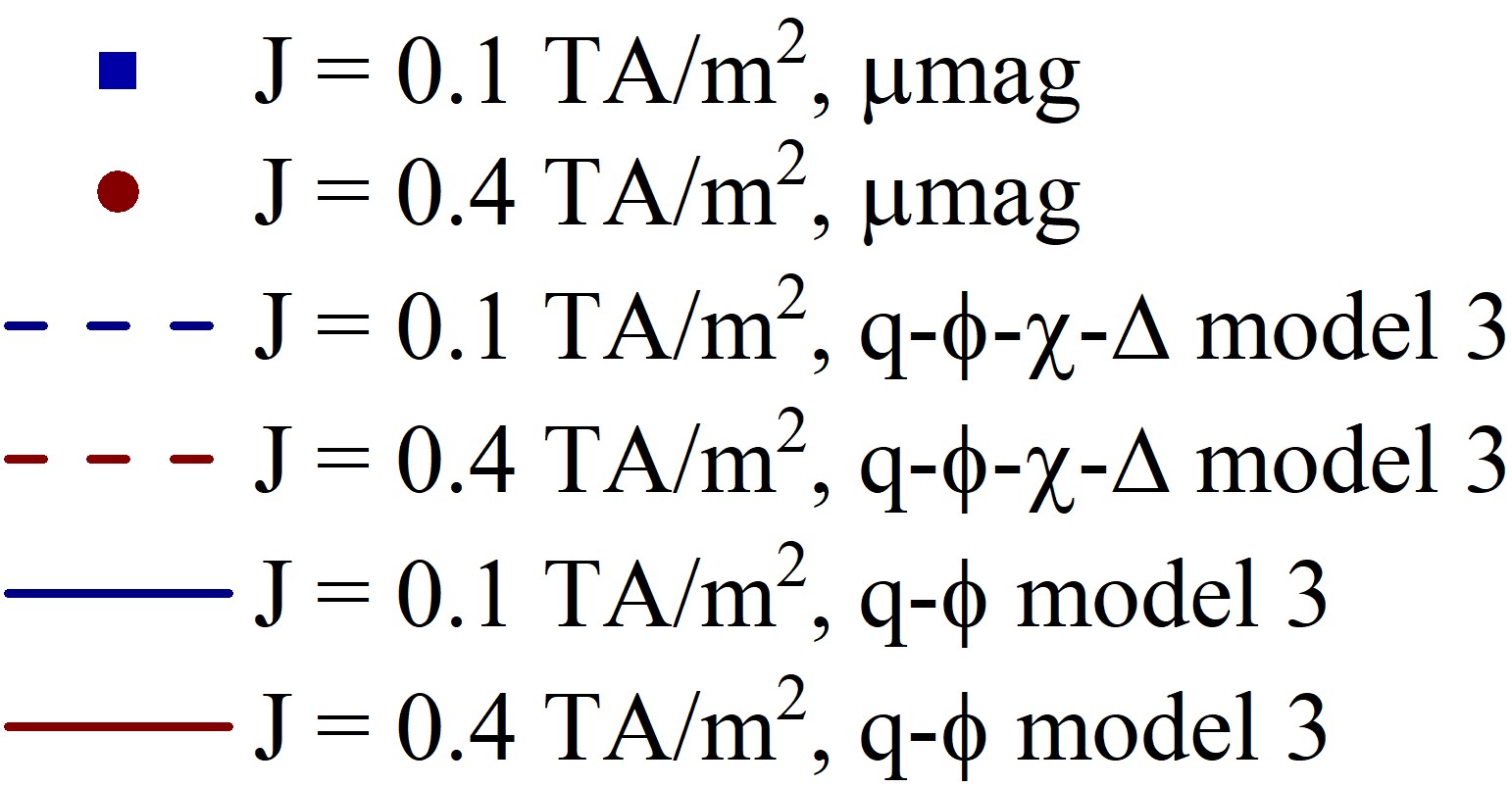}}
	\caption{Instantaneous steady state DW characteristics for SHE-driven DW motion in $Pt/CoFe/MgO$ with different currents and transverse fields applied. Model 3, model 1 unable to predict. Only the collective coordinate models with highest accuracy in predicting the DW velocity are shown.}
	\label{current_By_micromag}
\end{figure}

\begin{figure}
	\subfloat[DW velocity.]{\includegraphics[trim=1cm 1cm 1cm 2cm, clip=true,scale=0.32]{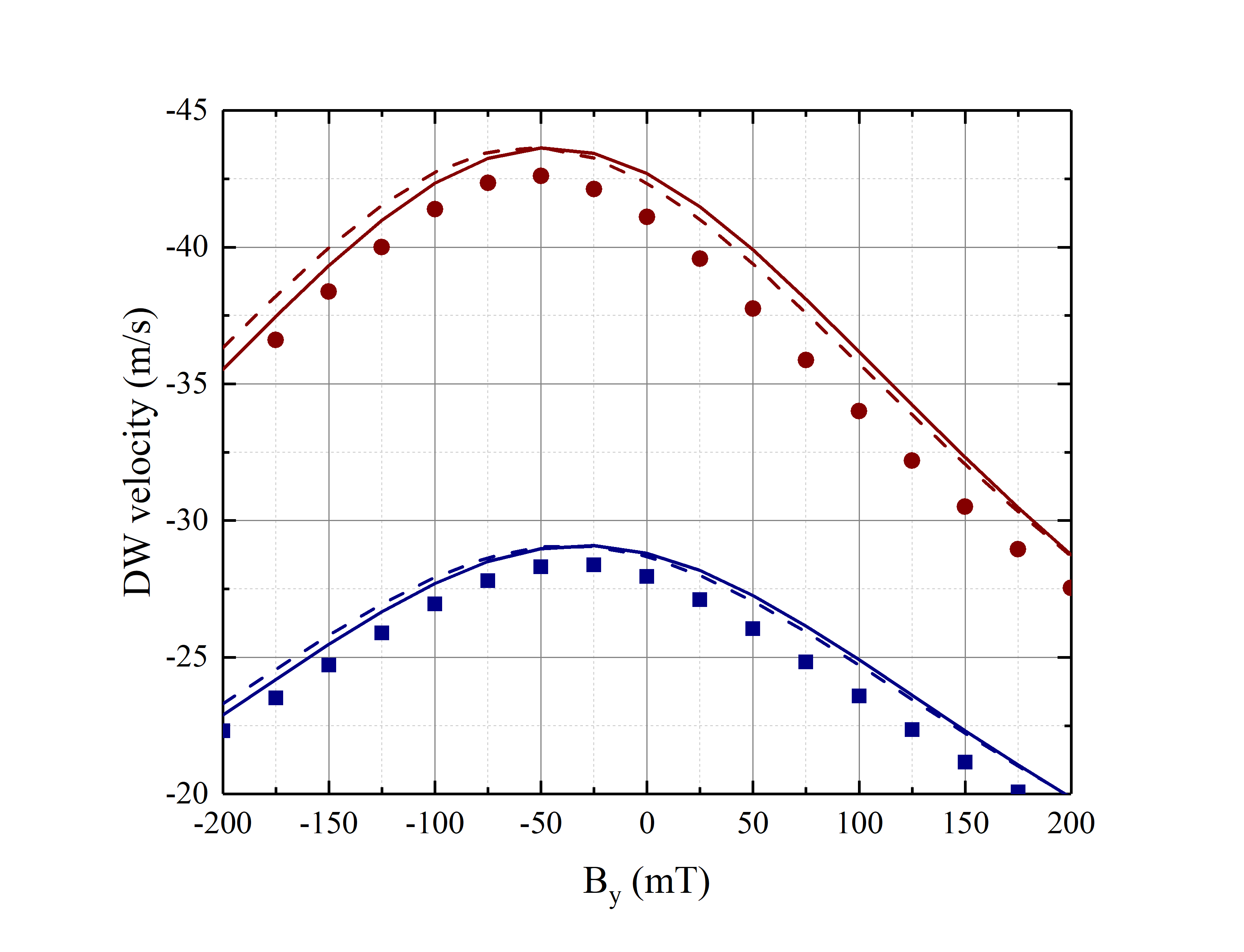}}
	\subfloat[DW width parameter ($\Delta$).]{\includegraphics[trim=1cm 1cm 1cm 2cm, clip=true,scale=0.32]{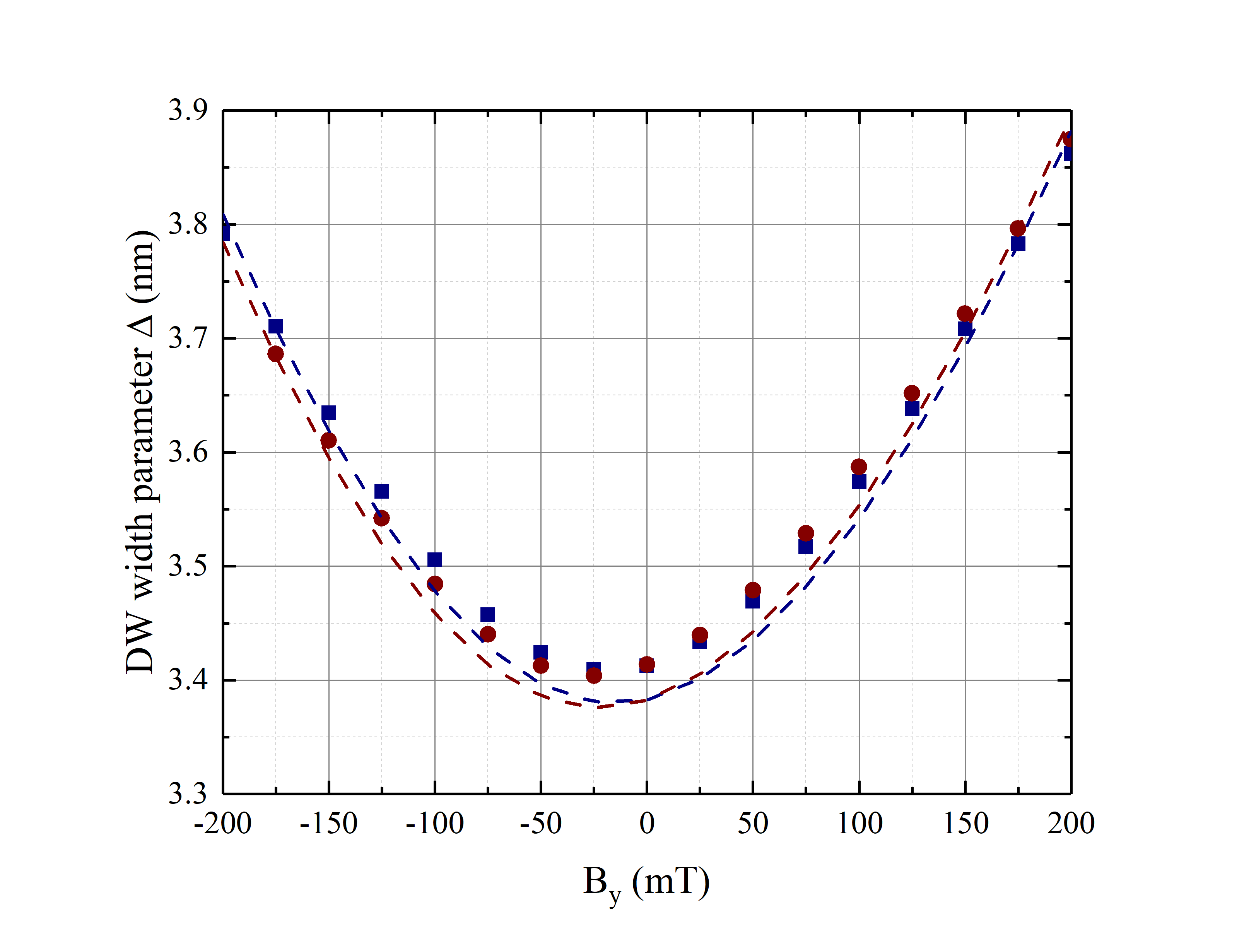}}\\
	\subfloat[$\phi-\chi$]{\includegraphics[trim=1cm 1cm 1cm 2cm, clip=true,scale=0.32]{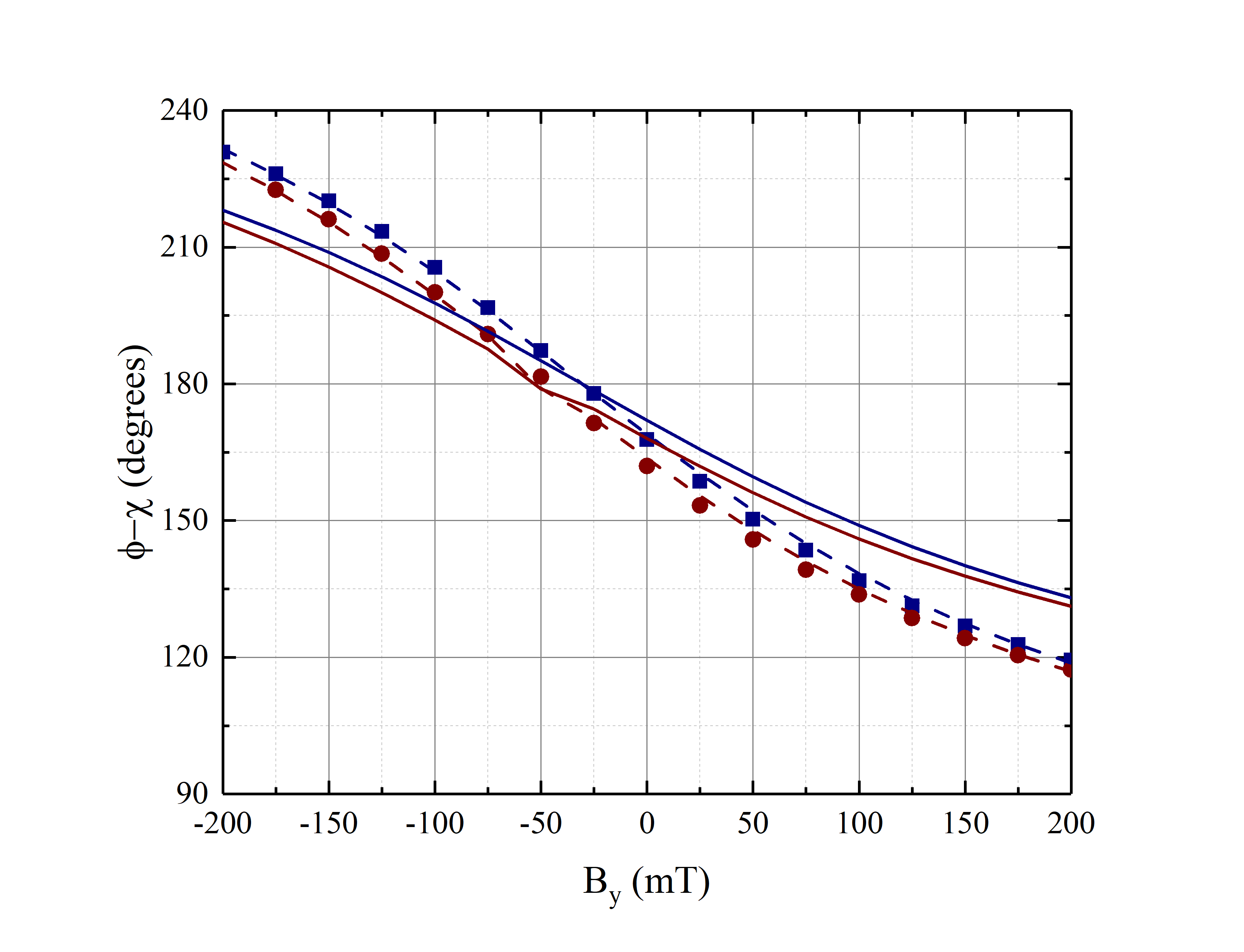}}
	\subfloat[DW magnetization angle ($\phi$).]{\includegraphics[trim=1cm 1cm 1cm 2cm, clip=true,scale=0.32]{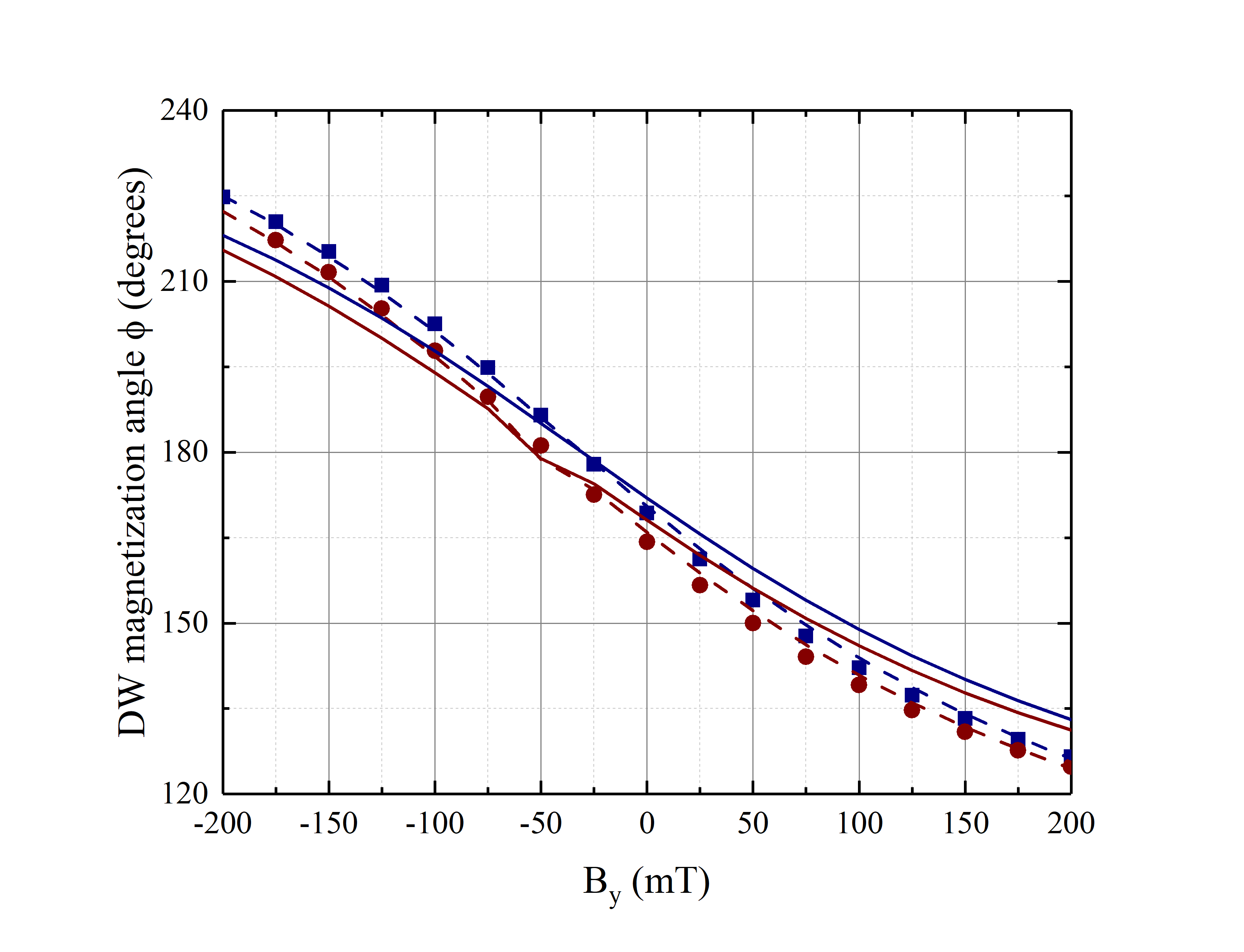}}\\
	\subfloat[DW tilting angle ($\chi$).]{\includegraphics[trim=1cm 1cm 1cm 2cm, clip=true,scale=0.32]{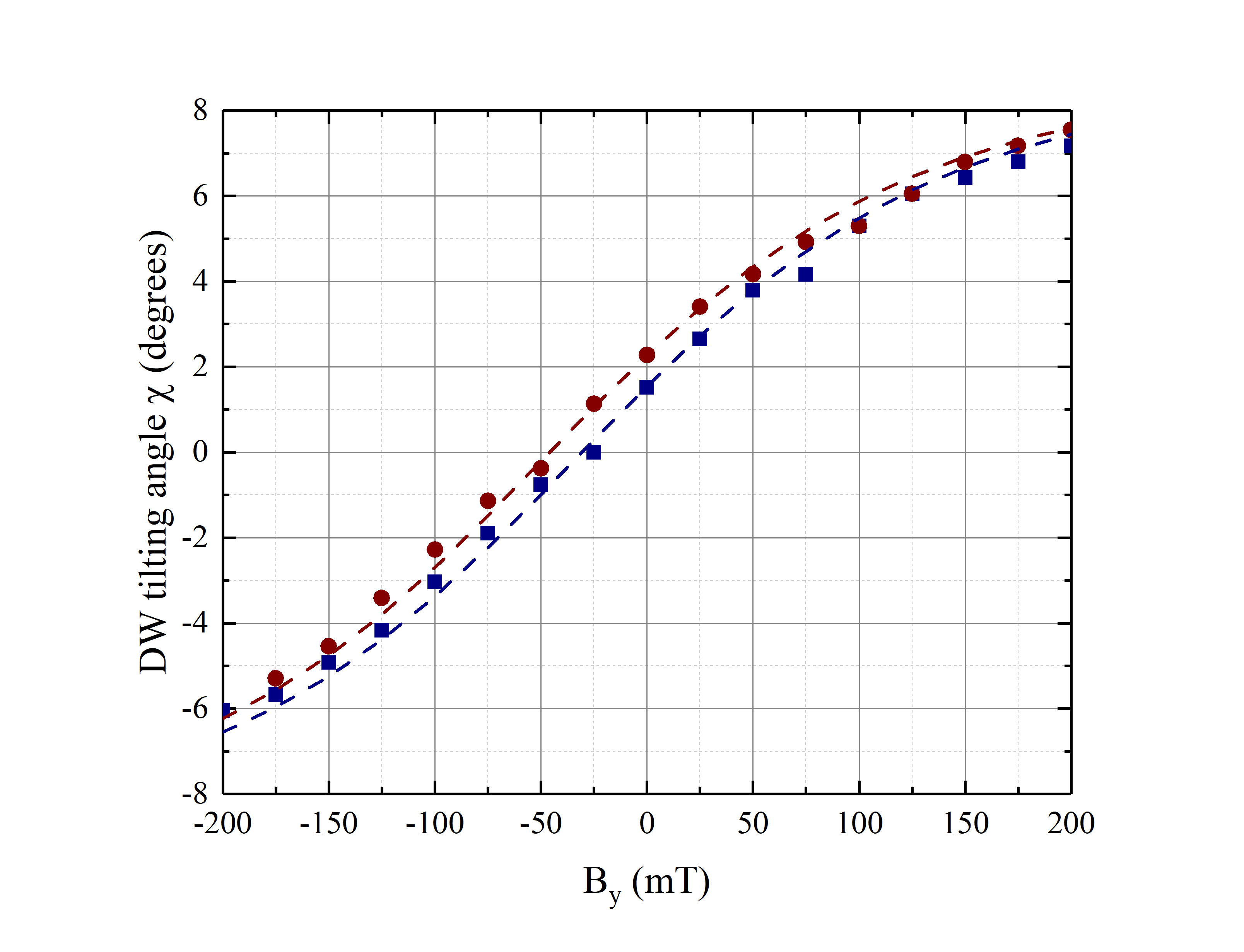}}
	\subfloat{\includegraphics[scale=0.4]{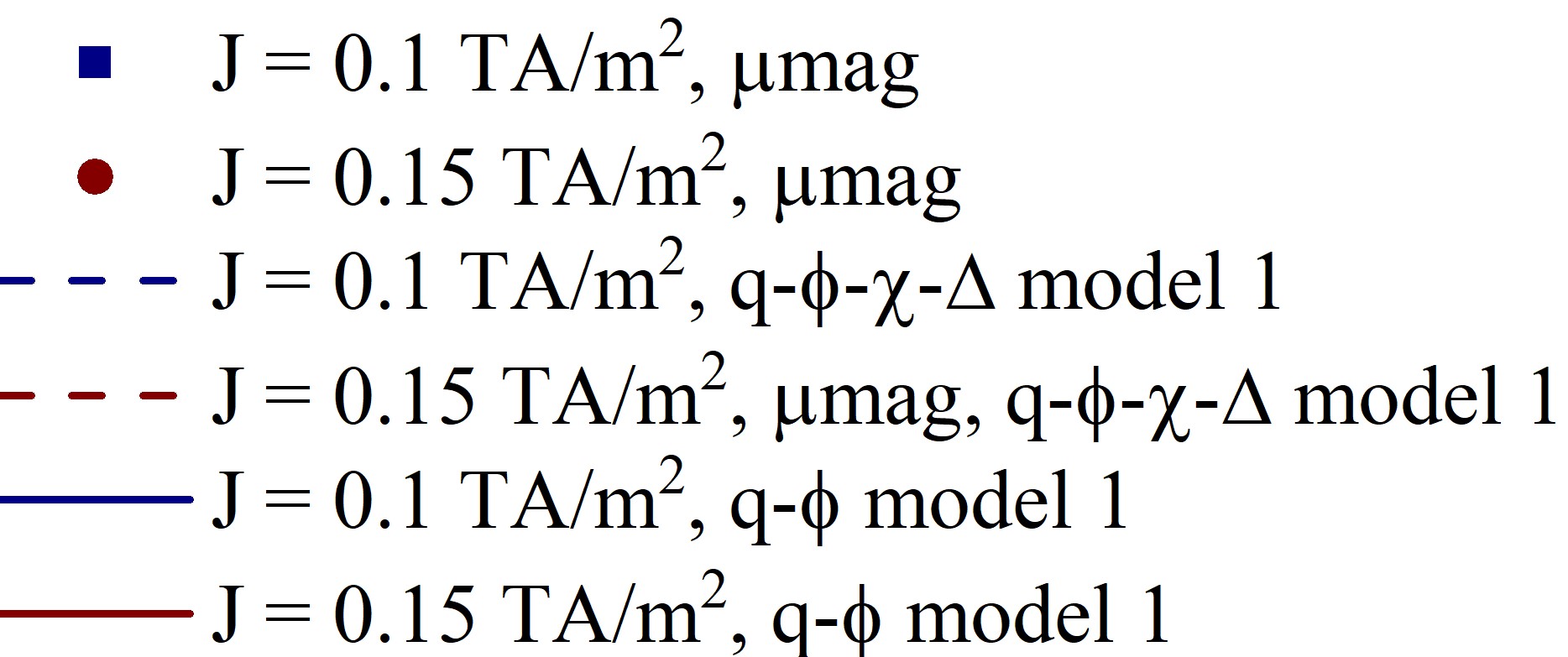}}
	\caption{Instantaneous steady state DW characteristics for SHE-driven DW motion in $Pt/Co/Ni/Co/MgO/Pt$ with different currents and transverse fields applied. Only the collective coordinate models with highest accuracy in predicting the DW velocity are shown.}
	\label{current_By_micromag_PtCoNiCoMgOPt}
\end{figure}

Figures \ref{fields_By_micromag} and \ref{fields_By_micromag_PtCoNiCoMgOPt} show the results of field-driven DW motion under the application of transverse fields in the materials under study, while figures \ref{current_By_micromag} and \ref{current_By_micromag_PtCoNiCoMgOPt} show the results of current-driven DW motion under transverse fields. 

As observed from Figures \ref{fields_By_micromag}.e, \ref{fields_By_micromag_PtCoNiCoMgOPt}.e, \ref{current_By_micromag}.e, and \ref{current_By_micromag_PtCoNiCoMgOPt}.e, the DW tilting angle changes dramatically under transverse fields with a behavior different compared to what was observed under longitudinal fields. While under longitudinal fields we only observed positive or negative tilting for a specific material, under transverse fields we can observe both types of angles; in a sense the transverse fields could be used to control the tilting of the DW. Obviously, this means that under these conditions the tilting of the DW is an important coordinate. Yet we see that the collective coordinate models are accurate in predicting the behavior of the DW, with and without the tilting included in the model.  

In the $Pt/CoFe/MgO$ sample (Figures \ref{fields_By_micromag}.c, and \ref{current_By_micromag}.c) , we can identify a critical transverse fields at which point $\phi \sim \chi \sim 0$. In the $Pt/Co/Ni/Co/MgO/Pt$ system (Figures \ref{fields_By_micromag_PtCoNiCoMgOPt},  and \ref{current_By_micromag_PtCoNiCoMgOPt}  panels c, d, and e), we instead have a point where $\chi = 0$ and $\phi = 180$. These points coincide with when the DW is fully N\'eel, with the difference in magnetization being due to the different chirality of the DW in the two systems.

\subsection{Analysis of the critical points}

In the micromagnetic simulations and collective coordinate results of Figures \ref{fields_Bx_micromag}-\ref{current_By_micromag_PtCoNiCoMgOPt}, we were able to identify several points where the DW behavior showed features that could be reproduced irrespective of material properties. These points could be used to derive further simplified forms of the DW dynamic equations.

We observed that in several cases (field- and current-driven DW motion under longitudinal and transverse fields in $Pt/CoFe/MgO$, and field-driven case under transverse fields and field- and current-driven cases under transverse field for $Pt/Co/Ni/Co/MgO/Pt$), an in-plane field exists for which $\phi -\chi \sim 0$ or $180^\circ$. In addition, under longitudinal fields this point is independent of the drive interaction. We observed in Figures \ref{fields_Bx_micromag}-\ref{current_By_micromag_PtCoNiCoMgOPt} that the CCMs can accurately predict the Dw behavior at this point.

From a CCM perspective, this is the point where the DW has a fully N\'eel like structure. This means the contributions from the DMI and magnetostatic terms become zero, and we have:

\begin{equation}
\left(I_2+ \alpha^2 \frac{I_1 I_4}{I_2} \right) I_1  \frac{\dot{q}}{\mu_0 \gamma p_w \Delta}  = 
\begin{cases} 
\alpha  \left(I_4 \frac{H_z}{\cos\chi} - \frac{I_3 I_4}{I_2} H_{SL} \left[\tan\chi u_{SOT,x} -   u_{SOT,y}\right]\right) \\
+ \left(\alpha \beta \frac{I_1 I_4}{I_2} + I_5 \right) \frac{u}{p_w \Delta} + I_6  \left[H_x \tan\chi - H_y \right]
&\mbox{for } \phi-\chi \equiv 0 \\ 
\\
\alpha  \left(I_4 \frac{H_z}{\cos\chi} + \frac{I_3 I_4}{I_2} H_{SL} \left[\tan\chi u_{SOT,x} -  u_{SOT,y}\right]\right) \\
+ \left(\alpha \beta \frac{I_1 I_4}{I_2} + I_5 \right) \frac{u}{p_w \Delta} - I_6  \left[H_x \tan\chi - H_y \right] 
&\mbox{for } \phi-\chi \equiv 180^\circ \\ 
\end{cases}
\label{vel_phi_n_chi_0_1}
\end{equation}

Under the application of transverse fields, the equations become slightly more simplified, as at the same time $\chi = 0$, which yields: 

\begin{equation}
\left(I_2+ \alpha^2 \frac{I_1 I_4}{I_2} \right) \frac{I_1}{\mu_0 \gamma p_w \Delta} \dot{q} = 
\begin{cases} \alpha  \left(I_4 H_z + \frac{I_3 I_4}{I_2} H_{SL}  u_{SOT,y}\right)\\
 + \left(\alpha \beta \frac{I_1 I_4}{I_2} + I_5 \right) \frac{u}{p_w \Delta} - I_6  H_y 
&\mbox{if } \phi-\chi \equiv 0 \, (D < 0) \\ 
\\
\alpha  \left(I_4 H_z - \frac{I_3 I_4}{I_2} H_{SL} u_{SOT,y} \right) \\
+ \left(\alpha \beta \frac{I_1 I_4}{I_2} + I_5 \right) \frac{u}{p_w \Delta} + I_6  H_y 
&\mbox{if } \phi-\chi \equiv 180^\circ \, (D > 0) \\ 
\end{cases}
\label{vel_phi_n_chi_0_2}
\end{equation}

Equations \ref{vel_phi_n_chi_0_1} and \ref{vel_phi_n_chi_0_2} are thought-provoking, as they connect measured properties of the DW (DW velocity and tilting) to parameters arising from material properties such as the DW width parameter. These equations can be used to measure specific properties of the DW. In an experimental setting, first a transverse field should be identified at which the DW tilting is negligible; in this condition one may assume based on our results that $\phi \sim \chi \sim 0$. Using field-driven DW motion measurements, equation \ref{vel_phi_n_chi_0_2} may be used to measure the DW width parameter, which can in turn help estimate the exchange constant through $\Delta = \sqrt{\frac{A}{K}}$. In a current-driven case, the same equation could be used to estimate the SHE angle.

 
In current-driven DW motion a longitudinal fields, we identified a longitudinal field for which DW velocity is zero. In $Pt/CoFe/MgO$, this field was about $B_x \sim -200 \, mT$, while in $Pt/CoFe/MgO$ it was $B_x \sim 200 \, mT$. We had shown in our previous work that this field is related to the DMI strength \cite{NAS-17}. Under these conditions, the DW velocity equation simplifies to:

\begin{equation}
\begin{split}
	\alpha \frac{I_3 I_4}{I_2} H_{SL} \left[\cos\phi u_{SOT,y} - \sin\phi u_{SOT,x}\right] + \left(\alpha \beta \frac{I_1 I_4}{I_2} + I_5 \right) \frac{u}{p_w \Delta} \cos\chi &=  \frac{1}{2} I_4  M_s \left(N_x - N_y \right) \sin 2(\phi-\chi)\\
	& + I_3 \frac{D}{\mu_0 M_s p_w \Delta} \sin(\phi - \chi) \\
	&- I_6  \left[H_x \sin\phi - H_y \cos\phi \right] 
\end{split}
\end{equation}

Looking at Figures \ref{current_Bx_micromag} and \ref{current_Bx_micromag_PtCoNiMgOPt}, we also observe that at this in-plane field $\phi$ and $\chi$ seem to be independent of the drive interaction, and $\phi \sim 90^\circ$. Using this assumption, we have:
\begin{equation}
-\alpha \frac{I_3 I_4}{I_2} H_{SL}  u_{SOT,x} + \left(\alpha \beta \frac{I_1 I_4}{I_2} + I_5 \right) \frac{u}{p_w \Delta} \cos\chi =  \frac{1}{2} I_4  M_s \left(N_x - N_y \right) \sin 2\chi + I_3 \frac{D}{\mu_0 M_s p_w \Delta} \cos\chi - I_6  H_x 
\label{vel_zero_final}
\end{equation}

Equation \ref{vel_zero_final} could be used to measure DMI under conditions which the DW is stationary under applied currents. Plugging observations from the micromagnetic simulations into a two coordinate form of model 3 (and assuming $J = 0.1 \, TA/m^2$), we predict a DMI strength of $D = -1.1 \, mJ/m^2$ for $Pt/CoFe/MgO$ and $D = 0.57 \, mJ/m^2$ for $Pt/Co/Ni/Co/MgO/Pt$, which are very close to the values used in the micromagnetic simulations.


In the $Pt/Co/Ni/Co/MgO/Pt$ sample, in field-driven DW motion ($B_z = 30 \, mT$), we observed initiation of Walker Breakdown at $B_x = 50 \, mT$ and cessation of this behavior at $B_x = 325 \, mT$. This scenario could be studied using our CCMs. Assuming small tilting for the DW (which is valid in this case), we can simplify the steady state equation as: 
\begin{equation}
\begin{split}
\cos\phi &= \frac{I_2 H_z + I_3 H_{SL} \cos\phi}{\alpha \frac{I_4^2}{I_2} M_s \left(N_y - N_x \right) \sin\phi}    
+ \alpha \frac{ \frac{ I_3 I_4}{I_2}  \frac{D}{\mu_0 M_s p_w \Delta}  + \frac{I_4 I_6}{I_2} \left(H_y \cot\phi - H_x\right) }{\alpha \frac{I_4^2}{I_2} M_s \left(N_y - N_x \right)} \\
& = \frac{1}{2} \left[\left(\frac{I_2}{I_4}\right)^2 \frac{H_z}{H_w \sin\phi} + \frac{I_2 I_3}{I_4^2} \frac{H_{SL}}{H_w} \cot\phi + \alpha \frac{I_3}{I_4} \frac{H_{DMI}}{H_w} - \alpha \frac{I_6}{I_4} \frac{H_x}{H_w} + \alpha \frac{I_6}{I_4} \frac{H_y}{H_w} \cot\phi \right]
\end{split}
\label{WB}
\end{equation}
where $H_w$ is the conventional Walker Breakdown field \cite{SCH-74}, and $H_{DMI}$ is the DMI field. Walker Breakdown happens when the right side of the equation above is larger than 1 or smaller than -1. While in systems without DMI, only the drive interaction and the demagnetizing field played a role in this solution, in a system with DMI and in-plane fields additional terms are introduced; the relevant strength of these terms compared to each other determined whether Walker breakdown will take place or not. Note that magnetocrystalline anisotropy plays a role in this through the $I_i$ values, as these values depend on $\theta_c$ which in turn depends on $K_u$.

Finally, we also observed in-plane fields that led to switching of the system through the elastic extension of the DW. The threshold for this switching field seems to relate to the canting angle reaching $\theta_c = 45 \deg$ at which point the DW will prefer not to maintain a rigid structure and aligning completely with the external field is preferred in the domains. This leads to $H_{x,s} = \frac{\sqrt{2}}{2}\left[\frac{2 K_u}{\mu_0 M_s} + M_s (N_x - N_z)\right]$ for the longitudinal switching field and $H_{y,s} = \frac{\sqrt{2}}{2} \left[\frac{2 K_u}{\mu_0 M_s} + M_s (N_y - N_z)\right]$ for transverse fields. These equations are expected to over-predict the switching field, as they do not take into account edge effects in the system and the effect of the drive interaction. For the $Pt/CoFe/MgO$ system under study in this manuscript, $B_{x,s} \approx \pm 354 mT$ which is only $25-50 mT$ higher than the field at which the system switched in micromagnetic simulations for positive longitudinal fields. For negative longitudinal fields, switching could not be observed due to the nucleation of a new DW. The transverse switching field was found to be $B_{y,s} \approx \pm 359 mT$ for $Pt/CoFe/MgO$ depending on the width of the system; however, we observed elongations in the DW prior to reaching such high fields, albeit these elongations were seen in conjunction with translational motion of the DW. The nature of these elongations and their modeling is beyond the scope of this work, as our CCMs assume the DW is a rigid object. We did not observe any of these effects for the $Pt/Co/Ni/Co/MgO/Pt$ cases; we verified that the switching field for this sample under both longitudinal and transverse fields is about $\pm 1.48 T$, well above the in-plane field values studied. The difference between $Pt/CoFe/MgO$ and  $Pt/Co/Ni/Co/MgO/Pt$ can be attributed to the higher uniaxial magnetic anisotropy of this system.



\pdfcomment{Compare predictinos from these equations tom icromagnetics}

\subsection{Selecting the Right CCM}
The results of the micromagnetic simulations presented in  Figures \ref{fields_Bx_micromag}-\ref{current_By_micromag_PtCoNiCoMgOPt} highlighted the importance of using the right CCM when studying different systems.

First, we found that canting in the domains should be included in the model only if canting in the domains is larger than about 10$^\circ$; otherwise its inclusion does not add value to the models and can overcomplicated the case being studied. Hence, we recommend using the canted models only when $\theta_c > 5^\circ$.

We also found that ansatz 3 is more applicable without the $\Delta$ degree of freedom, as it does not predict this parameter correctly and seems to be of a more rigid nature than the Bloch profile. This is understandable from a modeling perspective, as $\Delta$ in a way determines the transition from DW to domain, and canting impacts the domain's structure.

Finally, most of our models are adept at predicting the right tilting and magnetization angles at the critical in-plane fields identified. As such, use of these critical points when trying to identify material properties from the collective coordinate models is recommended.

Overall, it seems that when studying the velocity of the domain wall under longitudinal fields, use of the $q-\phi$ form of model 3 is sufficient, while under transverse fields or other cases where predicting the DW tilting is of interest, the $q-\phi-\chi$ form of model 3 or the $q-\phi-\chi-\Delta$ form of model 2 should be used.

\pdfcomment{What if you apply such a high field that the system is already in WB in the absence of an in-plane field; seems elastic behaviour happens when phi-chi-90}

\FloatBarrier

\section{Conclusion}
In this paper, we studied DW motion in PMA materials with DMI under the application of in-plane fields. We showed how the application of moderate in-plane fields could change the dynamics of domain walls by adjusting the internal structure of the DW (magnetization and DW width) along with the tilting of the DW. 

A new extended collective coordinate model was introduced taking into account the effect of canting in the domains and was compared to other models present in the literature. Canting was found to play an important role in some systems (depending on their anisotropy), and is a parameter that needs to be factored in any calculations involving collective coordinate models under in-plane fields. We observed that a two coordinate $q-\phi$ model including the effect of canting through a canted ansatz would suffice for studying the DW velocity, while a more complex $q-\phi-\chi-\Delta$ model using the Bloch profile should be used when studying the DW tilting angle is of interest.

Several critical in-plane fields were identified, under which the DW behaves in a predictable way. These critical fields simplify the collective coordinate models into simple analytical solutions which connect material properties to measurable DW features and could be used in estimating specific features of the materials under study.

\FloatBarrier

\section*{Acknowledgements}
This study was conducted as part of the Marie Currie ITN WALL project, which has received funding from the European Union's Seventh Framework Programme for research, technological development and demonstration under grant agreement no. 608031.
The work by E. M. was also supported by project MAT2014-52477-C5-4-P from the Spanish government, and projects SA282U14 and SA090U16 from the Junta de Castilla y Leon.

\bibliography{ref_old}
\bibliographystyle{unsrtnat}




\end{document}